\documentclass[apjl]{emulateapj}
\usepackage{comment}
\usepackage{ifthen}

\newcommand{\forloop}[5][1]%
{%
\setcounter{#2}{#3}%
\ifthenelse{#4}%
	{%
	#5%
	\addtocounter{#2}{#1}%
	\forloop[#1]{#2}{\value{#2}}{#4}{#5}%
	}%
	{%
	}%
}%

\newcommand{\ctbd}[1]{}

\newcommand{\lc}{light curve}
\newcommand{\lcs}{light curves}
\newcommand{\Lc}{Light curve}

\newcommand{\sband}[1]{\ensuremath{#1}}

\newcommand{\kms}{\ensuremath{\rm km\,s^{-1}}}
\newcommand{\ms}{\ensuremath{\rm m\,s^{-1}}}
\newcommand{\msd}{\ensuremath{\rm m\,s^{-1}\,d^{-1}}}

\newcommand{\gcmc}{\ensuremath{\rm g\,cm^{-3}}}
\newcommand{\ergscmsq}{\ensuremath{\rm erg\,s^{-1}\,cm^{-2}}}

\newcommand{\vsini}{\ensuremath{v \sin{i}}}
\newcommand{\feh}{\ensuremath{\rm [Fe/H]}}

\newcommand{\vmac}{\ensuremath{v_{\rm mac}}}
\newcommand{\vmic}{\ensuremath{v_{\rm mic}}}

\newcommand{\vic}{\ensuremath{V\!-\!I_C}}

\newcommand{\rsun}{\ensuremath{R_\sun}}
\newcommand{\msun}{\ensuremath{M_\sun}}
\newcommand{\lsun}{\ensuremath{L_\sun}}

\newcommand{\rstar}{\ensuremath{R_\star}}
\newcommand{\mstar}{\ensuremath{M_\star}}
\newcommand{\lstar}{\ensuremath{L_\star}}

\newcommand{\teffstar}{\ensuremath{T_{\rm eff\star}}}
\newcommand{\rhostar}{\ensuremath{\rho_\star}}
\newcommand{\loggstar}{\ensuremath{\log{g_{\star}}}}

\newcommand{\rpl}{\ensuremath{R_{p}}}
\newcommand{\mpl}{\ensuremath{M_{p}}}

\newcommand{\rhopl}{\ensuremath{\rho_{p}}}

\newcommand{\arstar}{\ensuremath{a/\rstar}}
\newcommand{\zrstar}{\ensuremath{\zeta/\rstar}}

\newcommand{\rjup}{\ensuremath{R_{\rm J}}}
\newcommand{\mjup}{\ensuremath{M_{\rm J}}}

\newcommand{\refsec}[1]{\mbox{\S\ \ref{sec:#1}}}

\newcommand{\reffigl}[1]{Figure~\ref{fig:#1}}
\newcommand{\refsecl}[1]{\mbox{Section \ref{sec:#1}}}

\newcommand{\reftabl}[1]{Table~\ref{tab:#1}}

\newcommand{\reffigls}[2]{Figures~\ref{fig:#1}--\ref{fig:#2}}
\newcommand{\reftabls}[2]{Tables~\ref{tab:#1}--\ref{tab:#2}}

\newcommand{\flwof}{\mbox{FLWO 1.2\,m}}

\newcommand{\loopand}{\ifnum\value{planetcounter}=4 and \else\fi}
\newcommand{\loopcomma}{\ifnum\value{planetcounter}<4 ,\else. \fi}
\newcommand{\loopcommanoperiod}{\ifnum\value{planetcounter}<4 ,\else \space\fi}
\newcommand{\loopcommanospace}{\ifnum\value{planetcounter}<4 ,\else \fi}

\newcommand{\hatcurhtrxxxxA}{HTR340-006}                               
\newcommand{\hatcurfieldxxxxA}{293}                                    
\newcommand{\hatcurCCraxxxxA}{\ensuremath{20^{\mathrm h}12^{\mathrm m}46.80{\mathrm s}}}                             
\newcommand{\hatcurCCdecxxxxA}{\ensuremath{+18{\arcdeg}06{\arcmin}17.5{\arcsec}}}                            
\newcommand{\hatcurCCmagxxxxA}{10.162}                                 
\newcommand{\hatcurCCtwomassxxxxA}{2MASS~20124688+1806175}             
\newcommand{\hatcurCCntwomassxxxxA}{20124688+1806175}             
\newcommand{\hatcurCCgscxxxxA}{GSC~1622-01261}                         
\newcommand{\hatcurCCngscxxxxA}{1622-01261}                         
\newcommand{\hatcurCCtassmvshortxxxxA}{\ensuremath{10.16}}                 \newcommand{\hatcurCCtassmvxxxxA}{\ensuremath{10.162\pm0.073}}                              
\newcommand{\hatcurCCtwomassJmagxxxxA}{\ensuremath{9.460\pm0.022}}     
\newcommand{\hatcurCCtwomassHmagxxxxA}{\ensuremath{9.322\pm0.030}}     
\newcommand{\hatcurCCtwomassKmagxxxxA}{\ensuremath{9.247\pm0.023}}     
\newcommand{\hatcurCCcitJmagxxxxA}{\ensuremath{9.485\pm0.022}}         
\newcommand{\hatcurCCcitHmagxxxxA}{\ensuremath{9.317\pm0.030}}         
\newcommand{\hatcurCCcitKmagxxxxA}{\ensuremath{9.271\pm0.023}}         
\newcommand{\hatcurCCbbJmagxxxxA}{\ensuremath{9.522\pm0.024}}          
\newcommand{\hatcurCCbbHmagxxxxA}{\ensuremath{9.338\pm0.031}}          
\newcommand{\hatcurCCbbKmagxxxxA}{\ensuremath{9.291\pm0.023}}          
\newcommand{\hatcurCCesoJmagxxxxA}{\ensuremath{9.522\pm0.025}}         
\newcommand{\hatcurCCesoHmagxxxxA}{\ensuremath{9.334\pm0.034}}         
\newcommand{\hatcurCCesoKmagxxxxA}{\ensuremath{9.291\pm0.024}}         
\newcommand{\hatcurCCesoJHmagxxxxA}{\ensuremath{0.187\pm0.040}}        
\newcommand{\hatcurCCesoJKmagxxxxA}{\ensuremath{0.232\pm0.034}}        
\newcommand{\hatcurCCesoHKmagxxxxA}{\ensuremath{0.044\pm0.042}}        
\newcommand{\hatcurLCdipxxxxA}{\ensuremath{7.9}}                       
\newcommand{\hatcurLCrprstarxxxxA}{\ensuremath{0.0801\pm0.0026}}       
\newcommand{\hatcurLCbsqxxxxA}{\ensuremath{0.113_{-0.062}^{+0.080}}}   
\newcommand{\hatcurLCimpxxxxA}{\ensuremath{0.336_{-0.128}^{+0.099}}}   
\newcommand{\hatcurLCzetaxxxxA}{\ensuremath{14.99\pm0.09}}             
\newcommand{\hatcurLCdurxxxxA}{\ensuremath{0.1455\pm0.0016}}           
\newcommand{\hatcurLCdurshortxxxxA}{\ensuremath{0.1455}}               
\newcommand{\hatcurLCdurhrxxxxA}{\ensuremath{3.493\pm0.039}}           
\newcommand{\hatcurLCdurhrshortxxxxA}{\ensuremath{3.493}}              
\newcommand{\hatcurLCqxxxxA}{\ensuremath{0.0267\pm0.0003}}             
\newcommand{\hatcurLCqshortxxxxA}{\ensuremath{0.027}}                  
\newcommand{\hatcurLCingdurxxxxA}{\ensuremath{0.0121\pm0.0013}}        
\newcommand{\hatcurLCPxxxxA}{\ensuremath{5.452654\pm0.000016}}         
\newcommand{\hatcurLCPprecxxxxA}{\ensuremath{5.4526536}}               
\newcommand{\hatcurLCPshortxxxxA}{\ensuremath{5.4527}}                 
\newcommand{\hatcurLCTxxxxA}{\ensuremath{2455431.59629\pm0.00055}}     
\newcommand{\hatcurLCTAxxxxA}{\ensuremath{2454351.97087\pm0.00316}}    
\newcommand{\hatcurLCTBxxxxA}{\ensuremath{2455480.67017\pm0.00061}}    
\newcommand{\hatcurLChatnetmAxxxxA}{\ensuremath{9.9968\pm0.0001}}      
\newcommand{\hatcurLCiblendAxxxxA}{\ensuremath{0.49\pm0.08}}           
\newcommand{\hatcurLChatnetmBxxxxA}{\ensuremath{10.2748\pm0.0001}}     
\newcommand{\hatcurLCiblendBxxxxA}{\ensuremath{0.87\pm0.08}}           
\newcommand{\hatcurSMEiteffxxxxA}{\ensuremath{6400\pm100}}             
\newcommand{\hatcurSMEizfehxxxxA}{\ensuremath{+0.21\pm0.1}}             
\newcommand{\hatcurSMEizfehshortxxxxA}{\ensuremath{+0.21}}              
\newcommand{\hatcurSMEiloggxxxxA}{\ensuremath{3.98\pm0.1}}            
\newcommand{\hatcurSMEivsinxxxxA}{\ensuremath{24.5\pm1.0}}             
\newcommand{\hatcurSMEivmacxxxxA}{\ensuremath{4.98}}                   
\newcommand{\hatcurSMEivmicxxxxA}{\ensuremath{0.85}}                   
\newcommand{\hatcurSMEiiteffxxxxA}{\ensuremath{6442\pm88}}             
\newcommand{\hatcurSMEiizfehxxxxA}{\ensuremath{+0.22\pm0.04}}           
\newcommand{\hatcurSMEiizfehshortxxxxA}{\ensuremath{+0.22}}             
\newcommand{\hatcurSMEiiloggxxxxA}{\ensuremath{4.21\pm0.06}}           
\newcommand{\hatcurSMEiivsinxxxxA}{\ensuremath{24.0\pm0.5}}            
\newcommand{\hatcurSMEiivmacxxxxA}{\ensuremath{5.05}}                  
\newcommand{\hatcurSMEiivmicxxxxA}{\ensuremath{0.85}}                  
\newcommand{\hatcurDSteffxxxxA}{\ensuremath{NULL\pmNULL}}              
\newcommand{\hatcurDSzfehxxxxA}{\ensuremath{NULL\pmNULL}}              
\newcommand{\hatcurDSloggxxxxA}{\ensuremath{NULL\pmNULL}}              
\newcommand{\hatcurDSvsinixxxxA}{\ensuremath{NULL\pmNULL}}             
\newcommand{\hatcurDSgammaxxxxA}{\ensuremath{NULL\pmNULL}}             
\newcommand{\hatcurDSnumspecxxxxA}{\ensuremath{0}}                     
\newcommand{\hatcurDSspanxxxxA}{\ensuremath{0}}                        
\newcommand{\hatcurDSrvrmsxxxxA}{\ensuremath{0.00}}                    
\newcommand{\hatcurTRESteffxxxxA}{\ensuremath{5900\pm300}}             
\newcommand{\hatcurTRESzfehxxxxA}{\ensuremath{0.0\pm0.0}}              
\newcommand{\hatcurTRESloggxxxxA}{\ensuremath{3.5\pm0.3}}              
\newcommand{\hatcurTRESvsinixxxxA}{\ensuremath{26.7\pm2.0}}            
\newcommand{\hatcurTRESgammaxxxxA}{\ensuremath{-49.26\pm0.30}}         
\newcommand{\hatcurTRESnumspecxxxxA}{\ensuremath{3}}                   
\newcommand{\hatcurTRESspanxxxxA}{\ensuremath{64}}                     
\newcommand{\hatcurTRESrvrmsxxxxA}{\ensuremath{0.77}}                  
\newcommand{\hatcurFIESteffxxxxA}{\ensuremath{NULL\pmNULL}}            
\newcommand{\hatcurFIESzfehxxxxA}{\ensuremath{NULL\pmNULL}}            
\newcommand{\hatcurFIESloggxxxxA}{\ensuremath{NULL\pmNULL}}            
\newcommand{\hatcurFIESvsinixxxxA}{\ensuremath{NULL\pmNULL}}           
\newcommand{\hatcurFIESgammaxxxxA}{\ensuremath{NULL\pmNULL}}           
\newcommand{\hatcurFIESnumspecxxxxA}{\ensuremath{0}}                   
\newcommand{\hatcurFIESspanxxxxA}{\ensuremath{0}}                      
\newcommand{\hatcurFIESrvrmsxxxxA}{\ensuremath{0.00}}                  
\newcommand{\hatcurLBizxxxxA}{\ensuremath{0.1269}}                     
\newcommand{\hatcurLBiizxxxxA}{\ensuremath{0.3728}}                    
\newcommand{\hatcurLBiixxxxA}{\ensuremath{0.1785}}                     
\newcommand{\hatcurLBiiixxxxA}{\ensuremath{0.3825}}                    
\newcommand{\hatcurLBiIxxxxA}{\ensuremath{0.1596}}                     
\newcommand{\hatcurLBiiIxxxxA}{\ensuremath{0.3807}}                    
\newcommand{\hatcurLBigxxxxA}{\ensuremath{0.4243}}                     
\newcommand{\hatcurLBiigxxxxA}{\ensuremath{0.3303}}                    
\newcommand{\hatcurLBikepxxxxA}{\ensuremath{}}                 
\newcommand{\hatcurLBiikepxxxxA}{\ensuremath{}}                
\newcommand{\hatcurISOmxxxxA}{\ensuremath{1.39\pm0.05}}                
\newcommand{\hatcurISOmshortxxxxA}{\ensuremath{1.39}}                  
\newcommand{\hatcurISOmlongxxxxA}{\ensuremath{1.392\pm0.047}}          
\newcommand{\hatcurISOrxxxxA}{\ensuremath{1.53_{-0.10}^{+0.13}}}       
\newcommand{\hatcurISOrshortxxxxA}{\ensuremath{1.53}}                  
\newcommand{\hatcurISOrlongxxxxA}{\ensuremath{1.535_{-0.102}^{+0.135}}} 
\newcommand{\hatcurISOrhoxxxxA}{\ensuremath{0.54\pm0.11}}              
\newcommand{\hatcurISOloggxxxxA}{\ensuremath{4.21\pm0.06}}             
\newcommand{\hatcurISOlumxxxxA}{\ensuremath{3.63_{-0.51}^{+0.75}}}     
\newcommand{\hatcurISOlumshortxxxxA}{\ensuremath{3.63}}                
\newcommand{\hatcurISOmvxxxxA}{\ensuremath{3.32\pm0.19}}               
\newcommand{\hatcurISOvixxxxA}{\ensuremath{0.513\pm0.022}}             
\newcommand{\hatcurISOagexxxxA}{\ensuremath{1.7_{-0.5}^{+0.4}}}        
\newcommand{\hatcurISOsigmaxxxxA}{\ensuremath{0.00190\pm0.00029}}      
\newcommand{\hatcurISOMJxxxxA}{\ensuremath{2.50\pm0.17}}               
\newcommand{\hatcurISOMHxxxxA}{\ensuremath{2.28\pm0.17}}               
\newcommand{\hatcurISOMKxxxxA}{\ensuremath{2.24\pm0.17}}               
\newcommand{\hatcurISOJKxxxxA}{\ensuremath{0.26\pm0.02}}               
\newcommand{\hatcurISOspecxxxxA}{F}                                    
\newcommand{\hatcurRVKxxxxA}{\ensuremath{343.1\pm21.3}}                
\newcommand{\hatcurRVkxxxxA}{\ensuremath{0.410\pm0.031}}               
\newcommand{\hatcurRVhxxxxA}{\ensuremath{0.156\pm0.052}}               
\newcommand{\hatcurRVtronexxxxA}{\ensuremath{0.8683\pm0.4719}}         
\newcommand{\hatcurRVtrtwoxxxxA}{\ensuremath{0.0000\pm0.0000}}         
\newcommand{\hatcurRVgammaAxxxxA}{\ensuremath{-0.8\pm19.8}}            
\newcommand{\hatcurRVjitterAxxxxA}{\ensuremath{56.0}}                  
\newcommand{\hatcurRVfitrmsAxxxxA}{\ensuremath{57.6}}                  
\newcommand{\hatcurRVgammaBxxxxA}{\ensuremath{219.7\pm32.0}}           
\newcommand{\hatcurRVjitterBxxxxA}{\ensuremath{32.0}}                  
\newcommand{\hatcurRVfitrmsBxxxxA}{\ensuremath{57.3}}                  
\newcommand{\hatcurRVgammaCxxxxA}{\ensuremath{14.1\pm23.2}}            
\newcommand{\hatcurRVjitterCxxxxA}{\ensuremath{0.0}}                   
\newcommand{\hatcurRVfitrmsCxxxxA}{\ensuremath{50.4}}                  
\newcommand{\hatcurRVjitterxxxxA}{\hatcurRVjitterAxxxxA}			   
\newcommand{\hatcurRVeccenxxxxA}{\ensuremath{0.441\pm0.032}}           
\newcommand{\hatcurRVomegaxxxxA}{\ensuremath{20\pm14}}                 
\newcommand{\hatcurPPixxxxA}{\ensuremath{87.1\pm1.2}}                  
\newcommand{\hatcurPPgxxxxA}{\ensuremath{57.4\pm10.1}}                 
\newcommand{\hatcurPPloggxxxxA}{\ensuremath{3.76\pm0.08}}              
\newcommand{\hatcurPParxxxxA}{\ensuremath{9.48\pm0.64}}                
\newcommand{\hatcurPParelxxxxA}{\ensuremath{0.0677\pm0.0008}}          
\newcommand{\hatcurPPrhoxxxxA}{\ensuremath{2.40\pm0.63}}               
\newcommand{\hatcurPPmxxxxA}{\ensuremath{3.33\pm0.21}}                 
\newcommand{\hatcurPPmshortxxxxA}{\ensuremath{3.33}}                   
\newcommand{\hatcurPPmlongxxxxA}{\ensuremath{3.328\pm0.211}}           
\newcommand{\hatcurPPmexxxxA}{\ensuremath{1057.7\pm66.9}}              
\newcommand{\hatcurPPmeshortxxxxA}{\ensuremath{1057.7}}                
\newcommand{\hatcurPPmelongxxxxA}{\ensuremath{1057.70\pm66.91}}        
\newcommand{\hatcurPPrxxxxA}{\ensuremath{1.20_{-0.09}^{+0.13}}}        
\newcommand{\hatcurPPrshortxxxxA}{\ensuremath{1.20}}                   
\newcommand{\hatcurPPrlongxxxxA}{\ensuremath{1.197_{-0.092}^{+0.128}}} 
\newcommand{\hatcurPPrexxxxA}{\ensuremath{13.4_{-1.0}^{+1.4}}}         
\newcommand{\hatcurPPreshortxxxxA}{\ensuremath{13.4}}                  
\newcommand{\hatcurPPrelongxxxxA}{\ensuremath{13.42_{-1.03}^{+1.44}}}  
\newcommand{\hatcurPPmrcorrxxxxA}{\ensuremath{0.23}}                   
\newcommand{\hatcurPPteffxxxxA}{\ensuremath{1520\pm60}}                
\newcommand{\hatcurPPthetaxxxxA}{\ensuremath{0.269\pm0.029}}           
\newcommand{\hatcurPPfluxperixxxxA}{\ensuremath{3.46_{-0.62}^{+1.05}}} 
\newcommand{\hatcurPPfluxperidimxxxxA}{\ensuremath{9}}                 
\newcommand{\hatcurPPfluxapxxxxA}{\ensuremath{5.23_{-0.61}^{+0.79}}}   
\newcommand{\hatcurPPfluxapdimxxxxA}{\ensuremath{8}}                   
\newcommand{\hatcurPPfluxavgxxxxA}{\ensuremath{1.21_{-0.16}^{+0.23}}}  
\newcommand{\hatcurPPfluxavgdimxxxxA}{\ensuremath{9}}                  
\newcommand{\hatcurXsecphasexxxxA}{\ensuremath{0.7565\pm0.0182}}       
\newcommand{\hatcurXsecondaryxxxxA}{\ensuremath{2455435.721\pm0.099}}  
\newcommand{\hatcurXsecdurxxxxA}{\ensuremath{0.1871\pm0.0170}}         
\newcommand{\hatcurXsecingdurxxxxA}{\ensuremath{0.0176\pm0.0052}}      
\newcommand{\hatcurPPphiconjxxxxA}{\ensuremath{0.0785\pm0.0134}}       
\newcommand{\hatcurPPperixxxxA}{\ensuremath{2455431.17\pm0.07}}        
\newcommand{\hatcurPPaequivxxxxA}{\ensuremath{0.0355\pm0.0026}}        
\newcommand{\hatcurPPtcircxxxxA}{\ensuremath{1947.6_{-787.8}^{+1529.1}}} 
\newcommand{\hatcurPPtinfallxxxxA}{\ensuremath{1662.1_{-455.4}^{+673.7}}} 
\newcommand{\hatcurXdistxxxxA}{\ensuremath{257_{-17}^{+22}}}           
\newcommand{\hatcurCCpmraxxxxA}{\ensuremath{16.6\pm0.5}}               
\newcommand{\hatcurCCpmdecxxxxA}{\ensuremath{-11.5\pm0.8}}             
\newcommand{\hatcurCCpmxxxxA}{\ensuremath{20.1943\pm0.943398}}         

\newcommand{\hatcurhtrxxxxB}{HTR364-003}                               
\newcommand{\hatcurfieldxxxxB}{364}                                    
\newcommand{\hatcurCCraxxxxB}{\ensuremath{08^{\mathrm h}13^{\mathrm m}00.19{\mathrm s}}}                             
\newcommand{\hatcurCCdecxxxxB}{\ensuremath{+04{\arcdeg}47{\arcmin}13.3{\arcsec}}}                             
\newcommand{\hatcurCCmagxxxxB}{12.46}                                  
\newcommand{\hatcurCCtwomassxxxxB}{2MASS~08130018+0447132}             
\newcommand{\hatcurCCntwomassxxxxB}{08130018+0447132}             
\newcommand{\hatcurCCgscxxxxB}{GSC~0203-01079}                         
\newcommand{\hatcurCCngscxxxxB}{0203-01079}                         
\newcommand{\hatcurCCtassmvshortxxxxB}{\ensuremath{12.5}}                   \newcommand{\hatcurCCtassmvxxxxB}{\ensuremath{12.46\pm0.11}}                               
\newcommand{\hatcurCCtwomassJmagxxxxB}{\ensuremath{11.358\pm0.024}}    
\newcommand{\hatcurCCtwomassHmagxxxxB}{\ensuremath{11.072\pm0.023}}    
\newcommand{\hatcurCCtwomassKmagxxxxB}{\ensuremath{11.030\pm0.021}}    
\newcommand{\hatcurCCcitJmagxxxxB}{\ensuremath{11.377\pm0.024}}        
\newcommand{\hatcurCCcitHmagxxxxB}{\ensuremath{11.067\pm0.023}}        
\newcommand{\hatcurCCcitKmagxxxxB}{\ensuremath{11.054\pm0.021}}        
\newcommand{\hatcurCCbbJmagxxxxB}{\ensuremath{11.423\pm0.026}}         
\newcommand{\hatcurCCbbHmagxxxxB}{\ensuremath{11.088\pm0.024}}         
\newcommand{\hatcurCCbbKmagxxxxB}{\ensuremath{11.074\pm0.021}}         
\newcommand{\hatcurCCesoJmagxxxxB}{\ensuremath{11.425\pm0.027}}        
\newcommand{\hatcurCCesoHmagxxxxB}{\ensuremath{11.082\pm0.027}}        
\newcommand{\hatcurCCesoKmagxxxxB}{\ensuremath{11.073\pm0.022}}        
\newcommand{\hatcurCCesoJHmagxxxxB}{\ensuremath{0.342\pm0.036}}        
\newcommand{\hatcurCCesoJKmagxxxxB}{\ensuremath{0.352\pm0.034}}        
\newcommand{\hatcurCCesoHKmagxxxxB}{\ensuremath{0.009\pm0.035}}        
\newcommand{\hatcurLCdipxxxxB}{\ensuremath{9.0}}                       
\newcommand{\hatcurLCrprstarxxxxB}{\ensuremath{0.0954\pm0.0027}}       
\newcommand{\hatcurLCbsqxxxxB}{\ensuremath{0.128_{-0.066}^{+0.078}}}   
\newcommand{\hatcurLCimpxxxxB}{\ensuremath{0.357_{-0.127}^{+0.092}}}   
\newcommand{\hatcurLCzetaxxxxB}{\ensuremath{13.52\pm0.09}}             
\newcommand{\hatcurLCdurxxxxB}{\ensuremath{0.1640\pm0.0018}}           
\newcommand{\hatcurLCdurshortxxxxB}{\ensuremath{0.1640}}               
\newcommand{\hatcurLCdurhrxxxxB}{\ensuremath{3.936\pm0.044}}           
\newcommand{\hatcurLCdurhrshortxxxxB}{\ensuremath{3.936}}              
\newcommand{\hatcurLCqxxxxB}{\ensuremath{0.0450\pm0.0005}}             
\newcommand{\hatcurLCqshortxxxxB}{\ensuremath{0.045}}                  
\newcommand{\hatcurLCingdurxxxxB}{\ensuremath{0.0162\pm0.0017}}        
\newcommand{\hatcurLCPxxxxB}{\ensuremath{3.646706\pm0.000021}}         
\newcommand{\hatcurLCPprecxxxxB}{\ensuremath{3.6467065}}               
\newcommand{\hatcurLCPshortxxxxB}{\ensuremath{3.6467}}                 
\newcommand{\hatcurLCTxxxxB}{\ensuremath{2455578.66081\pm0.00050}}     
\newcommand{\hatcurLCTAxxxxB}{\ensuremath{2454809.20573\pm0.00462}}    
\newcommand{\hatcurLCTBxxxxB}{\ensuremath{2455585.95422\pm0.00049}}    
\newcommand{\hatcurLChatnetmxxxxB}{\ensuremath{12.3835\pm0.0002}}      
\newcommand{\hatcurLCiblendxxxxB}{\ensuremath{0.77\pm0.08}}            
\newcommand{\hatcurSMEiteffxxxxB}{\ensuremath{5940\pm88}}              
\newcommand{\hatcurSMEizfehxxxxB}{\ensuremath{+0.01\pm0.08}}            
\newcommand{\hatcurSMEizfehshortxxxxB}{\ensuremath{+;0.01}}              
\newcommand{\hatcurSMEiloggxxxxB}{\ensuremath{3.98\pm0.1}}            
\newcommand{\hatcurSMEivsinxxxxB}{\ensuremath{0.5\pm0.5}}              
\newcommand{\hatcurSMEivmacxxxxB}{\ensuremath{4.28}}                   
\newcommand{\hatcurSMEivmicxxxxB}{\ensuremath{0.85}}                   
\newcommand{\hatcurSMEiiteffxxxxB}{\ensuremath{6096\pm88}}             
\newcommand{\hatcurSMEiizfehxxxxB}{\ensuremath{+0.11\pm0.08}}           
\newcommand{\hatcurSMEiizfehshortxxxxB}{\ensuremath{+0.11}}             
\newcommand{\hatcurSMEiiloggxxxxB}{\ensuremath{4.20\pm0.06}}           
\newcommand{\hatcurSMEiivsinxxxxB}{\ensuremath{0.5\pm0.5}}             
\newcommand{\hatcurSMEiivmacxxxxB}{\ensuremath{4.52}}                  
\newcommand{\hatcurSMEiivmicxxxxB}{\ensuremath{0.85}}                  
\newcommand{\hatcurDSteffxxxxB}{\ensuremath{NULL\pmNULL}}              
\newcommand{\hatcurDSzfehxxxxB}{\ensuremath{NULL\pmNULL}}              
\newcommand{\hatcurDSloggxxxxB}{\ensuremath{NULL\pmNULL}}              
\newcommand{\hatcurDSvsinixxxxB}{\ensuremath{NULL\pmNULL}}             
\newcommand{\hatcurDSgammaxxxxB}{\ensuremath{NULL\pmNULL}}             
\newcommand{\hatcurDSnumspecxxxxB}{\ensuremath{0}}                     
\newcommand{\hatcurDSspanxxxxB}{\ensuremath{0}}                        
\newcommand{\hatcurDSrvrmsxxxxB}{\ensuremath{0.00}}                    
\newcommand{\hatcurTRESteffxxxxB}{\ensuremath{6000\pm150}}             
\newcommand{\hatcurTRESzfehxxxxB}{\ensuremath{0.0\pm0.0}}              
\newcommand{\hatcurTRESloggxxxxB}{\ensuremath{4.0\pm0.5}}              
\newcommand{\hatcurTRESvsinixxxxB}{\ensuremath{4\pm0.5}}               
\newcommand{\hatcurTRESgammaxxxxB}{\ensuremath{40.95\pm0.20}}          
\newcommand{\hatcurTRESnumspecxxxxB}{\ensuremath{4}}                   
\newcommand{\hatcurTRESspanxxxxB}{\ensuremath{32}}                     
\newcommand{\hatcurTRESrvrmsxxxxB}{\ensuremath{0.36}}                  
\newcommand{\hatcurFIESteffxxxxB}{\ensuremath{NULL\pmNULL}}            
\newcommand{\hatcurFIESzfehxxxxB}{\ensuremath{NULL\pmNULL}}            
\newcommand{\hatcurFIESloggxxxxB}{\ensuremath{NULL\pmNULL}}            
\newcommand{\hatcurFIESvsinixxxxB}{\ensuremath{NULL\pmNULL}}           
\newcommand{\hatcurFIESgammaxxxxB}{\ensuremath{NULL\pmNULL}}           
\newcommand{\hatcurFIESnumspecxxxxB}{\ensuremath{0}}                   
\newcommand{\hatcurFIESspanxxxxB}{\ensuremath{0}}                      
\newcommand{\hatcurFIESrvrmsxxxxB}{\ensuremath{0.00}}                  
\newcommand{\hatcurLBizxxxxB}{\ensuremath{0.1650}}                     
\newcommand{\hatcurLBiizxxxxB}{\ensuremath{0.3515}}                    
\newcommand{\hatcurLBiixxxxB}{\ensuremath{0.2198}}                     
\newcommand{\hatcurLBiiixxxxB}{\ensuremath{0.3587}}                    
\newcommand{\hatcurLBiIxxxxB}{\ensuremath{0.2005}}                     
\newcommand{\hatcurLBiiIxxxxB}{\ensuremath{0.3575}}                    
\newcommand{\hatcurLBigxxxxB}{\ensuremath{0.4820}}                     
\newcommand{\hatcurLBiigxxxxB}{\ensuremath{0.2885}}                    
\newcommand{\hatcurLBikepxxxxB}{\ensuremath{}}                 
\newcommand{\hatcurLBiikepxxxxB}{\ensuremath{}}                
\newcommand{\hatcurISOmxxxxB}{\ensuremath{1.24\pm0.05}}                
\newcommand{\hatcurISOmshortxxxxB}{\ensuremath{1.24}}                  
\newcommand{\hatcurISOmlongxxxxB}{\ensuremath{1.236\pm0.048}}          
\newcommand{\hatcurISOrxxxxB}{\ensuremath{1.44\pm0.08}}                
\newcommand{\hatcurISOrshortxxxxB}{\ensuremath{1.44}}                  
\newcommand{\hatcurISOrlongxxxxB}{\ensuremath{1.435\pm0.084}}          
\newcommand{\hatcurISOrhoxxxxB}{\ensuremath{0.59\pm0.09}}              
\newcommand{\hatcurISOloggxxxxB}{\ensuremath{4.21\pm0.04}}             
\newcommand{\hatcurISOlumxxxxB}{\ensuremath{2.55_{-0.30}^{+0.40}}}     
\newcommand{\hatcurISOlumshortxxxxB}{\ensuremath{2.55}}                
\newcommand{\hatcurISOmvxxxxB}{\ensuremath{3.77\pm0.15}}               
\newcommand{\hatcurISOvixxxxB}{\ensuremath{0.605\pm0.024}}             
\newcommand{\hatcurISOagexxxxB}{\ensuremath{3.5_{-0.5}^{+0.8}}}        
\newcommand{\hatcurISOsigmaxxxxB}{\ensuremath{0.00070\pm0.00008}}      
\newcommand{\hatcurISOMJxxxxB}{\ensuremath{2.77\pm0.13}}               
\newcommand{\hatcurISOMHxxxxB}{\ensuremath{2.48\pm0.13}}               
\newcommand{\hatcurISOMKxxxxB}{\ensuremath{2.43\pm0.13}}               
\newcommand{\hatcurISOJKxxxxB}{\ensuremath{0.34\pm0.02}}               
\newcommand{\hatcurISOspecxxxxB}{G}                                    
\newcommand{\hatcurRVKxxxxB}{\ensuremath{120.7\pm2.2}}                 
\newcommand{\hatcurRVkxxxxB}{\ensuremath{-0.004\pm0.013}}              
\newcommand{\hatcurRVhxxxxB}{\ensuremath{-0.017\pm0.026}}              
\newcommand{\hatcurRVgammaxxxxB}{\ensuremath{-27.2\pm2.0}}             
\newcommand{\hatcurRVjitterxxxxB}{\ensuremath{3.7}}                    
\newcommand{\hatcurRVfitrmsxxxxB}{\ensuremath{4.5}}                    
\newcommand{\hatcurRVeccenxxxxB}{\ensuremath{0.025\pm0.018}}           
\newcommand{\hatcurRVomegaxxxxB}{\ensuremath{248\pm93}}                
\newcommand{\hatcurPPixxxxB}{\ensuremath{87.3\pm1.0}}                  
\newcommand{\hatcurPPgxxxxB}{\ensuremath{14.7\pm1.9}}                  
\newcommand{\hatcurPPloggxxxxB}{\ensuremath{3.17\pm0.06}}              
\newcommand{\hatcurPParxxxxB}{\ensuremath{7.45\pm0.37}}                
\newcommand{\hatcurPParelxxxxB}{\ensuremath{0.0498\pm0.0006}}          
\newcommand{\hatcurPPrhoxxxxB}{\ensuremath{0.55\pm0.11}}               
\newcommand{\hatcurPPmxxxxB}{\ensuremath{1.05\pm0.03}}                 
\newcommand{\hatcurPPmshortxxxxB}{\ensuremath{1.05}}                   
\newcommand{\hatcurPPmlongxxxxB}{\ensuremath{1.054\pm0.033}}           
\newcommand{\hatcurPPmexxxxB}{\ensuremath{334.8\pm10.3}}               
\newcommand{\hatcurPPmeshortxxxxB}{\ensuremath{334.8}}                 
\newcommand{\hatcurPPmelongxxxxB}{\ensuremath{334.80\pm10.35}}         
\newcommand{\hatcurPPrxxxxB}{\ensuremath{1.33\pm0.10}}                 
\newcommand{\hatcurPPrshortxxxxB}{\ensuremath{1.33}}                   
\newcommand{\hatcurPPrlongxxxxB}{\ensuremath{1.332\pm0.098}}           
\newcommand{\hatcurPPrexxxxB}{\ensuremath{14.9\pm1.1}}                 
\newcommand{\hatcurPPreshortxxxxB}{\ensuremath{14.9}}                  
\newcommand{\hatcurPPrelongxxxxB}{\ensuremath{14.93\pm1.10}}           
\newcommand{\hatcurPPmrcorrxxxxB}{\ensuremath{0.49}}                   
\newcommand{\hatcurPPteffxxxxB}{\ensuremath{1581\pm45}}                
\newcommand{\hatcurPPthetaxxxxB}{\ensuremath{0.064\pm0.005}}           
\newcommand{\hatcurPPfluxperixxxxB}{\ensuremath{1.49_{-0.14}^{+0.19}}} 
\newcommand{\hatcurPPfluxperidimxxxxB}{\ensuremath{9}}                 
\newcommand{\hatcurPPfluxapxxxxB}{\ensuremath{1.34\pm0.17}}            
\newcommand{\hatcurPPfluxapdimxxxxB}{\ensuremath{9}}                   
\newcommand{\hatcurPPfluxavgxxxxB}{\ensuremath{1.41_{-0.14}^{+0.19}}}  
\newcommand{\hatcurPPfluxavgdimxxxxB}{\ensuremath{9}}                  
\newcommand{\hatcurXsecphasexxxxB}{\ensuremath{0.4978\pm0.0081}}       
\newcommand{\hatcurXsecondaryxxxxB}{\ensuremath{2455580.476\pm0.030}}  
\newcommand{\hatcurXsecdurxxxxB}{\ensuremath{0.1596\pm0.0076}}         
\newcommand{\hatcurXsecingdurxxxxB}{\ensuremath{0.0156\pm0.0019}}      
\newcommand{\hatcurPPphiconjxxxxB}{\ensuremath{-0.2480_{-0.1859}^{+0.5085}}} 
\newcommand{\hatcurPPperixxxxB}{\ensuremath{2455579.56\pm1.40}}        
\newcommand{\hatcurPPaequivxxxxB}{\ensuremath{0.0312\pm0.0018}}        
\newcommand{\hatcurPPtcircxxxxB}{\ensuremath{237.3_{-64.8}^{+93.9}}}   
\newcommand{\hatcurPPtinfallxxxxB}{\ensuremath{929.8\pm223.9}}         
\newcommand{\hatcurXdistxxxxB}{\ensuremath{535\pm32}}                  
\newcommand{\hatcurCCpmraxxxxB}{\ensuremath{-11.7\pm2.9}}              
\newcommand{\hatcurCCpmdecxxxxB}{\ensuremath{-0.9\pm8.0}}              
\newcommand{\hatcurCCpmxxxxB}{\ensuremath{11.7346\pm8.50941}}          

\newcommand{\hatcurhtrxxxxC}{HTR143-001}                               
\newcommand{\hatcurfieldxxxxC}{143}                                    
\newcommand{\hatcurCCraxxxxC}{\ensuremath{12^{\mathrm h}33^{\mathrm m}03.96{\mathrm s}}}                             
\newcommand{\hatcurCCdecxxxxC}{\ensuremath{+44{\arcdeg}54{\arcmin}55.3{\arcsec}}}                            
\newcommand{\hatcurCCmagxxxxC}{12.262}                                 
\newcommand{\hatcurCCtwomassxxxxC}{2MASS~12330390+4454552}             
\newcommand{\hatcurCCntwomassxxxxC}{12330390+4454552}             
\newcommand{\hatcurCCgscxxxxC}{GSC~3020-02221}                         
\newcommand{\hatcurCCngscxxxxC}{3020-02221}                            
\newcommand{\hatcurCCtassmvshortxxxxC}{\ensuremath{12.26}}                 \newcommand{\hatcurCCtassmvxxxxC}{\ensuremath{12.262\pm0.068}}                              
\newcommand{\hatcurCCtwomassJmagxxxxC}{\ensuremath{11.046\pm0.027}}    
\newcommand{\hatcurCCtwomassHmagxxxxC}{\ensuremath{10.723\pm0.030}}    
\newcommand{\hatcurCCtwomassKmagxxxxC}{\ensuremath{10.603\pm0.021}}    
\newcommand{\hatcurCCcitJmagxxxxC}{\ensuremath{11.059\pm0.027}}        
\newcommand{\hatcurCCcitHmagxxxxC}{\ensuremath{10.716\pm0.030}}        
\newcommand{\hatcurCCcitKmagxxxxC}{\ensuremath{10.627\pm0.021}}        
\newcommand{\hatcurCCbbJmagxxxxC}{\ensuremath{11.114\pm0.029}}         
\newcommand{\hatcurCCbbHmagxxxxC}{\ensuremath{10.739\pm0.031}}         
\newcommand{\hatcurCCbbKmagxxxxC}{\ensuremath{10.647\pm0.021}}         
\newcommand{\hatcurCCesoJmagxxxxC}{\ensuremath{11.117\pm0.031}}        
\newcommand{\hatcurCCesoHmagxxxxC}{\ensuremath{10.736\pm0.036}}        
\newcommand{\hatcurCCesoKmagxxxxC}{\ensuremath{10.646\pm0.022}}        
\newcommand{\hatcurCCesoJHmagxxxxC}{\ensuremath{0.380\pm0.045}}        
\newcommand{\hatcurCCesoJKmagxxxxC}{\ensuremath{0.472\pm0.037}}        
\newcommand{\hatcurCCesoHKmagxxxxC}{\ensuremath{0.091\pm0.042}}        
\newcommand{\hatcurLCdipxxxxC}{\ensuremath{14.7}}                      
\newcommand{\hatcurLCrprstarxxxxC}{\ensuremath{0.1186\pm0.0012}}       
\newcommand{\hatcurLCbsqxxxxC}{\ensuremath{0.097_{-0.048}^{+0.057}}}   
\newcommand{\hatcurLCimpxxxxC}{\ensuremath{0.312_{-0.105}^{+0.078}}}   
\newcommand{\hatcurLCzetaxxxxC}{\ensuremath{24.51\pm0.14}}             
\newcommand{\hatcurLCdurxxxxC}{\ensuremath{0.0923\pm0.0007}}           
\newcommand{\hatcurLCdurshortxxxxC}{\ensuremath{0.0923}}               
\newcommand{\hatcurLCdurhrxxxxC}{\ensuremath{2.215\pm0.017}}           
\newcommand{\hatcurLCdurhrshortxxxxC}{\ensuremath{2.215}}              
\newcommand{\hatcurLCqxxxxC}{\ensuremath{0.0695\pm0.0005}}             
\newcommand{\hatcurLCqshortxxxxC}{\ensuremath{0.070}}                  
\newcommand{\hatcurLCingdurxxxxC}{\ensuremath{0.0107\pm0.0007}}        
\newcommand{\hatcurLCPxxxxC}{\ensuremath{1.327347\pm0.000003}}         
\newcommand{\hatcurLCPprecxxxxC}{\ensuremath{1.3273472}}               
\newcommand{\hatcurLCPshortxxxxC}{\ensuremath{1.3273}}                 
\newcommand{\hatcurLCTxxxxC}{\ensuremath{2455565.18144\pm0.00020}}     
\newcommand{\hatcurLCTAxxxxC}{\ensuremath{2455302.36670\pm0.00058}}    
\newcommand{\hatcurLCTBxxxxC}{\ensuremath{2455608.98390\pm0.00022}}    
\newcommand{\hatcurSMEiteffxxxxC}{\ensuremath{5850\pm100}}             
\newcommand{\hatcurSMEizfehxxxxC}{\ensuremath{+0.38\pm0.1}}            
\newcommand{\hatcurSMEizfehshortxxxxC}{\ensuremath{+0.38}}              
\newcommand{\hatcurSMEiloggxxxxC}{\ensuremath{4.73\pm0.17}}            
\newcommand{\hatcurSMEivsinxxxxC}{\ensuremath{2.86\pm0.5}}             
\newcommand{\hatcurSMEivmacxxxxC}{\ensuremath{0.00}}                   
\newcommand{\hatcurSMEivmicxxxxC}{\ensuremath{0.00}}                   
\newcommand{\hatcurSMEiiteffxxxxC}{\ensuremath{5560\pm100}}            
\newcommand{\hatcurSMEiizfehxxxxC}{\ensuremath{+0.26\pm0.10}}           
\newcommand{\hatcurSMEiizfehshortxxxxC}{\ensuremath{+0.26}}             
\newcommand{\hatcurSMEiiloggxxxxC}{\ensuremath{4.34\pm0.00}}           
\newcommand{\hatcurSMEiivsinxxxxC}{\ensuremath{3.58\pm0.5}}            
\newcommand{\hatcurSMEiivmacxxxxC}{\ensuremath{0.00}}                  
\newcommand{\hatcurSMEiivmicxxxxC}{\ensuremath{0.00}}                  
\newcommand{\hatcurDSteffxxxxC}{\ensuremath{NULL\pmNULL}}              
\newcommand{\hatcurDSzfehxxxxC}{\ensuremath{NULL\pmNULL}}              
\newcommand{\hatcurDSloggxxxxC}{\ensuremath{NULL\pmNULL}}              
\newcommand{\hatcurDSvsinixxxxC}{\ensuremath{NULL\pmNULL}}             
\newcommand{\hatcurDSgammaxxxxC}{\ensuremath{NULL\pmNULL}}             
\newcommand{\hatcurDSnumspecxxxxC}{\ensuremath{0}}                     
\newcommand{\hatcurDSspanxxxxC}{\ensuremath{0}}                        
\newcommand{\hatcurDSrvrmsxxxxC}{\ensuremath{0.00}}                    
\newcommand{\hatcurTRESteffxxxxC}{\ensuremath{5500\pm150}}             
\newcommand{\hatcurTRESzfehxxxxC}{\ensuremath{0.0\pm0.0}}              
\newcommand{\hatcurTRESloggxxxxC}{\ensuremath{4.5\pm0.5}}              
\newcommand{\hatcurTRESvsinixxxxC}{\ensuremath{4\pm0.5}}               
\newcommand{\hatcurTRESgammaxxxxC}{\ensuremath{-16.29\pm0.10}}         
\newcommand{\hatcurTRESnumspecxxxxC}{\ensuremath{1}}                   
\newcommand{\hatcurTRESspanxxxxC}{\ensuremath{1}}                      
\newcommand{\hatcurTRESrvrmsxxxxC}{\ensuremath{0.00}}                  
\newcommand{\hatcurFIESteffxxxxC}{\ensuremath{5875\pm100}}             
\newcommand{\hatcurFIESzfehxxxxC}{\ensuremath{0.0\pm0.0}}              
\newcommand{\hatcurFIESloggxxxxC}{\ensuremath{4.0\pm0.5}}              
\newcommand{\hatcurFIESvsinixxxxC}{\ensuremath{6.0\pm0.5}}             
\newcommand{\hatcurFIESgammaxxxxC}{\ensuremath{-21.075\pm0.06}}        
\newcommand{\hatcurFIESnumspecxxxxC}{\ensuremath{2}}                   
\newcommand{\hatcurFIESspanxxxxC}{\ensuremath{1}}                      
\newcommand{\hatcurFIESrvrmsxxxxC}{\ensuremath{0.06}}                  
\newcommand{\hatcurLBizxxxxC}{\ensuremath{0.2415}}                     
\newcommand{\hatcurLBiizxxxxC}{\ensuremath{0.3183}}                    
\newcommand{\hatcurLBiixxxxC}{\ensuremath{0.3142}}                     
\newcommand{\hatcurLBiiixxxxC}{\ensuremath{0.3113}}                    
\newcommand{\hatcurLBiIxxxxC}{\ensuremath{0.2900}}                     
\newcommand{\hatcurLBiiIxxxxC}{\ensuremath{0.3142}}                    
\newcommand{\hatcurLBigxxxxC}{\ensuremath{0.6364}}                     
\newcommand{\hatcurLBiigxxxxC}{\ensuremath{0.1742}}                    
\newcommand{\hatcurLBikepxxxxC}{\ensuremath{}}                 
\newcommand{\hatcurLBiikepxxxxC}{\ensuremath{}}                
\newcommand{\hatcurISOmxxxxC}{\ensuremath{1.02\pm0.05}}                
\newcommand{\hatcurISOmshortxxxxC}{\ensuremath{1.02}}                  
\newcommand{\hatcurISOmlongxxxxC}{\ensuremath{1.022\pm0.049}}          
\newcommand{\hatcurISOrxxxxC}{\ensuremath{1.10\pm0.06}}                
\newcommand{\hatcurISOrshortxxxxC}{\ensuremath{1.10}}                  
\newcommand{\hatcurISOrlongxxxxC}{\ensuremath{1.096\pm0.056}}          
\newcommand{\hatcurISOrhoxxxxC}{\ensuremath{1.09\pm0.15}}              
\newcommand{\hatcurISOloggxxxxC}{\ensuremath{4.37\pm0.04}}             
\newcommand{\hatcurISOlumxxxxC}{\ensuremath{1.03\pm0.15}}              
\newcommand{\hatcurISOlumshortxxxxC}{\ensuremath{1.03}}                
\newcommand{\hatcurISOmvxxxxC}{\ensuremath{4.83\pm0.17}}               
\newcommand{\hatcurISOvixxxxC}{\ensuremath{0.772\pm0.030}}             
\newcommand{\hatcurISOagexxxxC}{\ensuremath{6.6_{-1.8}^{+2.9}}}        
\newcommand{\hatcurISOsigmaxxxxC}{\ensuremath{0.00200\pm0.00024}}      
\newcommand{\hatcurISOMJxxxxC}{\ensuremath{3.58\pm0.13}}               
\newcommand{\hatcurISOMHxxxxC}{\ensuremath{3.20\pm0.12}}               
\newcommand{\hatcurISOMKxxxxC}{\ensuremath{3.14\pm0.12}}               
\newcommand{\hatcurISOJKxxxxC}{\ensuremath{0.44\pm0.03}}               
\newcommand{\hatcurISOspecxxxxC}{G}                                    
\newcommand{\hatcurRVKxxxxC}{\ensuremath{334.7\pm14.5}}                
\newcommand{\hatcurRVkxxxxC}{\ensuremath{-0.002\pm0.032}}              
\newcommand{\hatcurRVhxxxxC}{\ensuremath{0.051\pm0.040}}               
\newcommand{\hatcurRVgammaxxxxC}{\ensuremath{-266.1\pm13.1}}           
\newcommand{\hatcurRVjitterxxxxC}{\ensuremath{33.6}}                   
\newcommand{\hatcurRVfitrmsxxxxC}{\ensuremath{37.8}}                   
\newcommand{\hatcurRVeccenxxxxC}{\ensuremath{0.063\pm0.032}}           
\newcommand{\hatcurRVomegaxxxxC}{\ensuremath{95\pm63}}                 
\newcommand{\hatcurPPixxxxC}{\ensuremath{86.0\pm1.3}}                  
\newcommand{\hatcurPPgxxxxC}{\ensuremath{28.3\pm3.4}}                  
\newcommand{\hatcurPPloggxxxxC}{\ensuremath{3.45\pm0.05}}              
\newcommand{\hatcurPParxxxxC}{\ensuremath{4.66\pm0.22}}                
\newcommand{\hatcurPParelxxxxC}{\ensuremath{0.0238\pm0.0004}}          
\newcommand{\hatcurPPrhoxxxxC}{\ensuremath{1.12\pm0.19}}               
\newcommand{\hatcurPPmxxxxC}{\ensuremath{1.83\pm0.10}}                 
\newcommand{\hatcurPPmshortxxxxC}{\ensuremath{1.83}}                   
\newcommand{\hatcurPPmlongxxxxC}{\ensuremath{1.832\pm0.099}}           
\newcommand{\hatcurPPmexxxxC}{\ensuremath{582.3\pm31.4}}               
\newcommand{\hatcurPPmeshortxxxxC}{\ensuremath{582.3}}                 
\newcommand{\hatcurPPmelongxxxxC}{\ensuremath{582.34\pm31.36}}         
\newcommand{\hatcurPPrxxxxC}{\ensuremath{1.26\pm0.07}}                 
\newcommand{\hatcurPPrshortxxxxC}{\ensuremath{1.26}}                   
\newcommand{\hatcurPPrlongxxxxC}{\ensuremath{1.264\pm0.071}}           
\newcommand{\hatcurPPrexxxxC}{\ensuremath{14.2\pm0.8}}                 
\newcommand{\hatcurPPreshortxxxxC}{\ensuremath{14.2}}                  
\newcommand{\hatcurPPrelongxxxxC}{\ensuremath{14.17\pm0.80}}           
\newcommand{\hatcurPPmrcorrxxxxC}{\ensuremath{0.11}}                   
\newcommand{\hatcurPPteffxxxxC}{\ensuremath{1823\pm55}}                
\newcommand{\hatcurPPthetaxxxxC}{\ensuremath{0.067\pm0.005}}           
\newcommand{\hatcurPPfluxperixxxxC}{\ensuremath{2.82_{-0.38}^{+0.59}}} 
\newcommand{\hatcurPPfluxperidimxxxxC}{\ensuremath{9}}                 
\newcommand{\hatcurPPfluxapxxxxC}{\ensuremath{2.20\pm0.22}}            
\newcommand{\hatcurPPfluxapdimxxxxC}{\ensuremath{9}}                   
\newcommand{\hatcurPPfluxavgxxxxC}{\ensuremath{2.49\pm0.30}}           
\newcommand{\hatcurPPfluxavgdimxxxxC}{\ensuremath{9}}                  
\newcommand{\hatcurXsecphasexxxxC}{\ensuremath{0.4989\pm0.0202}}       
\newcommand{\hatcurXsecondaryxxxxC}{\ensuremath{2455565.844\pm0.027}}  
\newcommand{\hatcurXsecdurxxxxC}{\ensuremath{0.1013\pm0.0071}}         
\newcommand{\hatcurXsecingdurxxxxC}{\ensuremath{0.0120\pm0.0015}}      
\newcommand{\hatcurPPphiconjxxxxC}{\ensuremath{-0.0044\pm0.1363}}      
\newcommand{\hatcurPPperixxxxC}{\ensuremath{2455565.19\pm0.18}}        
\newcommand{\hatcurPPaequivxxxxC}{\ensuremath{0.0234\pm0.0014}}        
\newcommand{\hatcurPPtcircxxxxC}{\ensuremath{5.8_{-1.4}^{+2.0}}}       
\newcommand{\hatcurPPtinfallxxxxC}{\ensuremath{15.4_{-3.0}^{+4.3}}}    
\newcommand{\hatcurXdistxxxxC}{\ensuremath{317\pm17}}                  
\newcommand{\hatcurCCpmraxxxxC}{\ensuremath{-12.7\pm2.3}}              
\newcommand{\hatcurCCpmdecxxxxC}{\ensuremath{7.0\pm0.6}}               
\newcommand{\hatcurCCpmxxxxC}{\ensuremath{14.5014\pm2.37697}}          

\newcommand{\hatcurhtrxxxxD}{HTR114-005}                               
\newcommand{\hatcurfieldxxxxD}{115}                                    
\newcommand{\hatcurCCraxxxxD}{\ensuremath{18^{\mathrm h}57^{\mathrm m}11.16{\mathrm s}}}                             
\newcommand{\hatcurCCdecxxxxD}{\ensuremath{+51{\arcdeg}16{\arcmin}08.9{\arcsec}}}                            
\newcommand{\hatcurCCmagxxxxD}{99.999}                                 
\newcommand{\hatcurCCtwomassxxxxD}{2MASS~18571105+5116088}             
\newcommand{\hatcurCCntwomassxxxxD}{18571105+5116088}             
\newcommand{\hatcurCCgscxxxxD}{GSC~3553-00723}                         
\newcommand{\hatcurCCngscxxxxD}{3553-00723}                         
\newcommand{\hatcurCCtassmvshortxxxxD}{\ensuremath{13.2}}                   \newcommand{\hatcurCCtassmvxxxxD}{\ensuremath{13.23\pm0.32}}                              
\newcommand{\hatcurCCtwomassJmagxxxxD}{\ensuremath{12.092\pm0.027}}    
\newcommand{\hatcurCCtwomassHmagxxxxD}{\ensuremath{11.714\pm0.032}}    
\newcommand{\hatcurCCtwomassKmagxxxxD}{\ensuremath{11.667\pm0.020}}    
\newcommand{\hatcurCCcitJmagxxxxD}{\ensuremath{12.106\pm0.027}}        
\newcommand{\hatcurCCcitHmagxxxxD}{\ensuremath{11.710\pm0.032}}        
\newcommand{\hatcurCCcitKmagxxxxD}{\ensuremath{11.691\pm0.020}}        
\newcommand{\hatcurCCbbJmagxxxxD}{\ensuremath{12.160\pm0.029}}         
\newcommand{\hatcurCCbbHmagxxxxD}{\ensuremath{11.730\pm0.033}}         
\newcommand{\hatcurCCbbKmagxxxxD}{\ensuremath{11.711\pm0.020}}         
\newcommand{\hatcurCCesoJmagxxxxD}{\ensuremath{12.162\pm0.031}}        
\newcommand{\hatcurCCesoHmagxxxxD}{\ensuremath{11.724\pm0.036}}        
\newcommand{\hatcurCCesoKmagxxxxD}{\ensuremath{11.710\pm0.021}}        
\newcommand{\hatcurCCesoJHmagxxxxD}{\ensuremath{0.438\pm0.045}}        
\newcommand{\hatcurCCesoJKmagxxxxD}{\ensuremath{0.453\pm0.036}}        
\newcommand{\hatcurCCesoHKmagxxxxD}{\ensuremath{0.015\pm0.042}}        
\newcommand{\hatcurLCdipxxxxD}{\ensuremath{18.1}}                      
\newcommand{\hatcurLCrprstarxxxxD}{\ensuremath{0.1378\pm0.0030}}       
\newcommand{\hatcurLCbsqxxxxD}{\ensuremath{0.255_{-0.056}^{+0.044}}}   
\newcommand{\hatcurLCimpxxxxD}{\ensuremath{0.505_{-0.062}^{+0.041}}}   
\newcommand{\hatcurLCzetaxxxxD}{\ensuremath{24.33\pm0.18}}             
\newcommand{\hatcurLCdurxxxxD}{\ensuremath{0.0971\pm0.0015}}           
\newcommand{\hatcurLCdurshortxxxxD}{\ensuremath{0.0971}}               
\newcommand{\hatcurLCdurhrxxxxD}{\ensuremath{2.331\pm0.035}}           
\newcommand{\hatcurLCdurhrshortxxxxD}{\ensuremath{2.331}}              
\newcommand{\hatcurLCqxxxxD}{\ensuremath{0.0347\pm0.0005}}             
\newcommand{\hatcurLCqshortxxxxD}{\ensuremath{0.035}}                  
\newcommand{\hatcurLCingdurxxxxD}{\ensuremath{0.0153\pm0.0013}}        
\newcommand{\hatcurLCPxxxxD}{\ensuremath{2.797436\pm0.000007}}         
\newcommand{\hatcurLCPprecxxxxD}{\ensuremath{2.7974359}}               
\newcommand{\hatcurLCPshortxxxxD}{\ensuremath{2.7974}}                 
\newcommand{\hatcurLCTxxxxD}{\ensuremath{2455642.14318\pm0.00029}}     
\newcommand{\hatcurLCTAxxxxD}{\ensuremath{2454691.01497\pm0.00224}}    
\newcommand{\hatcurLCTBxxxxD}{\ensuremath{2455658.92779\pm0.00030}}    
\newcommand{\hatcurLChatnetmAxxxxD}{\ensuremath{13.2865\pm0.0002}}     
\newcommand{\hatcurLCiblendAxxxxD}{\ensuremath{0.80\pm0.06}}           
\newcommand{\hatcurLChatnetmBxxxxD}{\ensuremath{12.3855\pm0.0016}}     
\newcommand{\hatcurLCiblendBxxxxD}{\ensuremath{0.03\pm0.01}}           
\newcommand{\hatcurSMEiteffxxxxD}{\ensuremath{5570\pm100}}             
\newcommand{\hatcurSMEizfehxxxxD}{\ensuremath{+0.09\pm0.1}}             
\newcommand{\hatcurSMEizfehshortxxxxD}{\ensuremath{+0.09}}              
\newcommand{\hatcurSMEiloggxxxxD}{\ensuremath{4.67\pm0.1}}             
\newcommand{\hatcurSMEivsinxxxxD}{\ensuremath{2.95\pm0.5}}             
\newcommand{\hatcurSMEivmacxxxxD}{\ensuremath{0.00}}                   
\newcommand{\hatcurSMEivmicxxxxD}{\ensuremath{0.00}}                   
\newcommand{\hatcurSMEiiteffxxxxD}{\ensuremath{5500\pm100}}            
\newcommand{\hatcurSMEiizfehxxxxD}{\ensuremath{+0.03\pm0.10}}           
\newcommand{\hatcurSMEiizfehshortxxxxD}{\ensuremath{+0.03}}             
\newcommand{\hatcurSMEiiloggxxxxD}{\ensuremath{4.52\pm0.1}}            
\newcommand{\hatcurSMEiivsinxxxxD}{\ensuremath{3.07\pm0.5}}            
\newcommand{\hatcurSMEiivmacxxxxD}{\ensuremath{\ldots}}                  
\newcommand{\hatcurSMEiivmicxxxxD}{\ensuremath{\ldots}}                  
\newcommand{\hatcurDSteffxxxxD}{\ensuremath{NULL\pmNULL}}              
\newcommand{\hatcurDSzfehxxxxD}{\ensuremath{NULL\pmNULL}}              
\newcommand{\hatcurDSloggxxxxD}{\ensuremath{NULL\pmNULL}}              
\newcommand{\hatcurDSvsinixxxxD}{\ensuremath{NULL\pmNULL}}             
\newcommand{\hatcurDSgammaxxxxD}{\ensuremath{NULL\pmNULL}}             
\newcommand{\hatcurDSnumspecxxxxD}{\ensuremath{0}}                     
\newcommand{\hatcurDSspanxxxxD}{\ensuremath{0}}                        
\newcommand{\hatcurDSrvrmsxxxxD}{\ensuremath{0.00}}                    
\newcommand{\hatcurTRESteffxxxxD}{\ensuremath{5570\pm100}}             
\newcommand{\hatcurTRESzfehxxxxD}{\ensuremath{0.09\pm0.1}}             
\newcommand{\hatcurTRESloggxxxxD}{\ensuremath{4.67\pm0.1}}             
\newcommand{\hatcurTRESvsinixxxxD}{\ensuremath{2.95\pm0.5}}            
\newcommand{\hatcurTRESgammaxxxxD}{\ensuremath{-20.53\pm0.1}}           
\newcommand{\hatcurTRESnumspecxxxxD}{\ensuremath{13}}                  
\newcommand{\hatcurTRESspanxxxxD}{\ensuremath{52}}                     
\newcommand{\hatcurTRESrvrmsxxxxD}{\ensuremath{0.14}}                  
\newcommand{\hatcurFIESteffxxxxD}{\ensuremath{NULL\pmNULL}}            
\newcommand{\hatcurFIESzfehxxxxD}{\ensuremath{NULL\pmNULL}}            
\newcommand{\hatcurFIESloggxxxxD}{\ensuremath{NULL\pmNULL}}            
\newcommand{\hatcurFIESvsinixxxxD}{\ensuremath{NULL\pmNULL}}           
\newcommand{\hatcurFIESgammaxxxxD}{\ensuremath{NULL\pmNULL}}           
\newcommand{\hatcurFIESnumspecxxxxD}{\ensuremath{0}}                   
\newcommand{\hatcurFIESspanxxxxD}{\ensuremath{0}}                      
\newcommand{\hatcurFIESrvrmsxxxxD}{\ensuremath{0.00}}                  
\newcommand{\hatcurLBizxxxxD}{\ensuremath{0.2477}}                     
\newcommand{\hatcurLBiizxxxxD}{\ensuremath{0.3082}}                    
\newcommand{\hatcurLBiixxxxD}{\ensuremath{0.3156}}                     
\newcommand{\hatcurLBiiixxxxD}{\ensuremath{0.3032}}                    
\newcommand{\hatcurLBiIxxxxD}{\ensuremath{0.2932}}                     
\newcommand{\hatcurLBiiIxxxxD}{\ensuremath{0.3053}}                    
\newcommand{\hatcurLBigxxxxD}{\ensuremath{0.6258}}                     
\newcommand{\hatcurLBiigxxxxD}{\ensuremath{0.1808}}                    
\newcommand{\hatcurLBikepxxxxD}{\ensuremath{}}                 
\newcommand{\hatcurLBiikepxxxxD}{\ensuremath{}}                
\newcommand{\hatcurISOmxxxxD}{\ensuremath{0.93\pm0.04}}                
\newcommand{\hatcurISOmshortxxxxD}{\ensuremath{0.93}}                  
\newcommand{\hatcurISOmlongxxxxD}{\ensuremath{0.929\pm0.043}}          
\newcommand{\hatcurISOrxxxxD}{\ensuremath{0.88_{-0.04}^{+0.06}}}       
\newcommand{\hatcurISOrshortxxxxD}{\ensuremath{0.88}}                  
\newcommand{\hatcurISOrlongxxxxD}{\ensuremath{0.877_{-0.044}^{+0.059}}} 
\newcommand{\hatcurISOrhoxxxxD}{\ensuremath{1.95\pm0.30}}              
\newcommand{\hatcurISOloggxxxxD}{\ensuremath{4.52\pm0.05}}             
\newcommand{\hatcurISOlumxxxxD}{\ensuremath{0.62_{-0.09}^{+0.11}}}     
\newcommand{\hatcurISOlumshortxxxxD}{\ensuremath{0.62}}                
\newcommand{\hatcurISOmvxxxxD}{\ensuremath{5.41\pm0.19}}               
\newcommand{\hatcurISOvixxxxD}{\ensuremath{0.789\pm0.027}}             
\newcommand{\hatcurISOagexxxxD}{\ensuremath{3.6_{-2.2}^{+4.1}}}        
\newcommand{\hatcurISOsigmaxxxxD}{\ensuremath{0.00300\pm0.00079}}      
\newcommand{\hatcurISOMJxxxxD}{\ensuremath{4.11\pm0.15}}               
\newcommand{\hatcurISOMHxxxxD}{\ensuremath{3.71\pm0.14}}               
\newcommand{\hatcurISOMKxxxxD}{\ensuremath{3.64\pm0.14}}               
\newcommand{\hatcurISOJKxxxxD}{\ensuremath{0.47\pm0.03}}               
\newcommand{\hatcurISOspecxxxxD}{G}                                    
\newcommand{\hatcurRVKxxxxD}{\ensuremath{177.7\pm14.8}}                
\newcommand{\hatcurRVkxxxxD}{\ensuremath{-0.017\pm0.039}}              
\newcommand{\hatcurRVhxxxxD}{\ensuremath{0.007\pm0.060}}               
\newcommand{\hatcurRVgammaxxxxD}{\ensuremath{704.7\pm11.1}}            
\newcommand{\hatcurRVjitterxxxxD}{\ensuremath{25.8}}                   
\newcommand{\hatcurRVfitrmsxxxxD}{\ensuremath{35.4}}                   
\newcommand{\hatcurRVeccenxxxxD}{\ensuremath{0.058\pm0.038}}           
\newcommand{\hatcurRVomegaxxxxD}{\ensuremath{164\pm84}}                
\newcommand{\hatcurPPixxxxD}{\ensuremath{86.9_{-0.5}^{+0.4}}}          
\newcommand{\hatcurPPgxxxxD}{\ensuremath{21.0\pm3.2}}                  
\newcommand{\hatcurPPloggxxxxD}{\ensuremath{3.32\pm0.07}}              
\newcommand{\hatcurPParxxxxD}{\ensuremath{9.32_{-0.57}^{+0.42}}}       
\newcommand{\hatcurPParelxxxxD}{\ensuremath{0.0379\pm0.0006}}          
\newcommand{\hatcurPPrhoxxxxD}{\ensuremath{0.89\pm0.19}}               
\newcommand{\hatcurPPmxxxxD}{\ensuremath{1.17\pm0.10}}                 
\newcommand{\hatcurPPmshortxxxxD}{\ensuremath{1.17}}                   
\newcommand{\hatcurPPmlongxxxxD}{\ensuremath{1.169\pm0.103}}           
\newcommand{\hatcurPPmexxxxD}{\ensuremath{371.5\pm32.6}}               
\newcommand{\hatcurPPmeshortxxxxD}{\ensuremath{371.5}}                 
\newcommand{\hatcurPPmelongxxxxD}{\ensuremath{371.54\pm32.58}}         
\newcommand{\hatcurPPrxxxxD}{\ensuremath{1.18\pm0.08}}                 
\newcommand{\hatcurPPrshortxxxxD}{\ensuremath{1.18}}                   
\newcommand{\hatcurPPrlongxxxxD}{\ensuremath{1.178\pm0.077}}           
\newcommand{\hatcurPPrexxxxD}{\ensuremath{13.2\pm0.9}}                 
\newcommand{\hatcurPPreshortxxxxD}{\ensuremath{13.2}}                  
\newcommand{\hatcurPPrelongxxxxD}{\ensuremath{13.21\pm0.87}}           
\newcommand{\hatcurPPmrcorrxxxxD}{\ensuremath{0.02}}                   
\newcommand{\hatcurPPteffxxxxD}{\ensuremath{1271\pm47}}                
\newcommand{\hatcurPPthetaxxxxD}{\ensuremath{0.081\pm0.009}}           
\newcommand{\hatcurPPfluxperixxxxD}{\ensuremath{0.652_{-0.076}^{+0.188}}} 
\newcommand{\hatcurPPfluxperidimxxxxD}{\ensuremath{9}}                 
\newcommand{\hatcurPPfluxapxxxxD}{\ensuremath{0.528\pm0.075}}            
\newcommand{\hatcurPPfluxapdimxxxxD}{\ensuremath{9}}                   
\newcommand{\hatcurPPfluxavgxxxxD}{\ensuremath{0.589_{-0.075}^{+0.102}}}  
\newcommand{\hatcurPPfluxavgdimxxxxD}{\ensuremath{9}}                  
\newcommand{\hatcurXsecphasexxxxD}{\ensuremath{0.4890\pm0.0252}}       
\newcommand{\hatcurXsecondaryxxxxD}{\ensuremath{2455643.512\pm0.070}}  
\newcommand{\hatcurXsecdurxxxxD}{\ensuremath{0.0981\pm0.0083}}         
\newcommand{\hatcurXsecingdurxxxxD}{\ensuremath{0.0153\pm0.0029}}      
\newcommand{\hatcurPPphiconjxxxxD}{\ensuremath{-0.0957\pm0.2491}}      
\newcommand{\hatcurPPperixxxxD}{\ensuremath{2455642.41\pm0.70}}        
\newcommand{\hatcurPPaequivxxxxD}{\ensuremath{0.0482\pm0.0035}}        
\newcommand{\hatcurPPtcircxxxxD}{\ensuremath{127.6\pm41.1}}            
\newcommand{\hatcurPPtinfallxxxxD}{\ensuremath{1483.4\pm379.4}}        
\newcommand{\hatcurXdistxxxxD}{\ensuremath{411\pm26}}                  
\newcommand{\hatcurCCpmraxxxxD}{\ensuremath{-4.1\pm3.4}}               
\newcommand{\hatcurCCpmdecxxxxD}{\ensuremath{4.5\pm2.4}}               
\newcommand{\hatcurCCpmxxxxD}{\ensuremath{6.08769\pm4.16173}}          

\newcommand{\hatcurCCbbHmag}[1]{\ifnum#1=34 %
\hatcurCCbbHmagxxxxA
\else
\ifnum#1=35 %
\hatcurCCbbHmagxxxxB
\else
\ifnum#1=36 %
\hatcurCCbbHmagxxxxC
\else
\ifnum#1=37 %
\hatcurCCbbHmagxxxxD
\else
??????\fi
\fi
\fi
\fi
}
\newcommand{\hatcurCCbbJmag}[1]{\ifnum#1=34 %
\hatcurCCbbJmagxxxxA
\else
\ifnum#1=35 %
\hatcurCCbbJmagxxxxB
\else
\ifnum#1=36 %
\hatcurCCbbJmagxxxxC
\else
\ifnum#1=37 %
\hatcurCCbbJmagxxxxD
\else
??????\fi
\fi
\fi
\fi
}
\newcommand{\hatcurCCbbKmag}[1]{\ifnum#1=34 %
\hatcurCCbbKmagxxxxA
\else
\ifnum#1=35 %
\hatcurCCbbKmagxxxxB
\else
\ifnum#1=36 %
\hatcurCCbbKmagxxxxC
\else
\ifnum#1=37 %
\hatcurCCbbKmagxxxxD
\else
??????\fi
\fi
\fi
\fi
}
\newcommand{\hatcurCCcitHmag}[1]{\ifnum#1=34 %
\hatcurCCcitHmagxxxxA
\else
\ifnum#1=35 %
\hatcurCCcitHmagxxxxB
\else
\ifnum#1=36 %
\hatcurCCcitHmagxxxxC
\else
\ifnum#1=37 %
\hatcurCCcitHmagxxxxD
\else
??????\fi
\fi
\fi
\fi
}
\newcommand{\hatcurCCcitJmag}[1]{\ifnum#1=34 %
\hatcurCCcitJmagxxxxA
\else
\ifnum#1=35 %
\hatcurCCcitJmagxxxxB
\else
\ifnum#1=36 %
\hatcurCCcitJmagxxxxC
\else
\ifnum#1=37 %
\hatcurCCcitJmagxxxxD
\else
??????\fi
\fi
\fi
\fi
}
\newcommand{\hatcurCCcitKmag}[1]{\ifnum#1=34 %
\hatcurCCcitKmagxxxxA
\else
\ifnum#1=35 %
\hatcurCCcitKmagxxxxB
\else
\ifnum#1=36 %
\hatcurCCcitKmagxxxxC
\else
\ifnum#1=37 %
\hatcurCCcitKmagxxxxD
\else
??????\fi
\fi
\fi
\fi
}
\newcommand{\hatcurCCdec}[1]{\ifnum#1=34 %
\hatcurCCdecxxxxA
\else
\ifnum#1=35 %
\hatcurCCdecxxxxB
\else
\ifnum#1=36 %
\hatcurCCdecxxxxC
\else
\ifnum#1=37 %
\hatcurCCdecxxxxD
\else
??????\fi
\fi
\fi
\fi
}
\newcommand{\hatcurCCesoHKmag}[1]{\ifnum#1=34 %
\hatcurCCesoHKmagxxxxA
\else
\ifnum#1=35 %
\hatcurCCesoHKmagxxxxB
\else
\ifnum#1=36 %
\hatcurCCesoHKmagxxxxC
\else
\ifnum#1=37 %
\hatcurCCesoHKmagxxxxD
\else
??????\fi
\fi
\fi
\fi
}
\newcommand{\hatcurCCesoHmag}[1]{\ifnum#1=34 %
\hatcurCCesoHmagxxxxA
\else
\ifnum#1=35 %
\hatcurCCesoHmagxxxxB
\else
\ifnum#1=36 %
\hatcurCCesoHmagxxxxC
\else
\ifnum#1=37 %
\hatcurCCesoHmagxxxxD
\else
??????\fi
\fi
\fi
\fi
}
\newcommand{\hatcurCCesoJHmag}[1]{\ifnum#1=34 %
\hatcurCCesoJHmagxxxxA
\else
\ifnum#1=35 %
\hatcurCCesoJHmagxxxxB
\else
\ifnum#1=36 %
\hatcurCCesoJHmagxxxxC
\else
\ifnum#1=37 %
\hatcurCCesoJHmagxxxxD
\else
??????\fi
\fi
\fi
\fi
}
\newcommand{\hatcurCCesoJKmag}[1]{\ifnum#1=34 %
\hatcurCCesoJKmagxxxxA
\else
\ifnum#1=35 %
\hatcurCCesoJKmagxxxxB
\else
\ifnum#1=36 %
\hatcurCCesoJKmagxxxxC
\else
\ifnum#1=37 %
\hatcurCCesoJKmagxxxxD
\else
??????\fi
\fi
\fi
\fi
}
\newcommand{\hatcurCCesoJmag}[1]{\ifnum#1=34 %
\hatcurCCesoJmagxxxxA
\else
\ifnum#1=35 %
\hatcurCCesoJmagxxxxB
\else
\ifnum#1=36 %
\hatcurCCesoJmagxxxxC
\else
\ifnum#1=37 %
\hatcurCCesoJmagxxxxD
\else
??????\fi
\fi
\fi
\fi
}
\newcommand{\hatcurCCesoKmag}[1]{\ifnum#1=34 %
\hatcurCCesoKmagxxxxA
\else
\ifnum#1=35 %
\hatcurCCesoKmagxxxxB
\else
\ifnum#1=36 %
\hatcurCCesoKmagxxxxC
\else
\ifnum#1=37 %
\hatcurCCesoKmagxxxxD
\else
??????\fi
\fi
\fi
\fi
}
\newcommand{\hatcurCCgsc}[1]{\ifnum#1=34 %
\hatcurCCgscxxxxA
\else
\ifnum#1=35 %
\hatcurCCgscxxxxB
\else
\ifnum#1=36 %
\hatcurCCgscxxxxC
\else
\ifnum#1=37 %
\hatcurCCgscxxxxD
\else
??????\fi
\fi
\fi
\fi
}
\newcommand{\hatcurCCngsc}[1]{\ifnum#1=34 %
\hatcurCCngscxxxxA
\else
\ifnum#1=35 %
\hatcurCCngscxxxxB
\else
\ifnum#1=36 %
\hatcurCCngscxxxxC
\else
\ifnum#1=37 %
\hatcurCCngscxxxxD
\else
??????\fi
\fi
\fi
\fi
}
\newcommand{\hatcurCCmag}[1]{\ifnum#1=34 %
\hatcurCCmagxxxxA
\else
\ifnum#1=35 %
\hatcurCCmagxxxxB
\else
\ifnum#1=36 %
\hatcurCCmagxxxxC
\else
\ifnum#1=37 %
\hatcurCCmagxxxxD
\else
??????\fi
\fi
\fi
\fi
}
\newcommand{\hatcurCCpm}[1]{\ifnum#1=34 %
\hatcurCCpmxxxxA
\else
\ifnum#1=35 %
\hatcurCCpmxxxxB
\else
\ifnum#1=36 %
\hatcurCCpmxxxxC
\else
\ifnum#1=37 %
\hatcurCCpmxxxxD
\else
??????\fi
\fi
\fi
\fi
}
\newcommand{\hatcurCCpmdec}[1]{\ifnum#1=34 %
\hatcurCCpmdecxxxxA
\else
\ifnum#1=35 %
\hatcurCCpmdecxxxxB
\else
\ifnum#1=36 %
\hatcurCCpmdecxxxxC
\else
\ifnum#1=37 %
\hatcurCCpmdecxxxxD
\else
??????\fi
\fi
\fi
\fi
}
\newcommand{\hatcurCCpmra}[1]{\ifnum#1=34 %
\hatcurCCpmraxxxxA
\else
\ifnum#1=35 %
\hatcurCCpmraxxxxB
\else
\ifnum#1=36 %
\hatcurCCpmraxxxxC
\else
\ifnum#1=37 %
\hatcurCCpmraxxxxD
\else
??????\fi
\fi
\fi
\fi
}
\newcommand{\hatcurCCra}[1]{\ifnum#1=34 %
\hatcurCCraxxxxA
\else
\ifnum#1=35 %
\hatcurCCraxxxxB
\else
\ifnum#1=36 %
\hatcurCCraxxxxC
\else
\ifnum#1=37 %
\hatcurCCraxxxxD
\else
??????\fi
\fi
\fi
\fi
}
\newcommand{\hatcurCCtassmvshort}[1]{\ifnum#1=34 %
\hatcurCCtassmvshortxxxxA
\else
\ifnum#1=35 %
\hatcurCCtassmvshortxxxxB
\else
\ifnum#1=36 %
\hatcurCCtassmvshortxxxxC
\else
\ifnum#1=37 %
\hatcurCCtassmvshortxxxxD
\else
??????\fi
\fi
\fi
\fi
}
\newcommand{\hatcurCCtassmv}[1]{\ifnum#1=34 %
\hatcurCCtassmvxxxxA
\else
\ifnum#1=35 %
\hatcurCCtassmvxxxxB
\else
\ifnum#1=36 %
\hatcurCCtassmvxxxxC
\else
\ifnum#1=37 %
\hatcurCCtassmvxxxxD
\else
??????\fi
\fi
\fi
\fi
}
\newcommand{\hatcurCCtwomass}[1]{\ifnum#1=34 %
\hatcurCCtwomassxxxxA
\else
\ifnum#1=35 %
\hatcurCCtwomassxxxxB
\else
\ifnum#1=36 %
\hatcurCCtwomassxxxxC
\else
\ifnum#1=37 %
\hatcurCCtwomassxxxxD
\else
??????\fi
\fi
\fi
\fi
}
\newcommand{\hatcurCCntwomass}[1]{\ifnum#1=34 %
\hatcurCCntwomassxxxxA
\else
\ifnum#1=35 %
\hatcurCCntwomassxxxxB
\else
\ifnum#1=36 %
\hatcurCCntwomassxxxxC
\else
\ifnum#1=37 %
\hatcurCCntwomassxxxxD
\else
??????\fi
\fi
\fi
\fi
}
\newcommand{\hatcurCCtwomassHmag}[1]{\ifnum#1=34 %
\hatcurCCtwomassHmagxxxxA
\else
\ifnum#1=35 %
\hatcurCCtwomassHmagxxxxB
\else
\ifnum#1=36 %
\hatcurCCtwomassHmagxxxxC
\else
\ifnum#1=37 %
\hatcurCCtwomassHmagxxxxD
\else
??????\fi
\fi
\fi
\fi
}
\newcommand{\hatcurCCtwomassJmag}[1]{\ifnum#1=34 %
\hatcurCCtwomassJmagxxxxA
\else
\ifnum#1=35 %
\hatcurCCtwomassJmagxxxxB
\else
\ifnum#1=36 %
\hatcurCCtwomassJmagxxxxC
\else
\ifnum#1=37 %
\hatcurCCtwomassJmagxxxxD
\else
??????\fi
\fi
\fi
\fi
}
\newcommand{\hatcurCCtwomassKmag}[1]{\ifnum#1=34 %
\hatcurCCtwomassKmagxxxxA
\else
\ifnum#1=35 %
\hatcurCCtwomassKmagxxxxB
\else
\ifnum#1=36 %
\hatcurCCtwomassKmagxxxxC
\else
\ifnum#1=37 %
\hatcurCCtwomassKmagxxxxD
\else
??????\fi
\fi
\fi
\fi
}
\newcommand{\hatcurDSgamma}[1]{\ifnum#1=34 %
\hatcurDSgammaxxxxA
\else
\ifnum#1=35 %
\hatcurDSgammaxxxxB
\else
\ifnum#1=36 %
\hatcurDSgammaxxxxC
\else
\ifnum#1=37 %
\hatcurDSgammaxxxxD
\else
??????\fi
\fi
\fi
\fi
}
\newcommand{\hatcurDSlogg}[1]{\ifnum#1=34 %
\hatcurDSloggxxxxA
\else
\ifnum#1=35 %
\hatcurDSloggxxxxB
\else
\ifnum#1=36 %
\hatcurDSloggxxxxC
\else
\ifnum#1=37 %
\hatcurDSloggxxxxD
\else
??????\fi
\fi
\fi
\fi
}
\newcommand{\hatcurDSnumspec}[1]{\ifnum#1=34 %
\hatcurDSnumspecxxxxA
\else
\ifnum#1=35 %
\hatcurDSnumspecxxxxB
\else
\ifnum#1=36 %
\hatcurDSnumspecxxxxC
\else
\ifnum#1=37 %
\hatcurDSnumspecxxxxD
\else
??????\fi
\fi
\fi
\fi
}
\newcommand{\hatcurDSrvrms}[1]{\ifnum#1=34 %
\hatcurDSrvrmsxxxxA
\else
\ifnum#1=35 %
\hatcurDSrvrmsxxxxB
\else
\ifnum#1=36 %
\hatcurDSrvrmsxxxxC
\else
\ifnum#1=37 %
\hatcurDSrvrmsxxxxD
\else
??????\fi
\fi
\fi
\fi
}
\newcommand{\hatcurDSspan}[1]{\ifnum#1=34 %
\hatcurDSspanxxxxA
\else
\ifnum#1=35 %
\hatcurDSspanxxxxB
\else
\ifnum#1=36 %
\hatcurDSspanxxxxC
\else
\ifnum#1=37 %
\hatcurDSspanxxxxD
\else
??????\fi
\fi
\fi
\fi
}
\newcommand{\hatcurDSteff}[1]{\ifnum#1=34 %
\hatcurDSteffxxxxA
\else
\ifnum#1=35 %
\hatcurDSteffxxxxB
\else
\ifnum#1=36 %
\hatcurDSteffxxxxC
\else
\ifnum#1=37 %
\hatcurDSteffxxxxD
\else
??????\fi
\fi
\fi
\fi
}
\newcommand{\hatcurDSvsini}[1]{\ifnum#1=34 %
\hatcurDSvsinixxxxA
\else
\ifnum#1=35 %
\hatcurDSvsinixxxxB
\else
\ifnum#1=36 %
\hatcurDSvsinixxxxC
\else
\ifnum#1=37 %
\hatcurDSvsinixxxxD
\else
??????\fi
\fi
\fi
\fi
}
\newcommand{\hatcurDSzfeh}[1]{\ifnum#1=34 %
\hatcurDSzfehxxxxA
\else
\ifnum#1=35 %
\hatcurDSzfehxxxxB
\else
\ifnum#1=36 %
\hatcurDSzfehxxxxC
\else
\ifnum#1=37 %
\hatcurDSzfehxxxxD
\else
??????\fi
\fi
\fi
\fi
}
\newcommand{\hatcurfield}[1]{\ifnum#1=34 %
\hatcurfieldxxxxA
\else
\ifnum#1=35 %
\hatcurfieldxxxxB
\else
\ifnum#1=36 %
\hatcurfieldxxxxC
\else
\ifnum#1=37 %
\hatcurfieldxxxxD
\else
??????\fi
\fi
\fi
\fi
}
\newcommand{\hatcurFIESgamma}[1]{\ifnum#1=34 %
\hatcurFIESgammaxxxxA
\else
\ifnum#1=35 %
\hatcurFIESgammaxxxxB
\else
\ifnum#1=36 %
\hatcurFIESgammaxxxxC
\else
\ifnum#1=37 %
\hatcurFIESgammaxxxxD
\else
??????\fi
\fi
\fi
\fi
}
\newcommand{\hatcurFIESlogg}[1]{\ifnum#1=34 %
\hatcurFIESloggxxxxA
\else
\ifnum#1=35 %
\hatcurFIESloggxxxxB
\else
\ifnum#1=36 %
\hatcurFIESloggxxxxC
\else
\ifnum#1=37 %
\hatcurFIESloggxxxxD
\else
??????\fi
\fi
\fi
\fi
}
\newcommand{\hatcurFIESnumspec}[1]{\ifnum#1=34 %
\hatcurFIESnumspecxxxxA
\else
\ifnum#1=35 %
\hatcurFIESnumspecxxxxB
\else
\ifnum#1=36 %
\hatcurFIESnumspecxxxxC
\else
\ifnum#1=37 %
\hatcurFIESnumspecxxxxD
\else
??????\fi
\fi
\fi
\fi
}
\newcommand{\hatcurFIESrvrms}[1]{\ifnum#1=34 %
\hatcurFIESrvrmsxxxxA
\else
\ifnum#1=35 %
\hatcurFIESrvrmsxxxxB
\else
\ifnum#1=36 %
\hatcurFIESrvrmsxxxxC
\else
\ifnum#1=37 %
\hatcurFIESrvrmsxxxxD
\else
??????\fi
\fi
\fi
\fi
}
\newcommand{\hatcurFIESspan}[1]{\ifnum#1=34 %
\hatcurFIESspanxxxxA
\else
\ifnum#1=35 %
\hatcurFIESspanxxxxB
\else
\ifnum#1=36 %
\hatcurFIESspanxxxxC
\else
\ifnum#1=37 %
\hatcurFIESspanxxxxD
\else
??????\fi
\fi
\fi
\fi
}
\newcommand{\hatcurFIESteff}[1]{\ifnum#1=34 %
\hatcurFIESteffxxxxA
\else
\ifnum#1=35 %
\hatcurFIESteffxxxxB
\else
\ifnum#1=36 %
\hatcurFIESteffxxxxC
\else
\ifnum#1=37 %
\hatcurFIESteffxxxxD
\else
??????\fi
\fi
\fi
\fi
}
\newcommand{\hatcurFIESvsini}[1]{\ifnum#1=34 %
\hatcurFIESvsinixxxxA
\else
\ifnum#1=35 %
\hatcurFIESvsinixxxxB
\else
\ifnum#1=36 %
\hatcurFIESvsinixxxxC
\else
\ifnum#1=37 %
\hatcurFIESvsinixxxxD
\else
??????\fi
\fi
\fi
\fi
}
\newcommand{\hatcurFIESzfeh}[1]{\ifnum#1=34 %
\hatcurFIESzfehxxxxA
\else
\ifnum#1=35 %
\hatcurFIESzfehxxxxB
\else
\ifnum#1=36 %
\hatcurFIESzfehxxxxC
\else
\ifnum#1=37 %
\hatcurFIESzfehxxxxD
\else
??????\fi
\fi
\fi
\fi
}
\newcommand{\hatcurhtr}[1]{\ifnum#1=34 %
\hatcurhtrxxxxA
\else
\ifnum#1=35 %
\hatcurhtrxxxxB
\else
\ifnum#1=36 %
\hatcurhtrxxxxC
\else
\ifnum#1=37 %
\hatcurhtrxxxxD
\else
??????\fi
\fi
\fi
\fi
}
\newcommand{\hatcurISOage}[1]{\ifnum#1=34 %
\hatcurISOagexxxxA
\else
\ifnum#1=35 %
\hatcurISOagexxxxB
\else
\ifnum#1=36 %
\hatcurISOagexxxxC
\else
\ifnum#1=37 %
\hatcurISOagexxxxD
\else
??????\fi
\fi
\fi
\fi
}
\newcommand{\hatcurISOJK}[1]{\ifnum#1=34 %
\hatcurISOJKxxxxA
\else
\ifnum#1=35 %
\hatcurISOJKxxxxB
\else
\ifnum#1=36 %
\hatcurISOJKxxxxC
\else
\ifnum#1=37 %
\hatcurISOJKxxxxD
\else
??????\fi
\fi
\fi
\fi
}
\newcommand{\hatcurISOlogg}[1]{\ifnum#1=34 %
\hatcurISOloggxxxxA
\else
\ifnum#1=35 %
\hatcurISOloggxxxxB
\else
\ifnum#1=36 %
\hatcurISOloggxxxxC
\else
\ifnum#1=37 %
\hatcurISOloggxxxxD
\else
??????\fi
\fi
\fi
\fi
}
\newcommand{\hatcurISOlum}[1]{\ifnum#1=34 %
\hatcurISOlumxxxxA
\else
\ifnum#1=35 %
\hatcurISOlumxxxxB
\else
\ifnum#1=36 %
\hatcurISOlumxxxxC
\else
\ifnum#1=37 %
\hatcurISOlumxxxxD
\else
??????\fi
\fi
\fi
\fi
}
\newcommand{\hatcurISOlumshort}[1]{\ifnum#1=34 %
\hatcurISOlumshortxxxxA
\else
\ifnum#1=35 %
\hatcurISOlumshortxxxxB
\else
\ifnum#1=36 %
\hatcurISOlumshortxxxxC
\else
\ifnum#1=37 %
\hatcurISOlumshortxxxxD
\else
??????\fi
\fi
\fi
\fi
}
\newcommand{\hatcurISOm}[1]{\ifnum#1=34 %
\hatcurISOmxxxxA
\else
\ifnum#1=35 %
\hatcurISOmxxxxB
\else
\ifnum#1=36 %
\hatcurISOmxxxxC
\else
\ifnum#1=37 %
\hatcurISOmxxxxD
\else
??????\fi
\fi
\fi
\fi
}
\newcommand{\hatcurISOMH}[1]{\ifnum#1=34 %
\hatcurISOMHxxxxA
\else
\ifnum#1=35 %
\hatcurISOMHxxxxB
\else
\ifnum#1=36 %
\hatcurISOMHxxxxC
\else
\ifnum#1=37 %
\hatcurISOMHxxxxD
\else
??????\fi
\fi
\fi
\fi
}
\newcommand{\hatcurISOMJ}[1]{\ifnum#1=34 %
\hatcurISOMJxxxxA
\else
\ifnum#1=35 %
\hatcurISOMJxxxxB
\else
\ifnum#1=36 %
\hatcurISOMJxxxxC
\else
\ifnum#1=37 %
\hatcurISOMJxxxxD
\else
??????\fi
\fi
\fi
\fi
}
\newcommand{\hatcurISOMK}[1]{\ifnum#1=34 %
\hatcurISOMKxxxxA
\else
\ifnum#1=35 %
\hatcurISOMKxxxxB
\else
\ifnum#1=36 %
\hatcurISOMKxxxxC
\else
\ifnum#1=37 %
\hatcurISOMKxxxxD
\else
??????\fi
\fi
\fi
\fi
}
\newcommand{\hatcurISOmlong}[1]{\ifnum#1=34 %
\hatcurISOmlongxxxxA
\else
\ifnum#1=35 %
\hatcurISOmlongxxxxB
\else
\ifnum#1=36 %
\hatcurISOmlongxxxxC
\else
\ifnum#1=37 %
\hatcurISOmlongxxxxD
\else
??????\fi
\fi
\fi
\fi
}
\newcommand{\hatcurISOmshort}[1]{\ifnum#1=34 %
\hatcurISOmshortxxxxA
\else
\ifnum#1=35 %
\hatcurISOmshortxxxxB
\else
\ifnum#1=36 %
\hatcurISOmshortxxxxC
\else
\ifnum#1=37 %
\hatcurISOmshortxxxxD
\else
??????\fi
\fi
\fi
\fi
}
\newcommand{\hatcurISOmv}[1]{\ifnum#1=34 %
\hatcurISOmvxxxxA
\else
\ifnum#1=35 %
\hatcurISOmvxxxxB
\else
\ifnum#1=36 %
\hatcurISOmvxxxxC
\else
\ifnum#1=37 %
\hatcurISOmvxxxxD
\else
??????\fi
\fi
\fi
\fi
}
\newcommand{\hatcurISOr}[1]{\ifnum#1=34 %
\hatcurISOrxxxxA
\else
\ifnum#1=35 %
\hatcurISOrxxxxB
\else
\ifnum#1=36 %
\hatcurISOrxxxxC
\else
\ifnum#1=37 %
\hatcurISOrxxxxD
\else
??????\fi
\fi
\fi
\fi
}
\newcommand{\hatcurISOrho}[1]{\ifnum#1=34 %
\hatcurISOrhoxxxxA
\else
\ifnum#1=35 %
\hatcurISOrhoxxxxB
\else
\ifnum#1=36 %
\hatcurISOrhoxxxxC
\else
\ifnum#1=37 %
\hatcurISOrhoxxxxD
\else
??????\fi
\fi
\fi
\fi
}
\newcommand{\hatcurISOrlong}[1]{\ifnum#1=34 %
\hatcurISOrlongxxxxA
\else
\ifnum#1=35 %
\hatcurISOrlongxxxxB
\else
\ifnum#1=36 %
\hatcurISOrlongxxxxC
\else
\ifnum#1=37 %
\hatcurISOrlongxxxxD
\else
??????\fi
\fi
\fi
\fi
}
\newcommand{\hatcurISOrshort}[1]{\ifnum#1=34 %
\hatcurISOrshortxxxxA
\else
\ifnum#1=35 %
\hatcurISOrshortxxxxB
\else
\ifnum#1=36 %
\hatcurISOrshortxxxxC
\else
\ifnum#1=37 %
\hatcurISOrshortxxxxD
\else
??????\fi
\fi
\fi
\fi
}
\newcommand{\hatcurISOsigma}[1]{\ifnum#1=34 %
\hatcurISOsigmaxxxxA
\else
\ifnum#1=35 %
\hatcurISOsigmaxxxxB
\else
\ifnum#1=36 %
\hatcurISOsigmaxxxxC
\else
\ifnum#1=37 %
\hatcurISOsigmaxxxxD
\else
??????\fi
\fi
\fi
\fi
}
\newcommand{\hatcurISOspec}[1]{\ifnum#1=34 %
\hatcurISOspecxxxxA
\else
\ifnum#1=35 %
\hatcurISOspecxxxxB
\else
\ifnum#1=36 %
\hatcurISOspecxxxxC
\else
\ifnum#1=37 %
\hatcurISOspecxxxxD
\else
??????\fi
\fi
\fi
\fi
}
\newcommand{\hatcurISOvi}[1]{\ifnum#1=34 %
\hatcurISOvixxxxA
\else
\ifnum#1=35 %
\hatcurISOvixxxxB
\else
\ifnum#1=36 %
\hatcurISOvixxxxC
\else
\ifnum#1=37 %
\hatcurISOvixxxxD
\else
??????\fi
\fi
\fi
\fi
}
\newcommand{\hatcurLBig}[1]{\ifnum#1=34 %
\hatcurLBigxxxxA
\else
\ifnum#1=35 %
\hatcurLBigxxxxB
\else
\ifnum#1=36 %
\hatcurLBigxxxxC
\else
\ifnum#1=37 %
\hatcurLBigxxxxD
\else
??????\fi
\fi
\fi
\fi
}
\newcommand{\hatcurLBii}[1]{\ifnum#1=34 %
\hatcurLBiixxxxA
\else
\ifnum#1=35 %
\hatcurLBiixxxxB
\else
\ifnum#1=36 %
\hatcurLBiixxxxC
\else
\ifnum#1=37 %
\hatcurLBiixxxxD
\else
??????\fi
\fi
\fi
\fi
}
\newcommand{\hatcurLBiI}[1]{\ifnum#1=34 %
\hatcurLBiIxxxxA
\else
\ifnum#1=35 %
\hatcurLBiIxxxxB
\else
\ifnum#1=36 %
\hatcurLBiIxxxxC
\else
\ifnum#1=37 %
\hatcurLBiIxxxxD
\else
??????\fi
\fi
\fi
\fi
}
\newcommand{\hatcurLBiig}[1]{\ifnum#1=34 %
\hatcurLBiigxxxxA
\else
\ifnum#1=35 %
\hatcurLBiigxxxxB
\else
\ifnum#1=36 %
\hatcurLBiigxxxxC
\else
\ifnum#1=37 %
\hatcurLBiigxxxxD
\else
??????\fi
\fi
\fi
\fi
}
\newcommand{\hatcurLBiii}[1]{\ifnum#1=34 %
\hatcurLBiiixxxxA
\else
\ifnum#1=35 %
\hatcurLBiiixxxxB
\else
\ifnum#1=36 %
\hatcurLBiiixxxxC
\else
\ifnum#1=37 %
\hatcurLBiiixxxxD
\else
??????\fi
\fi
\fi
\fi
}
\newcommand{\hatcurLBiiI}[1]{\ifnum#1=34 %
\hatcurLBiiIxxxxA
\else
\ifnum#1=35 %
\hatcurLBiiIxxxxB
\else
\ifnum#1=36 %
\hatcurLBiiIxxxxC
\else
\ifnum#1=37 %
\hatcurLBiiIxxxxD
\else
??????\fi
\fi
\fi
\fi
}
\newcommand{\hatcurLBiikep}[1]{\ifnum#1=34 %
\hatcurLBiikepxxxxA
\else
\ifnum#1=35 %
\hatcurLBiikepxxxxB
\else
\ifnum#1=36 %
\hatcurLBiikepxxxxC
\else
\ifnum#1=37 %
\hatcurLBiikepxxxxD
\else
??????\fi
\fi
\fi
\fi
}
\newcommand{\hatcurLBiiz}[1]{\ifnum#1=34 %
\hatcurLBiizxxxxA
\else
\ifnum#1=35 %
\hatcurLBiizxxxxB
\else
\ifnum#1=36 %
\hatcurLBiizxxxxC
\else
\ifnum#1=37 %
\hatcurLBiizxxxxD
\else
??????\fi
\fi
\fi
\fi
}
\newcommand{\hatcurLBikep}[1]{\ifnum#1=34 %
\hatcurLBikepxxxxA
\else
\ifnum#1=35 %
\hatcurLBikepxxxxB
\else
\ifnum#1=36 %
\hatcurLBikepxxxxC
\else
\ifnum#1=37 %
\hatcurLBikepxxxxD
\else
??????\fi
\fi
\fi
\fi
}
\newcommand{\hatcurLBiz}[1]{\ifnum#1=34 %
\hatcurLBizxxxxA
\else
\ifnum#1=35 %
\hatcurLBizxxxxB
\else
\ifnum#1=36 %
\hatcurLBizxxxxC
\else
\ifnum#1=37 %
\hatcurLBizxxxxD
\else
??????\fi
\fi
\fi
\fi
}
\newcommand{\hatcurLCbsq}[1]{\ifnum#1=34 %
\hatcurLCbsqxxxxA
\else
\ifnum#1=35 %
\hatcurLCbsqxxxxB
\else
\ifnum#1=36 %
\hatcurLCbsqxxxxC
\else
\ifnum#1=37 %
\hatcurLCbsqxxxxD
\else
??????\fi
\fi
\fi
\fi
}
\newcommand{\hatcurLCdip}[1]{\ifnum#1=34 %
\hatcurLCdipxxxxA
\else
\ifnum#1=35 %
\hatcurLCdipxxxxB
\else
\ifnum#1=36 %
\hatcurLCdipxxxxC
\else
\ifnum#1=37 %
\hatcurLCdipxxxxD
\else
??????\fi
\fi
\fi
\fi
}
\newcommand{\hatcurLCdur}[1]{\ifnum#1=34 %
\hatcurLCdurxxxxA
\else
\ifnum#1=35 %
\hatcurLCdurxxxxB
\else
\ifnum#1=36 %
\hatcurLCdurxxxxC
\else
\ifnum#1=37 %
\hatcurLCdurxxxxD
\else
??????\fi
\fi
\fi
\fi
}
\newcommand{\hatcurLCdurhr}[1]{\ifnum#1=34 %
\hatcurLCdurhrxxxxA
\else
\ifnum#1=35 %
\hatcurLCdurhrxxxxB
\else
\ifnum#1=36 %
\hatcurLCdurhrxxxxC
\else
\ifnum#1=37 %
\hatcurLCdurhrxxxxD
\else
??????\fi
\fi
\fi
\fi
}
\newcommand{\hatcurLCdurhrshort}[1]{\ifnum#1=34 %
\hatcurLCdurhrshortxxxxA
\else
\ifnum#1=35 %
\hatcurLCdurhrshortxxxxB
\else
\ifnum#1=36 %
\hatcurLCdurhrshortxxxxC
\else
\ifnum#1=37 %
\hatcurLCdurhrshortxxxxD
\else
??????\fi
\fi
\fi
\fi
}
\newcommand{\hatcurLCdurshort}[1]{\ifnum#1=34 %
\hatcurLCdurshortxxxxA
\else
\ifnum#1=35 %
\hatcurLCdurshortxxxxB
\else
\ifnum#1=36 %
\hatcurLCdurshortxxxxC
\else
\ifnum#1=37 %
\hatcurLCdurshortxxxxD
\else
??????\fi
\fi
\fi
\fi
}
\newcommand{\hatcurLChatnetm}[1]{\ifnum#1=35 %
\hatcurLChatnetmxxxxB
\else
??????\fi
}
\newcommand{\hatcurLChatnetmA}[1]{\ifnum#1=34 %
\hatcurLChatnetmAxxxxA
\else
\ifnum#1=36 %
\hatcurLChatnetmAxxxxC
\else
\ifnum#1=37 %
\hatcurLChatnetmAxxxxD
\else
??????\fi
\fi
\fi
}
\newcommand{\hatcurLChatnetmB}[1]{\ifnum#1=34 %
\hatcurLChatnetmBxxxxA
\else
\ifnum#1=36 %
\hatcurLChatnetmBxxxxC
\else
\ifnum#1=37 %
\hatcurLChatnetmBxxxxD
\else
??????\fi
\fi
\fi
}
\newcommand{\hatcurLCiblend}[1]{\ifnum#1=35 %
\hatcurLCiblendxxxxB
\else
??????\fi
}
\newcommand{\hatcurLCiblendA}[1]{\ifnum#1=34 %
\hatcurLCiblendAxxxxA
\else
\ifnum#1=36 %
\hatcurLCiblendAxxxxC
\else
\ifnum#1=37 %
\hatcurLCiblendAxxxxD
\else
??????\fi
\fi
\fi
}
\newcommand{\hatcurLCiblendB}[1]{\ifnum#1=34 %
\hatcurLCiblendBxxxxA
\else
\ifnum#1=36 %
\hatcurLCiblendBxxxxC
\else
\ifnum#1=37 %
\hatcurLCiblendBxxxxD
\else
??????\fi
\fi
\fi
}
\newcommand{\hatcurLCimp}[1]{\ifnum#1=34 %
\hatcurLCimpxxxxA
\else
\ifnum#1=35 %
\hatcurLCimpxxxxB
\else
\ifnum#1=36 %
\hatcurLCimpxxxxC
\else
\ifnum#1=37 %
\hatcurLCimpxxxxD
\else
??????\fi
\fi
\fi
\fi
}
\newcommand{\hatcurLCingdur}[1]{\ifnum#1=34 %
\hatcurLCingdurxxxxA
\else
\ifnum#1=35 %
\hatcurLCingdurxxxxB
\else
\ifnum#1=36 %
\hatcurLCingdurxxxxC
\else
\ifnum#1=37 %
\hatcurLCingdurxxxxD
\else
??????\fi
\fi
\fi
\fi
}
\newcommand{\hatcurLCP}[1]{\ifnum#1=34 %
\hatcurLCPxxxxA
\else
\ifnum#1=35 %
\hatcurLCPxxxxB
\else
\ifnum#1=36 %
\hatcurLCPxxxxC
\else
\ifnum#1=37 %
\hatcurLCPxxxxD
\else
??????\fi
\fi
\fi
\fi
}
\newcommand{\hatcurLCPprec}[1]{\ifnum#1=34 %
\hatcurLCPprecxxxxA
\else
\ifnum#1=35 %
\hatcurLCPprecxxxxB
\else
\ifnum#1=36 %
\hatcurLCPprecxxxxC
\else
\ifnum#1=37 %
\hatcurLCPprecxxxxD
\else
??????\fi
\fi
\fi
\fi
}
\newcommand{\hatcurLCPshort}[1]{\ifnum#1=34 %
\hatcurLCPshortxxxxA
\else
\ifnum#1=35 %
\hatcurLCPshortxxxxB
\else
\ifnum#1=36 %
\hatcurLCPshortxxxxC
\else
\ifnum#1=37 %
\hatcurLCPshortxxxxD
\else
??????\fi
\fi
\fi
\fi
}
\newcommand{\hatcurLCq}[1]{\ifnum#1=34 %
\hatcurLCqxxxxA
\else
\ifnum#1=35 %
\hatcurLCqxxxxB
\else
\ifnum#1=36 %
\hatcurLCqxxxxC
\else
\ifnum#1=37 %
\hatcurLCqxxxxD
\else
??????\fi
\fi
\fi
\fi
}
\newcommand{\hatcurLCqshort}[1]{\ifnum#1=34 %
\hatcurLCqshortxxxxA
\else
\ifnum#1=35 %
\hatcurLCqshortxxxxB
\else
\ifnum#1=36 %
\hatcurLCqshortxxxxC
\else
\ifnum#1=37 %
\hatcurLCqshortxxxxD
\else
??????\fi
\fi
\fi
\fi
}
\newcommand{\hatcurLCrprstar}[1]{\ifnum#1=34 %
\hatcurLCrprstarxxxxA
\else
\ifnum#1=35 %
\hatcurLCrprstarxxxxB
\else
\ifnum#1=36 %
\hatcurLCrprstarxxxxC
\else
\ifnum#1=37 %
\hatcurLCrprstarxxxxD
\else
??????\fi
\fi
\fi
\fi
}
\newcommand{\hatcurLCT}[1]{\ifnum#1=34 %
\hatcurLCTxxxxA
\else
\ifnum#1=35 %
\hatcurLCTxxxxB
\else
\ifnum#1=36 %
\hatcurLCTxxxxC
\else
\ifnum#1=37 %
\hatcurLCTxxxxD
\else
??????\fi
\fi
\fi
\fi
}
\newcommand{\hatcurLCTA}[1]{\ifnum#1=34 %
\hatcurLCTAxxxxA
\else
\ifnum#1=35 %
\hatcurLCTAxxxxB
\else
\ifnum#1=36 %
\hatcurLCTAxxxxC
\else
\ifnum#1=37 %
\hatcurLCTAxxxxD
\else
??????\fi
\fi
\fi
\fi
}
\newcommand{\hatcurLCTB}[1]{\ifnum#1=34 %
\hatcurLCTBxxxxA
\else
\ifnum#1=35 %
\hatcurLCTBxxxxB
\else
\ifnum#1=36 %
\hatcurLCTBxxxxC
\else
\ifnum#1=37 %
\hatcurLCTBxxxxD
\else
??????\fi
\fi
\fi
\fi
}
\newcommand{\hatcurLCzeta}[1]{\ifnum#1=34 %
\hatcurLCzetaxxxxA
\else
\ifnum#1=35 %
\hatcurLCzetaxxxxB
\else
\ifnum#1=36 %
\hatcurLCzetaxxxxC
\else
\ifnum#1=37 %
\hatcurLCzetaxxxxD
\else
??????\fi
\fi
\fi
\fi
}
\newcommand{\hatcurPPaequiv}[1]{\ifnum#1=34 %
\hatcurPPaequivxxxxA
\else
\ifnum#1=35 %
\hatcurPPaequivxxxxB
\else
\ifnum#1=36 %
\hatcurPPaequivxxxxC
\else
\ifnum#1=37 %
\hatcurPPaequivxxxxD
\else
??????\fi
\fi
\fi
\fi
}
\newcommand{\hatcurPPar}[1]{\ifnum#1=34 %
\hatcurPParxxxxA
\else
\ifnum#1=35 %
\hatcurPParxxxxB
\else
\ifnum#1=36 %
\hatcurPParxxxxC
\else
\ifnum#1=37 %
\hatcurPParxxxxD
\else
??????\fi
\fi
\fi
\fi
}
\newcommand{\hatcurPParel}[1]{\ifnum#1=34 %
\hatcurPParelxxxxA
\else
\ifnum#1=35 %
\hatcurPParelxxxxB
\else
\ifnum#1=36 %
\hatcurPParelxxxxC
\else
\ifnum#1=37 %
\hatcurPParelxxxxD
\else
??????\fi
\fi
\fi
\fi
}
\newcommand{\hatcurPPfluxap}[1]{\ifnum#1=34 %
\hatcurPPfluxapxxxxA
\else
\ifnum#1=35 %
\hatcurPPfluxapxxxxB
\else
\ifnum#1=36 %
\hatcurPPfluxapxxxxC
\else
\ifnum#1=37 %
\hatcurPPfluxapxxxxD
\else
??????\fi
\fi
\fi
\fi
}
\newcommand{\hatcurPPfluxapdim}[1]{\ifnum#1=34 %
\hatcurPPfluxapdimxxxxA
\else
\ifnum#1=35 %
\hatcurPPfluxapdimxxxxB
\else
\ifnum#1=36 %
\hatcurPPfluxapdimxxxxC
\else
\ifnum#1=37 %
\hatcurPPfluxapdimxxxxD
\else
??????\fi
\fi
\fi
\fi
}
\newcommand{\hatcurPPfluxavg}[1]{\ifnum#1=34 %
\hatcurPPfluxavgxxxxA
\else
\ifnum#1=35 %
\hatcurPPfluxavgxxxxB
\else
\ifnum#1=36 %
\hatcurPPfluxavgxxxxC
\else
\ifnum#1=37 %
\hatcurPPfluxavgxxxxD
\else
??????\fi
\fi
\fi
\fi
}
\newcommand{\hatcurPPfluxavgdim}[1]{\ifnum#1=34 %
\hatcurPPfluxavgdimxxxxA
\else
\ifnum#1=35 %
\hatcurPPfluxavgdimxxxxB
\else
\ifnum#1=36 %
\hatcurPPfluxavgdimxxxxC
\else
\ifnum#1=37 %
\hatcurPPfluxavgdimxxxxD
\else
??????\fi
\fi
\fi
\fi
}
\newcommand{\hatcurPPfluxperi}[1]{\ifnum#1=34 %
\hatcurPPfluxperixxxxA
\else
\ifnum#1=35 %
\hatcurPPfluxperixxxxB
\else
\ifnum#1=36 %
\hatcurPPfluxperixxxxC
\else
\ifnum#1=37 %
\hatcurPPfluxperixxxxD
\else
??????\fi
\fi
\fi
\fi
}
\newcommand{\hatcurPPfluxperidim}[1]{\ifnum#1=34 %
\hatcurPPfluxperidimxxxxA
\else
\ifnum#1=35 %
\hatcurPPfluxperidimxxxxB
\else
\ifnum#1=36 %
\hatcurPPfluxperidimxxxxC
\else
\ifnum#1=37 %
\hatcurPPfluxperidimxxxxD
\else
??????\fi
\fi
\fi
\fi
}
\newcommand{\hatcurPPg}[1]{\ifnum#1=34 %
\hatcurPPgxxxxA
\else
\ifnum#1=35 %
\hatcurPPgxxxxB
\else
\ifnum#1=36 %
\hatcurPPgxxxxC
\else
\ifnum#1=37 %
\hatcurPPgxxxxD
\else
??????\fi
\fi
\fi
\fi
}
\newcommand{\hatcurPPi}[1]{\ifnum#1=34 %
\hatcurPPixxxxA
\else
\ifnum#1=35 %
\hatcurPPixxxxB
\else
\ifnum#1=36 %
\hatcurPPixxxxC
\else
\ifnum#1=37 %
\hatcurPPixxxxD
\else
??????\fi
\fi
\fi
\fi
}
\newcommand{\hatcurPPlogg}[1]{\ifnum#1=34 %
\hatcurPPloggxxxxA
\else
\ifnum#1=35 %
\hatcurPPloggxxxxB
\else
\ifnum#1=36 %
\hatcurPPloggxxxxC
\else
\ifnum#1=37 %
\hatcurPPloggxxxxD
\else
??????\fi
\fi
\fi
\fi
}
\newcommand{\hatcurPPm}[1]{\ifnum#1=34 %
\hatcurPPmxxxxA
\else
\ifnum#1=35 %
\hatcurPPmxxxxB
\else
\ifnum#1=36 %
\hatcurPPmxxxxC
\else
\ifnum#1=37 %
\hatcurPPmxxxxD
\else
??????\fi
\fi
\fi
\fi
}
\newcommand{\hatcurPPme}[1]{\ifnum#1=34 %
\hatcurPPmexxxxA
\else
\ifnum#1=35 %
\hatcurPPmexxxxB
\else
\ifnum#1=36 %
\hatcurPPmexxxxC
\else
\ifnum#1=37 %
\hatcurPPmexxxxD
\else
??????\fi
\fi
\fi
\fi
}
\newcommand{\hatcurPPmelong}[1]{\ifnum#1=34 %
\hatcurPPmelongxxxxA
\else
\ifnum#1=35 %
\hatcurPPmelongxxxxB
\else
\ifnum#1=36 %
\hatcurPPmelongxxxxC
\else
\ifnum#1=37 %
\hatcurPPmelongxxxxD
\else
??????\fi
\fi
\fi
\fi
}
\newcommand{\hatcurPPmeshort}[1]{\ifnum#1=34 %
\hatcurPPmeshortxxxxA
\else
\ifnum#1=35 %
\hatcurPPmeshortxxxxB
\else
\ifnum#1=36 %
\hatcurPPmeshortxxxxC
\else
\ifnum#1=37 %
\hatcurPPmeshortxxxxD
\else
??????\fi
\fi
\fi
\fi
}
\newcommand{\hatcurPPmlong}[1]{\ifnum#1=34 %
\hatcurPPmlongxxxxA
\else
\ifnum#1=35 %
\hatcurPPmlongxxxxB
\else
\ifnum#1=36 %
\hatcurPPmlongxxxxC
\else
\ifnum#1=37 %
\hatcurPPmlongxxxxD
\else
??????\fi
\fi
\fi
\fi
}
\newcommand{\hatcurPPmrcorr}[1]{\ifnum#1=34 %
\hatcurPPmrcorrxxxxA
\else
\ifnum#1=35 %
\hatcurPPmrcorrxxxxB
\else
\ifnum#1=36 %
\hatcurPPmrcorrxxxxC
\else
\ifnum#1=37 %
\hatcurPPmrcorrxxxxD
\else
??????\fi
\fi
\fi
\fi
}
\newcommand{\hatcurPPmshort}[1]{\ifnum#1=34 %
\hatcurPPmshortxxxxA
\else
\ifnum#1=35 %
\hatcurPPmshortxxxxB
\else
\ifnum#1=36 %
\hatcurPPmshortxxxxC
\else
\ifnum#1=37 %
\hatcurPPmshortxxxxD
\else
??????\fi
\fi
\fi
\fi
}
\newcommand{\hatcurPPperi}[1]{\ifnum#1=34 %
\hatcurPPperixxxxA
\else
\ifnum#1=35 %
\hatcurPPperixxxxB
\else
\ifnum#1=36 %
\hatcurPPperixxxxC
\else
\ifnum#1=37 %
\hatcurPPperixxxxD
\else
??????\fi
\fi
\fi
\fi
}
\newcommand{\hatcurPPphiconj}[1]{\ifnum#1=34 %
\hatcurPPphiconjxxxxA
\else
\ifnum#1=35 %
\hatcurPPphiconjxxxxB
\else
\ifnum#1=36 %
\hatcurPPphiconjxxxxC
\else
\ifnum#1=37 %
\hatcurPPphiconjxxxxD
\else
??????\fi
\fi
\fi
\fi
}
\newcommand{\hatcurPPr}[1]{\ifnum#1=34 %
\hatcurPPrxxxxA
\else
\ifnum#1=35 %
\hatcurPPrxxxxB
\else
\ifnum#1=36 %
\hatcurPPrxxxxC
\else
\ifnum#1=37 %
\hatcurPPrxxxxD
\else
??????\fi
\fi
\fi
\fi
}
\newcommand{\hatcurPPre}[1]{\ifnum#1=34 %
\hatcurPPrexxxxA
\else
\ifnum#1=35 %
\hatcurPPrexxxxB
\else
\ifnum#1=36 %
\hatcurPPrexxxxC
\else
\ifnum#1=37 %
\hatcurPPrexxxxD
\else
??????\fi
\fi
\fi
\fi
}
\newcommand{\hatcurPPrelong}[1]{\ifnum#1=34 %
\hatcurPPrelongxxxxA
\else
\ifnum#1=35 %
\hatcurPPrelongxxxxB
\else
\ifnum#1=36 %
\hatcurPPrelongxxxxC
\else
\ifnum#1=37 %
\hatcurPPrelongxxxxD
\else
??????\fi
\fi
\fi
\fi
}
\newcommand{\hatcurPPreshort}[1]{\ifnum#1=34 %
\hatcurPPreshortxxxxA
\else
\ifnum#1=35 %
\hatcurPPreshortxxxxB
\else
\ifnum#1=36 %
\hatcurPPreshortxxxxC
\else
\ifnum#1=37 %
\hatcurPPreshortxxxxD
\else
??????\fi
\fi
\fi
\fi
}
\newcommand{\hatcurPPrho}[1]{\ifnum#1=34 %
\hatcurPPrhoxxxxA
\else
\ifnum#1=35 %
\hatcurPPrhoxxxxB
\else
\ifnum#1=36 %
\hatcurPPrhoxxxxC
\else
\ifnum#1=37 %
\hatcurPPrhoxxxxD
\else
??????\fi
\fi
\fi
\fi
}
\newcommand{\hatcurPPrlong}[1]{\ifnum#1=34 %
\hatcurPPrlongxxxxA
\else
\ifnum#1=35 %
\hatcurPPrlongxxxxB
\else
\ifnum#1=36 %
\hatcurPPrlongxxxxC
\else
\ifnum#1=37 %
\hatcurPPrlongxxxxD
\else
??????\fi
\fi
\fi
\fi
}
\newcommand{\hatcurPPrshort}[1]{\ifnum#1=34 %
\hatcurPPrshortxxxxA
\else
\ifnum#1=35 %
\hatcurPPrshortxxxxB
\else
\ifnum#1=36 %
\hatcurPPrshortxxxxC
\else
\ifnum#1=37 %
\hatcurPPrshortxxxxD
\else
??????\fi
\fi
\fi
\fi
}
\newcommand{\hatcurPPtcirc}[1]{\ifnum#1=34 %
\hatcurPPtcircxxxxA
\else
\ifnum#1=35 %
\hatcurPPtcircxxxxB
\else
\ifnum#1=36 %
\hatcurPPtcircxxxxC
\else
\ifnum#1=37 %
\hatcurPPtcircxxxxD
\else
??????\fi
\fi
\fi
\fi
}
\newcommand{\hatcurPPteff}[1]{\ifnum#1=34 %
\hatcurPPteffxxxxA
\else
\ifnum#1=35 %
\hatcurPPteffxxxxB
\else
\ifnum#1=36 %
\hatcurPPteffxxxxC
\else
\ifnum#1=37 %
\hatcurPPteffxxxxD
\else
??????\fi
\fi
\fi
\fi
}
\newcommand{\hatcurPPtheta}[1]{\ifnum#1=34 %
\hatcurPPthetaxxxxA
\else
\ifnum#1=35 %
\hatcurPPthetaxxxxB
\else
\ifnum#1=36 %
\hatcurPPthetaxxxxC
\else
\ifnum#1=37 %
\hatcurPPthetaxxxxD
\else
??????\fi
\fi
\fi
\fi
}
\newcommand{\hatcurPPtinfall}[1]{\ifnum#1=34 %
\hatcurPPtinfallxxxxA
\else
\ifnum#1=35 %
\hatcurPPtinfallxxxxB
\else
\ifnum#1=36 %
\hatcurPPtinfallxxxxC
\else
\ifnum#1=37 %
\hatcurPPtinfallxxxxD
\else
??????\fi
\fi
\fi
\fi
}
\newcommand{\hatcurRVeccen}[1]{\ifnum#1=34 %
\hatcurRVeccenxxxxA
\else
\ifnum#1=35 %
\hatcurRVeccenxxxxB
\else
\ifnum#1=36 %
\hatcurRVeccenxxxxC
\else
\ifnum#1=37 %
\hatcurRVeccenxxxxD
\else
??????\fi
\fi
\fi
\fi
}
\newcommand{\hatcurRVfitrms}[1]{\ifnum#1=35 %
\hatcurRVfitrmsxxxxB
\else
\ifnum#1=36 %
\hatcurRVfitrmsxxxxC
\else
\ifnum#1=37 %
\hatcurRVfitrmsxxxxD
\else
??????\fi
\fi
\fi
}
\newcommand{\hatcurRVfitrmsA}[1]{\ifnum#1=34 %
\hatcurRVfitrmsAxxxxA
\else
??????\fi
}
\newcommand{\hatcurRVfitrmsB}[1]{\ifnum#1=34 %
\hatcurRVfitrmsBxxxxA
\else
??????\fi
}
\newcommand{\hatcurRVfitrmsC}[1]{\ifnum#1=34 %
\hatcurRVfitrmsCxxxxA
\else
??????\fi
}
\newcommand{\hatcurRVgamma}[1]{\ifnum#1=35 %
\hatcurRVgammaxxxxB
\else
\ifnum#1=36 %
\hatcurRVgammaxxxxC
\else
\ifnum#1=37 %
\hatcurRVgammaxxxxD
\else
??????\fi
\fi
\fi
}
\newcommand{\hatcurRVgammaA}[1]{\ifnum#1=34 %
\hatcurRVgammaAxxxxA
\else
??????\fi
}
\newcommand{\hatcurRVgammaB}[1]{\ifnum#1=34 %
\hatcurRVgammaBxxxxA
\else
??????\fi
}
\newcommand{\hatcurRVgammaC}[1]{\ifnum#1=34 %
\hatcurRVgammaCxxxxA
\else
??????\fi
}
\newcommand{\hatcurRVh}[1]{\ifnum#1=34 %
\hatcurRVhxxxxA
\else
\ifnum#1=35 %
\hatcurRVhxxxxB
\else
\ifnum#1=36 %
\hatcurRVhxxxxC
\else
\ifnum#1=37 %
\hatcurRVhxxxxD
\else
??????\fi
\fi
\fi
\fi
}
\newcommand{\hatcurRVjitter}[1]{\ifnum#1=34 %
\hatcurRVjitterxxxxA
\else
\ifnum#1=35 %
\hatcurRVjitterxxxxB
\else
\ifnum#1=36 %
\hatcurRVjitterxxxxC
\else
\ifnum#1=37 %
\hatcurRVjitterxxxxD
\else
??????\fi
\fi
\fi
\fi
}
\newcommand{\hatcurRVjitterA}[1]{\ifnum#1=34 %
\hatcurRVjitterAxxxxA
\else
??????\fi
}
\newcommand{\hatcurRVjitterB}[1]{\ifnum#1=34 %
\hatcurRVjitterBxxxxA
\else
??????\fi
}
\newcommand{\hatcurRVjitterC}[1]{\ifnum#1=34 %
\hatcurRVjitterCxxxxA
\else
??????\fi
}
\newcommand{\hatcurRVk}[1]{\ifnum#1=34 %
\hatcurRVkxxxxA
\else
\ifnum#1=35 %
\hatcurRVkxxxxB
\else
\ifnum#1=36 %
\hatcurRVkxxxxC
\else
\ifnum#1=37 %
\hatcurRVkxxxxD
\else
??????\fi
\fi
\fi
\fi
}
\newcommand{\hatcurRVK}[1]{\ifnum#1=34 %
\hatcurRVKxxxxA
\else
\ifnum#1=35 %
\hatcurRVKxxxxB
\else
\ifnum#1=36 %
\hatcurRVKxxxxC
\else
\ifnum#1=37 %
\hatcurRVKxxxxD
\else
??????\fi
\fi
\fi
\fi
}
\newcommand{\hatcurRVomega}[1]{\ifnum#1=34 %
\hatcurRVomegaxxxxA
\else
\ifnum#1=35 %
\hatcurRVomegaxxxxB
\else
\ifnum#1=36 %
\hatcurRVomegaxxxxC
\else
\ifnum#1=37 %
\hatcurRVomegaxxxxD
\else
??????\fi
\fi
\fi
\fi
}
\newcommand{\hatcurRVtrone}[1]{\ifnum#1=34 %
\hatcurRVtronexxxxA
\else
??????\fi
}
\newcommand{\hatcurRVtrtwo}[1]{\ifnum#1=34 %
\hatcurRVtrtwoxxxxA
\else
??????\fi
}
\newcommand{\hatcurSMEiilogg}[1]{\ifnum#1=34 %
\hatcurSMEiiloggxxxxA
\else
\ifnum#1=35 %
\hatcurSMEiiloggxxxxB
\else
\ifnum#1=36 %
\hatcurSMEiiloggxxxxC
\else
\ifnum#1=37 %
\hatcurSMEiiloggxxxxD
\else
??????\fi
\fi
\fi
\fi
}
\newcommand{\hatcurSMEiiteff}[1]{\ifnum#1=34 %
\hatcurSMEiiteffxxxxA
\else
\ifnum#1=35 %
\hatcurSMEiiteffxxxxB
\else
\ifnum#1=36 %
\hatcurSMEiiteffxxxxC
\else
\ifnum#1=37 %
\hatcurSMEiiteffxxxxD
\else
??????\fi
\fi
\fi
\fi
}
\newcommand{\hatcurSMEiivmac}[1]{\ifnum#1=34 %
\hatcurSMEiivmacxxxxA
\else
\ifnum#1=35 %
\hatcurSMEiivmacxxxxB
\else
\ifnum#1=36 %
\hatcurSMEiivmacxxxxC
\else
\ifnum#1=37 %
\hatcurSMEiivmacxxxxD
\else
??????\fi
\fi
\fi
\fi
}
\newcommand{\hatcurSMEiivmic}[1]{\ifnum#1=34 %
\hatcurSMEiivmicxxxxA
\else
\ifnum#1=35 %
\hatcurSMEiivmicxxxxB
\else
\ifnum#1=36 %
\hatcurSMEiivmicxxxxC
\else
\ifnum#1=37 %
\hatcurSMEiivmicxxxxD
\else
??????\fi
\fi
\fi
\fi
}
\newcommand{\hatcurSMEiivsin}[1]{\ifnum#1=34 %
\hatcurSMEiivsinxxxxA
\else
\ifnum#1=35 %
\hatcurSMEiivsinxxxxB
\else
\ifnum#1=36 %
\hatcurSMEiivsinxxxxC
\else
\ifnum#1=37 %
\hatcurSMEiivsinxxxxD
\else
??????\fi
\fi
\fi
\fi
}
\newcommand{\hatcurSMEiizfeh}[1]{\ifnum#1=34 %
\hatcurSMEiizfehxxxxA
\else
\ifnum#1=35 %
\hatcurSMEiizfehxxxxB
\else
\ifnum#1=36 %
\hatcurSMEiizfehxxxxC
\else
\ifnum#1=37 %
\hatcurSMEiizfehxxxxD
\else
??????\fi
\fi
\fi
\fi
}
\newcommand{\hatcurSMEiizfehshort}[1]{\ifnum#1=34 %
\hatcurSMEiizfehshortxxxxA
\else
\ifnum#1=35 %
\hatcurSMEiizfehshortxxxxB
\else
\ifnum#1=36 %
\hatcurSMEiizfehshortxxxxC
\else
\ifnum#1=37 %
\hatcurSMEiizfehshortxxxxD
\else
??????\fi
\fi
\fi
\fi
}
\newcommand{\hatcurSMEilogg}[1]{\ifnum#1=34 %
\hatcurSMEiloggxxxxA
\else
\ifnum#1=35 %
\hatcurSMEiloggxxxxB
\else
\ifnum#1=36 %
\hatcurSMEiloggxxxxC
\else
\ifnum#1=37 %
\hatcurSMEiloggxxxxD
\else
??????\fi
\fi
\fi
\fi
}
\newcommand{\hatcurSMEiteff}[1]{\ifnum#1=34 %
\hatcurSMEiteffxxxxA
\else
\ifnum#1=35 %
\hatcurSMEiteffxxxxB
\else
\ifnum#1=36 %
\hatcurSMEiteffxxxxC
\else
\ifnum#1=37 %
\hatcurSMEiteffxxxxD
\else
??????\fi
\fi
\fi
\fi
}
\newcommand{\hatcurSMEivmac}[1]{\ifnum#1=34 %
\hatcurSMEivmacxxxxA
\else
\ifnum#1=35 %
\hatcurSMEivmacxxxxB
\else
\ifnum#1=36 %
\hatcurSMEivmacxxxxC
\else
\ifnum#1=37 %
\hatcurSMEivmacxxxxD
\else
??????\fi
\fi
\fi
\fi
}
\newcommand{\hatcurSMEivmic}[1]{\ifnum#1=34 %
\hatcurSMEivmicxxxxA
\else
\ifnum#1=35 %
\hatcurSMEivmicxxxxB
\else
\ifnum#1=36 %
\hatcurSMEivmicxxxxC
\else
\ifnum#1=37 %
\hatcurSMEivmicxxxxD
\else
??????\fi
\fi
\fi
\fi
}
\newcommand{\hatcurSMEivsin}[1]{\ifnum#1=34 %
\hatcurSMEivsinxxxxA
\else
\ifnum#1=35 %
\hatcurSMEivsinxxxxB
\else
\ifnum#1=36 %
\hatcurSMEivsinxxxxC
\else
\ifnum#1=37 %
\hatcurSMEivsinxxxxD
\else
??????\fi
\fi
\fi
\fi
}
\newcommand{\hatcurSMEizfeh}[1]{\ifnum#1=34 %
\hatcurSMEizfehxxxxA
\else
\ifnum#1=35 %
\hatcurSMEizfehxxxxB
\else
\ifnum#1=36 %
\hatcurSMEizfehxxxxC
\else
\ifnum#1=37 %
\hatcurSMEizfehxxxxD
\else
??????\fi
\fi
\fi
\fi
}
\newcommand{\hatcurSMEizfehshort}[1]{\ifnum#1=34 %
\hatcurSMEizfehshortxxxxA
\else
\ifnum#1=35 %
\hatcurSMEizfehshortxxxxB
\else
\ifnum#1=36 %
\hatcurSMEizfehshortxxxxC
\else
\ifnum#1=37 %
\hatcurSMEizfehshortxxxxD
\else
??????\fi
\fi
\fi
\fi
}
\newcommand{\hatcurTRESgamma}[1]{\ifnum#1=34 %
\hatcurTRESgammaxxxxA
\else
\ifnum#1=35 %
\hatcurTRESgammaxxxxB
\else
\ifnum#1=36 %
\hatcurTRESgammaxxxxC
\else
\ifnum#1=37 %
\hatcurTRESgammaxxxxD
\else
??????\fi
\fi
\fi
\fi
}
\newcommand{\hatcurTRESlogg}[1]{\ifnum#1=34 %
\hatcurTRESloggxxxxA
\else
\ifnum#1=35 %
\hatcurTRESloggxxxxB
\else
\ifnum#1=36 %
\hatcurTRESloggxxxxC
\else
\ifnum#1=37 %
\hatcurTRESloggxxxxD
\else
??????\fi
\fi
\fi
\fi
}
\newcommand{\hatcurTRESnumspec}[1]{\ifnum#1=34 %
\hatcurTRESnumspecxxxxA
\else
\ifnum#1=35 %
\hatcurTRESnumspecxxxxB
\else
\ifnum#1=36 %
\hatcurTRESnumspecxxxxC
\else
\ifnum#1=37 %
\hatcurTRESnumspecxxxxD
\else
??????\fi
\fi
\fi
\fi
}
\newcommand{\hatcurTRESrvrms}[1]{\ifnum#1=34 %
\hatcurTRESrvrmsxxxxA
\else
\ifnum#1=35 %
\hatcurTRESrvrmsxxxxB
\else
\ifnum#1=36 %
\hatcurTRESrvrmsxxxxC
\else
\ifnum#1=37 %
\hatcurTRESrvrmsxxxxD
\else
??????\fi
\fi
\fi
\fi
}
\newcommand{\hatcurTRESspan}[1]{\ifnum#1=34 %
\hatcurTRESspanxxxxA
\else
\ifnum#1=35 %
\hatcurTRESspanxxxxB
\else
\ifnum#1=36 %
\hatcurTRESspanxxxxC
\else
\ifnum#1=37 %
\hatcurTRESspanxxxxD
\else
??????\fi
\fi
\fi
\fi
}
\newcommand{\hatcurTRESteff}[1]{\ifnum#1=34 %
\hatcurTRESteffxxxxA
\else
\ifnum#1=35 %
\hatcurTRESteffxxxxB
\else
\ifnum#1=36 %
\hatcurTRESteffxxxxC
\else
\ifnum#1=37 %
\hatcurTRESteffxxxxD
\else
??????\fi
\fi
\fi
\fi
}
\newcommand{\hatcurTRESvsini}[1]{\ifnum#1=34 %
\hatcurTRESvsinixxxxA
\else
\ifnum#1=35 %
\hatcurTRESvsinixxxxB
\else
\ifnum#1=36 %
\hatcurTRESvsinixxxxC
\else
\ifnum#1=37 %
\hatcurTRESvsinixxxxD
\else
??????\fi
\fi
\fi
\fi
}
\newcommand{\hatcurTRESzfeh}[1]{\ifnum#1=34 %
\hatcurTRESzfehxxxxA
\else
\ifnum#1=35 %
\hatcurTRESzfehxxxxB
\else
\ifnum#1=36 %
\hatcurTRESzfehxxxxC
\else
\ifnum#1=37 %
\hatcurTRESzfehxxxxD
\else
??????\fi
\fi
\fi
\fi
}
\newcommand{\hatcurXdist}[1]{\ifnum#1=34 %
\hatcurXdistxxxxA
\else
\ifnum#1=35 %
\hatcurXdistxxxxB
\else
\ifnum#1=36 %
\hatcurXdistxxxxC
\else
\ifnum#1=37 %
\hatcurXdistxxxxD
\else
??????\fi
\fi
\fi
\fi
}
\newcommand{\hatcurXsecdur}[1]{\ifnum#1=34 %
\hatcurXsecdurxxxxA
\else
\ifnum#1=35 %
\hatcurXsecdurxxxxB
\else
\ifnum#1=36 %
\hatcurXsecdurxxxxC
\else
\ifnum#1=37 %
\hatcurXsecdurxxxxD
\else
??????\fi
\fi
\fi
\fi
}
\newcommand{\hatcurXsecingdur}[1]{\ifnum#1=34 %
\hatcurXsecingdurxxxxA
\else
\ifnum#1=35 %
\hatcurXsecingdurxxxxB
\else
\ifnum#1=36 %
\hatcurXsecingdurxxxxC
\else
\ifnum#1=37 %
\hatcurXsecingdurxxxxD
\else
??????\fi
\fi
\fi
\fi
}
\newcommand{\hatcurXsecondary}[1]{\ifnum#1=34 %
\hatcurXsecondaryxxxxA
\else
\ifnum#1=35 %
\hatcurXsecondaryxxxxB
\else
\ifnum#1=36 %
\hatcurXsecondaryxxxxC
\else
\ifnum#1=37 %
\hatcurXsecondaryxxxxD
\else
??????\fi
\fi
\fi
\fi
}
\newcommand{\hatcurXsecphase}[1]{\ifnum#1=34 %
\hatcurXsecphasexxxxA
\else
\ifnum#1=35 %
\hatcurXsecphasexxxxB
\else
\ifnum#1=36 %
\hatcurXsecphasexxxxC
\else
\ifnum#1=37 %
\hatcurXsecphasexxxxD
\else
??????\fi
\fi
\fi
\fi
}

\newcommand{\hatcurxxxxA}{HAT-P-34}
\newcommand{\hatcurbxxxxA}{HAT-P-34b}
\newcommand{\hatcurcxxxxA}{HAT-P-34c}

\newcommand{\hatcurplanetnumxxxxA}{34}

\newcommand{\hatcurRVgammaabsxxxxA}{\hatcurDSgamma{\hatcurplanetnumxxxxA}}                           

\newcommand{\hatcurRVgammarelxxxxA}{\hatcurRVgamma{\hatcurplanetnumxxxxA}}                           

\newcommand{\hatcurCCtassvixxxxA}{\ensuremath{0.557\pm0.12}}                  

\newcommand{\hatcurSMEversionxxxxA}{ii}                                       

\newcommand{\hatcurisoshortxxxxA}{YY}
\newcommand{\hatcurisofullxxxxA}{Yonsei-Yale (YY)}
\newcommand{\hatcurisocitexxxxA}{yi:2001}

\newcommand{\hatcurlumindxxxxA}{\arstar}

\newcommand{\hatcurjhkfilsetxxxxA}{ESO}

\newcommand{\hatcurSMEteffxxxxA}{\ifthenelse{\equal{\hatcurSMEversionxxxxA}{i}}{\hatcurSMEiteff{\hatcurplanetnumxxxxA}}{\hatcurSMEiiteff{\hatcurplanetnumxxxxA}}}
\newcommand{\hatcurSMEzfehxxxxA}{\ifthenelse{\equal{\hatcurSMEversionxxxxA}{i}}{\hatcurSMEizfeh{\hatcurplanetnumxxxxA}}{\hatcurSMEiizfeh{\hatcurplanetnumxxxxA}}}
\newcommand{\hatcurSMEzfehshortxxxxA}{\ifthenelse{\equal{\hatcurSMEversionxxxxA}{i}}{\hatcurSMEizfehshort{\hatcurplanetnumxxxxA}}{\hatcurSMEiizfehshort{\hatcurplanetnumxxxxA}}}
\newcommand{\hatcurSMEloggxxxxA}{\ifthenelse{\equal{\hatcurSMEversionxxxxA}{i}}{\hatcurSMEilogg{\hatcurplanetnumxxxxA}}{\hatcurSMEiilogg{\hatcurplanetnumxxxxA}}}
\newcommand{\hatcurSMEvsinxxxxA}{\ifthenelse{\equal{\hatcurSMEversionxxxxA}{i}}{\hatcurSMEivsin{\hatcurplanetnumxxxxA}}{\hatcurSMEiivsin{\hatcurplanetnumxxxxA}}}
\newcommand{\hatcurSMEvmacxxxxA}{\ifthenelse{\equal{\hatcurSMEversionxxxxA}{i}}{\hatcurSMEivmac{\hatcurplanetnumxxxxA}}{\hatcurSMEiivmac{\hatcurplanetnumxxxxA}}}
\newcommand{\hatcurSMEvmicxxxxA}{\ifthenelse{\equal{\hatcurSMEversionxxxxA}{i}}{\hatcurSMEivmic{\hatcurplanetnumxxxxA}}{\hatcurSMEiivmic{\hatcurplanetnumxxxxA}}}

\newcommand{\hatcurxxxxB}{HAT-P-35}
\newcommand{\hatcurbxxxxB}{HAT-P-35b}
\newcommand{\hatcurcxxxxB}{HAT-P-35c}

\newcommand{\hatcurplanetnumxxxxB}{35}

\newcommand{\hatcurRVgammaabsxxxxB}{\hatcurDSgamma{\hatcurplanetnumxxxxB}}                           

\newcommand{\hatcurRVgammarelxxxxB}{\hatcurRVgamma{\hatcurplanetnumxxxxB}}                           

\newcommand{\hatcurCCtassvixxxxB}{\ensuremath{0.662\pm0.12}}                  

\newcommand{\hatcurSMEversionxxxxB}{ii}                                       

\newcommand{\hatcurisoshortxxxxB}{YY}
\newcommand{\hatcurisofullxxxxB}{Yonsei-Yale (YY)}
\newcommand{\hatcurisocitexxxxB}{yi:2001}

\newcommand{\hatcurlumindxxxxB}{\arstar}

\newcommand{\hatcurjhkfilsetxxxxB}{ESO}

\newcommand{\hatcurSMEteffxxxxB}{\ifthenelse{\equal{\hatcurSMEversionxxxxB}{i}}{\hatcurSMEiteff{\hatcurplanetnumxxxxB}}{\hatcurSMEiiteff{\hatcurplanetnumxxxxB}}}
\newcommand{\hatcurSMEzfehxxxxB}{\ifthenelse{\equal{\hatcurSMEversionxxxxB}{i}}{\hatcurSMEizfeh{\hatcurplanetnumxxxxB}}{\hatcurSMEiizfeh{\hatcurplanetnumxxxxB}}}
\newcommand{\hatcurSMEzfehshortxxxxB}{\ifthenelse{\equal{\hatcurSMEversionxxxxB}{i}}{\hatcurSMEizfehshort{\hatcurplanetnumxxxxB}}{\hatcurSMEiizfehshort{\hatcurplanetnumxxxxB}}}
\newcommand{\hatcurSMEloggxxxxB}{\ifthenelse{\equal{\hatcurSMEversionxxxxB}{i}}{\hatcurSMEilogg{\hatcurplanetnumxxxxB}}{\hatcurSMEiilogg{\hatcurplanetnumxxxxB}}}
\newcommand{\hatcurSMEvsinxxxxB}{\ifthenelse{\equal{\hatcurSMEversionxxxxB}{i}}{\hatcurSMEivsin{\hatcurplanetnumxxxxB}}{\hatcurSMEiivsin{\hatcurplanetnumxxxxB}}}
\newcommand{\hatcurSMEvmacxxxxB}{\ifthenelse{\equal{\hatcurSMEversionxxxxB}{i}}{\hatcurSMEivmac{\hatcurplanetnumxxxxB}}{\hatcurSMEiivmac{\hatcurplanetnumxxxxB}}}
\newcommand{\hatcurSMEvmicxxxxB}{\ifthenelse{\equal{\hatcurSMEversionxxxxB}{i}}{\hatcurSMEivmic{\hatcurplanetnumxxxxB}}{\hatcurSMEiivmic{\hatcurplanetnumxxxxB}}}

\newcommand{\hatcurxxxxC}{HAT-P-36}
\newcommand{\hatcurbxxxxC}{HAT-P-36b}
\newcommand{\hatcurcxxxxC}{HAT-P-36c}

\newcommand{\hatcurplanetnumxxxxC}{36}

\newcommand{\hatcurRVgammaabsxxxxC}{\hatcurDSgamma{\hatcurplanetnumxxxxC}}                           

\newcommand{\hatcurRVgammarelxxxxC}{\hatcurRVgamma{\hatcurplanetnumxxxxC}}                           

\newcommand{\hatcurCCtassvixxxxC}{\ensuremath{0.760\pm0.13}}                  

\newcommand{\hatcurSMEversionxxxxC}{ii}                                       

\newcommand{\hatcurisoshortxxxxC}{YY}
\newcommand{\hatcurisofullxxxxC}{Yonsei-Yale (YY)}
\newcommand{\hatcurisocitexxxxC}{yi:2001}

\newcommand{\hatcurlumindxxxxC}{\arstar}

\newcommand{\hatcurjhkfilsetxxxxC}{ESO}

\newcommand{\hatcurSMEteffxxxxC}{\ifthenelse{\equal{\hatcurSMEversionxxxxC}{i}}{\hatcurSMEiteff{\hatcurplanetnumxxxxC}}{\hatcurSMEiiteff{\hatcurplanetnumxxxxC}}}
\newcommand{\hatcurSMEzfehxxxxC}{\ifthenelse{\equal{\hatcurSMEversionxxxxC}{i}}{\hatcurSMEizfeh{\hatcurplanetnumxxxxC}}{\hatcurSMEiizfeh{\hatcurplanetnumxxxxC}}}
\newcommand{\hatcurSMEzfehshortxxxxC}{\ifthenelse{\equal{\hatcurSMEversionxxxxC}{i}}{\hatcurSMEizfehshort{\hatcurplanetnumxxxxC}}{\hatcurSMEiizfehshort{\hatcurplanetnumxxxxC}}}
\newcommand{\hatcurSMEloggxxxxC}{\ifthenelse{\equal{\hatcurSMEversionxxxxC}{i}}{\hatcurSMEilogg{\hatcurplanetnumxxxxC}}{\hatcurSMEiilogg{\hatcurplanetnumxxxxC}}}
\newcommand{\hatcurSMEvsinxxxxC}{\ifthenelse{\equal{\hatcurSMEversionxxxxC}{i}}{\hatcurSMEivsin{\hatcurplanetnumxxxxC}}{\hatcurSMEiivsin{\hatcurplanetnumxxxxC}}}
\newcommand{\hatcurSMEvmacxxxxC}{\ifthenelse{\equal{\hatcurSMEversionxxxxC}{i}}{\hatcurSMEivmac{\hatcurplanetnumxxxxC}}{\hatcurSMEiivmac{\hatcurplanetnumxxxxC}}}
\newcommand{\hatcurSMEvmicxxxxC}{\ifthenelse{\equal{\hatcurSMEversionxxxxC}{i}}{\hatcurSMEivmic{\hatcurplanetnumxxxxC}}{\hatcurSMEiivmic{\hatcurplanetnumxxxxC}}}

\newcommand{\hatcurxxxxD}{HAT-P-37}
\newcommand{\hatcurbxxxxD}{HAT-P-37b}
\newcommand{\hatcurcxxxxD}{HAT-P-37c}

\newcommand{\hatcurplanetnumxxxxD}{37}

\newcommand{\hatcurRVgammaabsxxxxD}{\hatcurDSgamma{\hatcurplanetnumxxxxD}}                           

\newcommand{\hatcurRVgammarelxxxxD}{\hatcurRVgamma{\hatcurplanetnumxxxxD}}                           

\newcommand{\hatcurCCtassvixxxxD}{\ensuremath{\cdots}}                  

\newcommand{\hatcurSMEversionxxxxD}{ii}                                       

\newcommand{\hatcurisoshortxxxxD}{YY}
\newcommand{\hatcurisofullxxxxD}{Yonsei-Yale (YY)}
\newcommand{\hatcurisocitexxxxD}{yi:2001}

\newcommand{\hatcurlumindxxxxD}{\arstar}

\newcommand{\hatcurjhkfilsetxxxxD}{ESO}

\newcommand{\hatcurSMEteffxxxxD}{\ifthenelse{\equal{\hatcurSMEversionxxxxD}{i}}{\hatcurSMEiteff{\hatcurplanetnumxxxxD}}{\hatcurSMEiiteff{\hatcurplanetnumxxxxD}}}
\newcommand{\hatcurSMEzfehxxxxD}{\ifthenelse{\equal{\hatcurSMEversionxxxxD}{i}}{\hatcurSMEizfeh{\hatcurplanetnumxxxxD}}{\hatcurSMEiizfeh{\hatcurplanetnumxxxxD}}}
\newcommand{\hatcurSMEzfehshortxxxxD}{\ifthenelse{\equal{\hatcurSMEversionxxxxD}{i}}{\hatcurSMEizfehshort{\hatcurplanetnumxxxxD}}{\hatcurSMEiizfehshort{\hatcurplanetnumxxxxD}}}
\newcommand{\hatcurSMEloggxxxxD}{\ifthenelse{\equal{\hatcurSMEversionxxxxD}{i}}{\hatcurSMEilogg{\hatcurplanetnumxxxxD}}{\hatcurSMEiilogg{\hatcurplanetnumxxxxD}}}
\newcommand{\hatcurSMEvsinxxxxD}{\ifthenelse{\equal{\hatcurSMEversionxxxxD}{i}}{\hatcurSMEivsin{\hatcurplanetnumxxxxD}}{\hatcurSMEiivsin{\hatcurplanetnumxxxxD}}}
\newcommand{\hatcurSMEvmacxxxxD}{\ifthenelse{\equal{\hatcurSMEversionxxxxD}{i}}{\hatcurSMEivmac{\hatcurplanetnumxxxxD}}{\hatcurSMEiivmac{\hatcurplanetnumxxxxD}}}
\newcommand{\hatcurSMEvmicxxxxD}{\ifthenelse{\equal{\hatcurSMEversionxxxxD}{i}}{\hatcurSMEivmic{\hatcurplanetnumxxxxD}}{\hatcurSMEiivmic{\hatcurplanetnumxxxxD}}}

\newcommand{\hatcur}[1]{\ifnum#1=34 %
\hatcurxxxxA
\else
\ifnum#1=35 %
\hatcurxxxxB
\else
\ifnum#1=36 %
\hatcurxxxxC
\else
\ifnum#1=37 %
\hatcurxxxxD
\else
??????\fi
\fi
\fi
\fi
}
\newcommand{\hatcurb}[1]{\ifnum#1=34 %
\hatcurbxxxxA
\else
\ifnum#1=35 %
\hatcurbxxxxB
\else
\ifnum#1=36 %
\hatcurbxxxxC
\else
\ifnum#1=37 %
\hatcurbxxxxD
\else
??????\fi
\fi
\fi
\fi
}
\newcommand{\hatcurc}[1]{\ifnum#1=34 %
\hatcurcxxxxA
\else
\ifnum#1=35 %
\hatcurcxxxxB
\else
\ifnum#1=36 %
\hatcurcxxxxC
\else
\ifnum#1=37 %
\hatcurcxxxxD
\else
??????\fi
\fi
\fi
\fi
}
\newcommand{\hatcurCCtassvi}[1]{\ifnum#1=34 %
\hatcurCCtassvixxxxA
\else
\ifnum#1=35 %
\hatcurCCtassvixxxxB
\else
\ifnum#1=36 %
\hatcurCCtassvixxxxC
\else
\ifnum#1=37 %
\hatcurCCtassvixxxxD
\else
??????\fi
\fi
\fi
\fi
}
\newcommand{\hatcurisocite}[1]{\ifnum#1=34 %
\hatcurisocitexxxxA
\else
\ifnum#1=35 %
\hatcurisocitexxxxB
\else
\ifnum#1=36 %
\hatcurisocitexxxxC
\else
\ifnum#1=37 %
\hatcurisocitexxxxD
\else
??????\fi
\fi
\fi
\fi
}
\newcommand{\hatcurisofull}[1]{\ifnum#1=34 %
\hatcurisofullxxxxA
\else
\ifnum#1=35 %
\hatcurisofullxxxxB
\else
\ifnum#1=36 %
\hatcurisofullxxxxC
\else
\ifnum#1=37 %
\hatcurisofullxxxxD
\else
??????\fi
\fi
\fi
\fi
}
\newcommand{\hatcurisoshort}[1]{\ifnum#1=34 %
\hatcurisoshortxxxxA
\else
\ifnum#1=35 %
\hatcurisoshortxxxxB
\else
\ifnum#1=36 %
\hatcurisoshortxxxxC
\else
\ifnum#1=37 %
\hatcurisoshortxxxxD
\else
??????\fi
\fi
\fi
\fi
}
\newcommand{\hatcurjhkfilset}[1]{\ifnum#1=34 %
\hatcurjhkfilsetxxxxA
\else
\ifnum#1=35 %
\hatcurjhkfilsetxxxxB
\else
\ifnum#1=36 %
\hatcurjhkfilsetxxxxC
\else
\ifnum#1=37 %
\hatcurjhkfilsetxxxxD
\else
??????\fi
\fi
\fi
\fi
}
\newcommand{\hatcurlumind}[1]{\ifnum#1=34 %
\hatcurlumindxxxxA
\else
\ifnum#1=35 %
\hatcurlumindxxxxB
\else
\ifnum#1=36 %
\hatcurlumindxxxxC
\else
\ifnum#1=37 %
\hatcurlumindxxxxD
\else
??????\fi
\fi
\fi
\fi
}
\newcommand{\hatcurplanetnum}[1]{\ifnum#1=34 %
\hatcurplanetnumxxxxA
\else
\ifnum#1=35 %
\hatcurplanetnumxxxxB
\else
\ifnum#1=36 %
\hatcurplanetnumxxxxC
\else
\ifnum#1=37 %
\hatcurplanetnumxxxxD
\else
??????\fi
\fi
\fi
\fi
}
\newcommand{\hatcurRVgammaabs}[1]{\ifnum#1=34 %
\hatcurRVgammaabsxxxxA
\else
\ifnum#1=35 %
\hatcurRVgammaabsxxxxB
\else
\ifnum#1=36 %
\hatcurRVgammaabsxxxxC
\else
\ifnum#1=37 %
\hatcurRVgammaabsxxxxD
\else
??????\fi
\fi
\fi
\fi
}
\newcommand{\hatcurRVgammarel}[1]{\ifnum#1=34 %
\hatcurRVgammarelxxxxA
\else
\ifnum#1=35 %
\hatcurRVgammarelxxxxB
\else
\ifnum#1=36 %
\hatcurRVgammarelxxxxC
\else
\ifnum#1=37 %
\hatcurRVgammarelxxxxD
\else
??????\fi
\fi
\fi
\fi
}
\newcommand{\hatcurSMElogg}[1]{\ifnum#1=34 %
\hatcurSMEloggxxxxA
\else
\ifnum#1=35 %
\hatcurSMEloggxxxxB
\else
\ifnum#1=36 %
\hatcurSMEloggxxxxC
\else
\ifnum#1=37 %
\hatcurSMEloggxxxxD
\else
??????\fi
\fi
\fi
\fi
}
\newcommand{\hatcurSMEteff}[1]{\ifnum#1=34 %
\hatcurSMEteffxxxxA
\else
\ifnum#1=35 %
\hatcurSMEteffxxxxB
\else
\ifnum#1=36 %
\hatcurSMEteffxxxxC
\else
\ifnum#1=37 %
\hatcurSMEteffxxxxD
\else
??????\fi
\fi
\fi
\fi
}
\newcommand{\hatcurSMEversion}[1]{\ifnum#1=34 %
\hatcurSMEversionxxxxA
\else
\ifnum#1=35 %
\hatcurSMEversionxxxxB
\else
\ifnum#1=36 %
\hatcurSMEversionxxxxC
\else
\ifnum#1=37 %
\hatcurSMEversionxxxxD
\else
??????\fi
\fi
\fi
\fi
}
\newcommand{\hatcurSMEvmac}[1]{\ifnum#1=34 %
\hatcurSMEvmacxxxxA
\else
\ifnum#1=35 %
\hatcurSMEvmacxxxxB
\else
\ifnum#1=36 %
\hatcurSMEvmacxxxxC
\else
\ifnum#1=37 %
\hatcurSMEvmacxxxxD
\else
??????\fi
\fi
\fi
\fi
}
\newcommand{\hatcurSMEvmic}[1]{\ifnum#1=34 %
\hatcurSMEvmicxxxxA
\else
\ifnum#1=35 %
\hatcurSMEvmicxxxxB
\else
\ifnum#1=36 %
\hatcurSMEvmicxxxxC
\else
\ifnum#1=37 %
\hatcurSMEvmicxxxxD
\else
??????\fi
\fi
\fi
\fi
}
\newcommand{\hatcurSMEvsin}[1]{\ifnum#1=34 %
\hatcurSMEvsinxxxxA
\else
\ifnum#1=35 %
\hatcurSMEvsinxxxxB
\else
\ifnum#1=36 %
\hatcurSMEvsinxxxxC
\else
\ifnum#1=37 %
\hatcurSMEvsinxxxxD
\else
??????\fi
\fi
\fi
\fi
}
\newcommand{\hatcurSMEzfeh}[1]{\ifnum#1=34 %
\hatcurSMEzfehxxxxA
\else
\ifnum#1=35 %
\hatcurSMEzfehxxxxB
\else
\ifnum#1=36 %
\hatcurSMEzfehxxxxC
\else
\ifnum#1=37 %
\hatcurSMEzfehxxxxD
\else
??????\fi
\fi
\fi
\fi
}
\newcommand{\hatcurSMEzfehshort}[1]{\ifnum#1=34 %
\hatcurSMEzfehshortxxxxA
\else
\ifnum#1=35 %
\hatcurSMEzfehshortxxxxB
\else
\ifnum#1=36 %
\hatcurSMEzfehshortxxxxC
\else
\ifnum#1=37 %
\hatcurSMEzfehshortxxxxD
\else
??????\fi
\fi
\fi
\fi
}

\newcounter{planetcounter}

\newboolean{emulateapj}
\setboolean{emulateapj}{true}

\newboolean{rvtablelong}
\setboolean{rvtablelong}{true}

\newboolean{astroph}
\setboolean{astroph}{true}

\shortauthors{Bakos et al.}
\shorttitle{
\setcounter{planetcounter}{1}
\loopand\hatcur{34}\lowercase{b}\loopcommanospace
\setcounter{planetcounter}{2}
\loopand\hatcur{35}\lowercase{b}\loopcommanospace
\setcounter{planetcounter}{3}
\loopand\hatcur{36}\lowercase{b}\loopcommanospace
\setcounter{planetcounter}{4}
\loopand\hatcur{37}\lowercase{b}\loopcommanospace
}
\ifthenelse{\boolean{emulateapj}}{
    \newcommand{\titledag}{$\dagger$}
}{
    \newcommand{\titledag}{\dagger}
}

\ifthenelse{\boolean{emulateapj}}{
    \newcommand{\titlestar}{$\star$}
}{
    \newcommand{\titlestar}{\star}
}

\begin{document}

\title{
\hatcur{34}\lowercase{b} --- \hatcur{37}\lowercase{b}: 
Four Transiting Planets More Massive Than Jupiter Orbiting Moderately
Bright Stars\altaffilmark{\titledag}
}

\author{
   G.~\'A.~Bakos\altaffilmark{1,2,\titlestar},
   J.~D.~Hartman\altaffilmark{1,2},
   G.~Torres\altaffilmark{2},
   B.~B\'eky\altaffilmark{2},
   D.~W.~Latham\altaffilmark{2},
   L.~A.~Buchhave\altaffilmark{3},
   Z.~Csubry\altaffilmark{1,2},
   G.~Kov\'acs\altaffilmark{4},
   A.~Bieryla\altaffilmark{2},
   S.~Quinn\altaffilmark{2},
   T.~Szklen\'ar\altaffilmark{2},
   G.~A.~Esquerdo\altaffilmark{2},
   A.~Shporer\altaffilmark{5},
   R.~W.~Noyes\altaffilmark{2},
   D.~A.~Fischer\altaffilmark{6},
   J.~A.~Johnson\altaffilmark{7},
   A.~W.~Howard\altaffilmark{8},
   G.~W.~Marcy\altaffilmark{8},
   B.~Sato\altaffilmark{9},
   K.~Penev\altaffilmark{1,2},
   M.~Everett\altaffilmark{2},
   D.~D.~Sasselov\altaffilmark{2},
   G.~F\H{u}r\'esz\altaffilmark{2},
   R.~P.~Stefanik\altaffilmark{2},
   J.~L\'az\'ar\altaffilmark{10}, 
   I.~Papp\altaffilmark{10} \& 
   P.~S\'ari\altaffilmark{10}
}

\altaffiltext{1}{Department of Astrophysical Sciences,
	Princeton University, Princeton, NJ 08544; 
	email: gbakos@astro.princeton.edu}

\altaffiltext{2}{Harvard-Smithsonian Center for Astrophysics}

\altaffiltext{3}{
	Niels Bohr Institute, University of Copenhagen, DK-2100, Denmark,
	and Centre for Star and Planet Formation, 
	Natural History Museum of Denmark, DK-1350 Copenhagen
}

\altaffiltext{4}{Konkoly Observatory, Budapest, Hungary}

\altaffiltext{5}{
    LCOGT, 6740 Cortona Drive, Santa Barbara, CA, \& Department of Physics,
    Broida Hall, UC Santa Barbara, CA}

\altaffiltext{6}{Astronomy Department, Yale University,
    New Haven, CT}

\altaffiltext{7}{California Institute of Technology, Department of
    Astrophysics, MC 249-17, Pasadena, CA}

\altaffiltext{8}{Department of Astronomy, University of California,
    Berkeley, CA}

\altaffiltext{9}{
	Department of Earth and Planetary Sciences, Tokyo Institute of Technology,
	2-12-1 Ookayama, Meguro-ku, Tokyo 152-8551, Japan}

\altaffiltext{10}{
	Hungarian Astronomical Association, Budapest, Hungary}

\altaffiltext{$\star$}{
	Alfred.~P.~Sloan Research Fellow.
}

\altaffiltext{$\dagger$}{
    Based in part on observations obtained at the W.~M.~Keck
    Observatory, which is operated by the University of California and
    the California Institute of Technology. Keck time has been granted
    by NOAO (A289Hr) and NASA (N167Hr, N029Hr). Based in part on data
    collected at Subaru Telescope, which is operated by the National
    Astronomical Observatory of Japan. Based in part on observations
    made with the Nordic Optical Telescope, operated on the island of
    La Palma jointly by Denmark, Finland, Iceland, Norway, and Sweden,
    in the Spanish Observatorio del Roque de los Muchachos of the
    Instituto de Astrofisica de Canarias.
}


\begin{abstract}

\setcounter{footnote}{10}
We report the discovery of four transiting extrasolar planets
(\hatcurb{34}--\hatcurb{37}) with masses ranging from
\hatcurPPmshort{35} to \hatcurPPmshort{34}\,\mjup\ and periods from
1.33 to 5.45\,days.
These planets orbit
relatively bright F and G dwarf stars (from $V=\hatcurCCtassmvshort{34}$ to
$V=\hatcurCCtassmvshort{37}$). Of particular interest is \hatcurb{34} which
is moderately massive (\hatcurPPmshort{34}\,\mjup), 
has a high eccentricity of $e = \hatcurRVeccen{34}$ at $P=\hatcurLCP{34}\,d$
period,  and shows hints of an outer component. The other three
planets have properties that are typical of hot Jupiters.
\setcounter{footnote}{0}
\end{abstract}

\keywords{
    planetary systems ---
    stars: individual
\setcounter{planetcounter}{1}
(\hatcur{34},
\hatcurCCgsc{34}\loopcommanoperiod
\setcounter{planetcounter}{2}
\hatcur{35},
\hatcurCCgsc{35}\loopcommanoperiod
\setcounter{planetcounter}{3}
\hatcur{36},
\hatcurCCgsc{36}\loopcommanoperiod
\setcounter{planetcounter}{4}
\hatcur{37},
\hatcurCCgsc{37}\loopcommanoperiod)
    techniques --- spectroscopic, photometric
}


\section{Introduction}
\label{sec:introduction}

Transiting extrasolar planets (TEPs) provide unique opportunities to
study the properties of planetary objects outside of the Solar
System. To date
well over 100 such planets have been discovered and
characterized\footnote{See e.g.~http://exoplanets.org 
\citep{wright:2011} for the list of published planets, or
www.exoplanet.eu \citep{schneider:2011} for a more extended compilation, 
including unpublished results.}, 
leading to numerous
insights into the physical properties of planetary systems
\citep[e.g.~see the recent review by][]{rauer:2011}. 
In addition, over a thousand
strong candidates from {\em Kepler} have been identified \citep{borucki:2011},
greatly expanding our understanding of several aspects of planetary systems, 
such as the properties of multi-planet systems
\citep{latham:2011,lissauer:2011},
and the distribution of planetary radii \citep{howard:2011}.
However, due to the large number of important variables which
influence the physical properties of a planet (e.g.~its mass,
composition, age, irradiation, and tides, to name a few), we are still
far from an empirically tested, comprehensive understanding of the
formation and evolution of planetary systems. 

Here we present the
discovery of four new TEPs identified by the Hungarian-made Automated
Telescope Network \citep[HATNet;][]{bakos:2004} survey which
contribute to the rapidly-growing sample of TEPs. These planets
transit relatively bright stars facilitating detailed characterization
of their properties, such as measurements of their masses via radial
velocity (RV) observations of the host star, or measuring their
orbital tilt via the Rossiter-McLaughlin effect.

The HATNet survey for TEPs around bright stars ($9\lesssim r \lesssim
14.5$) operates six wide-field instruments: four at the Fred Lawrence
Whipple Observatory (FLWO) in Arizona (HAT-5, -6, -7, and -10), and
two on the roof of the hangar servicing the Smithsonian Astrophysical
Observatory's Submillimeter Array, in Hawaii (HAT-8 and -9). Since
2006, HATNet has announced and published 33 TEPs
\citep[e.g.][]{johnson:2011}. In this work we report
our thirty-fourth through thirty-seventh discoveries, around the
stars
\setcounter{planetcounter}{1}
\loopand\hatcurCCgsc{34}\loopcomma
\setcounter{planetcounter}{2}
\loopand\hatcurCCgsc{35}\loopcomma
\setcounter{planetcounter}{3}
\loopand\hatcurCCgsc{36}\loopcomma
\setcounter{planetcounter}{4}
\loopand\hatcurCCgsc{37}\loopcomma

In \refsecl{obs} we summarize the detection of the photometric transit
signals and the subsequent spectroscopic and photometric observations
of each star to confirm the planets. In \refsecl{analysis} we analyze
the data to determine the stellar and planetary parameters. The
properties of these planets are briefly discussed in
\refsecl{discussion}.

\ifthenelse{\boolean{emulateapj}}{
    \begin{deluxetable*}{lrrrrrrr}
}{
    \begin{deluxetable}{lrrrrrrr}
}
\tablewidth{0pc}
\tabletypesize{\scriptsize}
\tablecaption{
    Summary of discovery data.
    \label{tab:disco}
}
\tablehead{
    \colhead{~~~~~Planet Host~~~~~}  &
    \colhead{GSC} &
    \colhead{2MASS} &
    \colhead{RA} &
    \colhead{DEC} &
	\colhead{V\tablenotemark{a}} &
	\colhead{Depth\tablenotemark{b}} &
	\colhead{Period}\\
    \colhead{}  &
    \colhead{} &
    \colhead{} &
    \colhead{HH:MM:SS} &
    \colhead{DD:MM:SS} &
	\colhead{mag} &
	\colhead{mmag} &
	\colhead{days}
}
\startdata
\hatcur{34} & \hatcurCCngsc{34} & \hatcurCCntwomass{34} & \hatcurCCra{34} &
	\hatcurCCdec{34} & \hatcurCCtassmv{34} & \hatcurLCdip{34} &
	\hatcurLCPshort{34} \\
\hatcur{35} & \hatcurCCngsc{35} & \hatcurCCntwomass{35} & \hatcurCCra{35} &
	\hatcurCCdec{35} & \hatcurCCtassmv{35} & \hatcurLCdip{35} &
	\hatcurLCPshort{35} \\
\hatcur{36} & \hatcurCCngsc{36} & \hatcurCCntwomass{36} & \hatcurCCra{36} &
	\hatcurCCdec{36} & \hatcurCCtassmv{36} & \hatcurLCdip{36} &
	\hatcurLCPshort{36} \\
\hatcur{37} & \hatcurCCngsc{37} & \hatcurCCntwomass{37} & \hatcurCCra{37} &
	\hatcurCCdec{37} & \hatcurCCtassmv{37}\tablenotemark{c} & \hatcurLCdip{37} &
	\hatcurLCPshort{37}
\enddata
\tablenotetext{a}{From \citealp{droege:2006}.}
\tablenotetext{b}{
	Note that the apparent depth of the HATNet transit for all four targets
    is shallower than the true transit depth due to blending with unresolved
    neighbors in the low spatial resolution HATNet images (the median
    full-width at half maximum of the point-spread function at the center of
    a HATNet image is $\sim 25\arcsec$). Also, we applied the trend filtering
    procedure in non signal-reconstructive mode, which reduces
    the transit depth while increasing the signal-to-noise ratio of the
    detection.  For each system the ratio of the planet and stellar radii,
    which is related to the true transit depth, is determined in
    \refsecl{globmod} using the higher spatial-resolution photometric
    follow-up observations described in \refsecl{phot}.}
\tablenotetext{c}{From \citealp{lasker:2008}.}
\ifthenelse{\boolean{emulateapj}}{
    \end{deluxetable*}
}{
    \end{deluxetable}
}

\section{Observations}
\label{sec:obs}

The observational procedure employed by HATNet to discover TEPs has
been described in detail in several previous discovery papers
\citep[e.g.][]{bakos:2010,latham:2009}. In the following subsections
we highlight specific details of this procedure that are pertinent to
the discoveries of the four planets presented in this paper.

\subsection{Photometric detection}
\label{sec:detection}

\reftabl{photobs} summarizes the HATNet discovery observations of each
new planetary system. The HATNet images were processed and reduced to
trend-filtered light curves following the procedure described by
\cite{bakos:2010} and \cite{pal:2009b}.  The \lcs{} were searched for 
periodic box-shaped
signals using the Box Least-Squares \citep[BLS; see][]{kovacs:2002}
method. We detected significant signals in the \lcs\ of the stars
summarized in \reftabl{disco}.


\begin{figure*}[!ht]
\plottwo{img/\hatcurhtr{34}-hatnet.eps}{img/\hatcurhtr{35}-hatnet.eps}
\plottwo{img/\hatcurhtr{36}-hatnet.eps}{img/\hatcurhtr{37}-hatnet.eps}
\caption[]{
    HATNet \lcs{} of \hatcur{34} through \hatcur{37}. 
    See \reftabl{photobs} for a summary of the
    observations. For each planet we show two panels. The top panel
    shows the unbinned light curve folded with the period resulting
    from the global fit described in \refsecl{analysis}. The solid
    line shows the model fit to the light curve
    (\refsecl{globmod}). The bottom panel shows the region zoomed-in
    on the transit. The dark filled circles show the light curve
    binned in phase with a binsize of 0.002.
\label{fig:hatnet}}
\end{figure*}

\ifthenelse{\boolean{emulateapj}}{
    \begin{deluxetable*}{llrrr}
}{
    \begin{deluxetable}{llrrr}
}
\tablewidth{0pc}
\tabletypesize{\scriptsize}
\tablecaption{
    Summary of photometric observations
    \label{tab:photobs}
}
\tablehead{
    \colhead{~~~~~~~~Instrument/Field~~~~~~~~}  &
    \colhead{Date(s)} &
    \colhead{Number of Images} &
    \colhead{Cadence (sec)} &
    \colhead{Filter}
}
\startdata
\sidehead{\textbf{\hatcur{34}}}
~~~~HAT-7/G293        & 2008 Oct--2009 May & 755 & $330$ & \sband{r} \\
~~~~HAT-8/G293        & 2008 Sep--2008 Dec & 2611 & $330$ & \sband{r} \\
~~~~HAT-6/G341        & 2007 Sep--2007 Dec & 1949 & $330$ & \sband{R} \\
~~~~HAT-9/G341        & 2007 Sep--2007 Nov & 2379 & $330$ & \sband{R} \\
~~~~KeplerCam         & 2010 May 21        & 263 & $60$ & \sband{z} \\
~~~~KeplerCam         & 2010 Oct 10        & 530 & $30$ & \sband{i} \\
\sidehead{\textbf{\hatcur{35}}}
~~~~HAT-5/G364 & 2009 May & 21 & $330$ & \sband{r} \\
~~~~HAT-9/G364 & 2008 Dec--2009 May & 3155 & $330$ & \sband{r} \\
~~~~KeplerCam         & 2011 Jan 16        & 110 & $100$ & \sband{i} \\
~~~~FTN               & 2011 Jan 23        & 185 & $45$ & \sband{i} \\
~~~~KeplerCam         & 2011 Mar 08        & 268 & $60$ & \sband{i} \\
\sidehead{\textbf{\hatcur{36}}}
~~~~HAT-5/G143 & 2010 Apr--2010 Jul & 4471 & $210$ & \sband{r} \\
~~~~HAT-8/G143 & 2010 Apr--2010 Jul & 6262 & $210$ & \sband{r} \\
~~~~KeplerCam         & 2010 Dec 24        & 131 & $130$ & \sband{i} \\
~~~~KeplerCam         & 2011 Feb 03        & 101 & $100$ & \sband{i} \\
~~~~KeplerCam         & 2011 Feb 07        & 105 & $100$ & \sband{i} \\
~~~~KeplerCam         & 2011 Feb 15        & 186 & $60$ & \sband{i} \\
\sidehead{\textbf{\hatcur{37}}}
~~~~HAT-7/G115 & 2009 Sep--2010 Jul & 7102 & $210$ & \sband{r} \\
~~~~HAT-9/G115 & 2008 Aug--2008 Sep & 2293 & $330$ & \sband{R} \\
~~~~KeplerCam         & 2011 Feb 23        & 37  & $165$ & \sband{i} \\
~~~~KeplerCam         & 2011 Mar 23        & 73  & $134$ & \sband{i} \\
~~~~KeplerCam         & 2011 Apr 06        & 102 & $134$ & \sband{i}
\enddata
\ifthenelse{\boolean{emulateapj}}{
    \end{deluxetable*}
}{
    \end{deluxetable}
}

\subsection{Reconnaissance Spectroscopy}
\label{sec:recspec}

High-resolution, low-S/N ``reconnaissance'' spectra were obtained for
\hatcur{34} and \hatcur{35} using the Tillinghast Reflector Echelle
Spectrograph \citep[TRES;][]{furesz:2008} on the 1.5\,m Tillinghast
Reflector at FLWO. These observations were reduced and analyzed
following the procedure described by \cite{quinn:2010} and
\cite{buchhave:2010}; the results are listed in
\reftabl{reconspecobs}. For both objects the spectra were
single-lined, and showed radial velocity (RV) variations on
the order of $\sim 100$\,\ms. Proper phasing of the RV
with the photometric ephemeris gives confidence in acquiring further, high
signal-to-noise spectroscopic observations to refine the orbit (see 
\refsec{hispec}). While
for \hatcur{34} the variations initially
did not appear to phase with the photometric ephemeris, we entertained
the possibility of a very significant non-zero eccentricity, and pursued
follow-up of the target. For \hatcur{35} the
variations were in phase with the photometric ephemeris indicating a
$\sim 2.7\,\mjup$ companion.
For both \hatcur{36} and \hatcur{37} we obtained two TRES spectra near
each of the predicted quadrature phases. For both objects the spectra
were single-lined. For \hatcur{36} the resulting RV measurements
showed $\sim400$\,\ms\ variation in phase with the photometric
ephemeris, while for \hatcur{37} the RV measurements showed $\sim
260$\,\ms\ variation in phase with the ephemeris. We opted to continue
observing both of these objects using TRES with the aim of confirming
the planets. The TRES observations of \hatcur{36} and \hatcur{37} are
discussed further in the following subsection.

\ifthenelse{\boolean{emulateapj}}{
    \begin{deluxetable*}{lrrr}
}{
    \begin{deluxetable}{lrrr}
}
\tablewidth{0pc}
\tabletypesize{\scriptsize}
\tablecaption{
    Summary of reconnaissance spectroscopy observations\tablenotemark{a}
    \label{tab:reconspecobs}
}
\tablehead{
    \multicolumn{1}{c}{Instrument}          &
    \multicolumn{1}{c}{HJD}             &
    \multicolumn{1}{c}{$\gamma_{\rm RV}$\tablenotemark{b}} &
    \multicolumn{1}{c}{CC Peak\tablenotemark{c}} \\
    &
    &
    \multicolumn{1}{c}{(\kms)}              &
}
\startdata
\sidehead{\textbf{\hatcur{34}}}
~~~~TRES                & $2454935.00839$ &  $-47.81$ & $0.726$ \\
~~~~TRES                & $2454966.97204$ &  $-49.19$ & $0.939$ \\
~~~~TRES                & $2454998.97956$ &  $-49.12$ & $0.819$ \\
\sidehead{\textbf{\hatcur{35}}}
~~~~TRES                & $2455289.64284$ &  $41.24$ & $0.863$ \\
~~~~TRES                & $2455291.62482$ &  $40.54$ & $0.883$ \\
~~~~TRES                & $2455320.64182$ & $40.83$ & $0.935$ \\
~~~~TRES                & $2455321.64383$ & $41.08$ & $0.940$
\enddata                                    
\tablenotetext{a}{                          
    For \hatcur{36} and \hatcur{37}, which were confirmed using the
    TRES spectrograph, there is no clear distinction between
    reconnaissance and high-precision observations. We do not list the
    results from the analysis of the TRES spectra for these targets
    here, these are instead described in \refsecl{hispec}.
}                                           
\tablenotetext{b}{                          
    The heliocentric RV of the target in the IAU system, 
	and corrected for the orbital motion of the planet.
}
\tablenotetext{c}{
  The peak value of the cross-correlation function between the observed
  spectrum and the best-matching synthetic template spectrum
  (normalized to be between 0 and 1). Observations with a peak height
  closer to $1.0$ generally correspond to higher S/N spectra.
}
\ifthenelse{\boolean{emulateapj}}{
    \end{deluxetable*}
}{
    \end{deluxetable}
}

\subsection{High resolution, high S/N spectroscopy}
\label{sec:hispec}

We proceeded with the follow-up of each candidate by obtaining
high-resolution, high-S/N spectra to characterize the RV variations,
and to refine the determination of the stellar parameters. These
observations are summarized in \reftabl{highsnspecobs}. The RV
measurements and uncertainties for \hatcur{34} through \hatcur{37}
are given in the Appendix, and in \reftabls{rvs34}{rvs37}, respectively. 
The period-folded
data, along with our best fit described below in \refsecl{analysis},
are displayed in \reffigls{rvbis34}{rvbis37}.

Four facilities were used in the confirmation of these planets
(including three separate facilities used for \hatcur{34}). These
facilities are HIRES \citep{vogt:1994} on the 10\,m Keck~I telescope
in Hawaii, the High-Dispersion Spectrograph
\citep[HDS;][]{noguchi:2002} on the 8.3\,m Subaru telescope in Hawaii,
the FIbre-fed \'{E}chelle Spectrograph (FIES) on the 2.5\,m Nordic
Optical Telescope (NOT) at La Palma, Spain \citep{djupvik:2010}, and
TRES on the FLWO~1.5\,m telescope.

The HIRES and HDS observations made use of the iodine-cell method
\citep{marcy:1992,butler:1996} for precise wavelength calibration and
relative RV determination, while the FIES and TRES observations made
use of Th-Ar lamp spectra obtained before and after the science
exposures. The HIRES observations were reduced to relative RVs in the
barycentric frame following \cite{butler:1996}, \cite{johnson:2009},
and \cite{howard:2010}, the HDS observations were reduced following
\cite{sato:2002,sato:2005}, and the FIES and TRES observations were
reduced following \cite{buchhave:2010}. 

We found that for all four systems the RV residuals from the best-fit
models, described below in \refsecl{globmod}, exhibit excess scatter
over what is expected based on the formal measurement
uncertainties. 
Such excess scatter, or ``jitter'' has been well known for stars, and can
stem for multiple sources.  The excess is in the residuals of the
observations with respect to a physical (and possibly instrumental) model. 
If this model is not adequate, the residuals can be larger than expected. 
For example, in the case of HAT-P-34b, ignoring the linear trend in the RVs
would lead to a much increased ``jitter''.  Additional planets may cause
jitter, as the limited number of RV observations is not enough to uniquely
identify and model such systems.  The typical source of the jitter, however,
is the star itself, namely inhomogeneities (spots, flares, plages, etc.) on
the stellar surface \citep[e.g.][]{makarov:2009,martinez:2010} causing jitters up
to 100\,ms.  Granulation and stellar oscillations contribute at a smaller
scale, but are present for non-active stars that are outside the instability
strip.  A recent publication by \citet{cegla:2012} discusses the stellar
jitter due to variable gravitational redshift of the star, as the stellar
radius changes due to oscillations ($\Delta R$ of $10^{-4}$ causing
$\sim0.1\,\ms$).  And, of course, systematics in the instrument further
inflate the jitter.  A review of RV jitter of stars observed by the Keck
telescope are given in \citet{wright:2005}.

In order to ensure realistic estimates of the system
parameter uncertainties we add in quadrature an RV jitter to the
formal RV measurement uncertainties such that $\chi^{2}$ per degree of
freedom is unity for the best-fit model for each planet. We adopt an
independent jitter for the observations made by each instrument of
each planet. The RV uncertainties given in \reftabls{rvs34}{rvs37} do
{\em not} include this jitter; we do include the jitter in
\reffigls{rvbis34}{rvbis37}.

\ifthenelse{\boolean{emulateapj}}{
    \begin{deluxetable*}{llrr}
}{
    \begin{deluxetable}{llrr}
}
\tablewidth{0pc}
\tabletypesize{\scriptsize}
\tablecaption{
    Summary of high-SN spectroscopic observations used in measuring the orbits
    \label{tab:highsnspecobs}
}
\tablehead{
    \multicolumn{1}{c}{Instrument}  &
    \multicolumn{1}{c}{Date(s)}     &
    \multicolumn{1}{c}{Number of}   \\
    &
    &
    \multicolumn{1}{c}{RV obs.}
}
\startdata
\sidehead{\textbf{\hatcur{34}}}
~~~~Subaru/HDS           & 2010 May & 6 \\
~~~~Keck/HIRES           & 2010 Jun--2010 Sep & 14 \\
~~~~NOT/FIES             & 2010 Jul--2010 Aug & 10 \\
\sidehead{\textbf{\hatcur{35}}}
~~~~Keck/HIRES           & 2010 Sep--2010 Dec & 7 \\
~~~~NOT/FIES             & 2010 Oct & 5\tablenotemark{a} \\
\sidehead{\textbf{\hatcur{36}}}
~~~~FLWO~1.5/TRES        & 2010 Dec--2011 Jan & 12 \\
\sidehead{\textbf{\hatcur{37}}}
~~~~FLWO~1.5/TRES        & 2011 Mar--2011 May & 13
\enddata 
\tablenotetext{a}{
One of the NOT/FIES spectra of \hatcur{35} was aborted early due to
morning twilight and high humidity, another exposure was obtained
partly during transit and may be affected by the Rossiter-McLaughlin
effect. The remaining three NOT/FIES spectra do not provide sufficient
phase coverage to constrain the orbit. We therefore do not include the
velocities measured from these spectra in the analysis of \hatcur{35}.
} \ifthenelse{\boolean{emulateapj}}{
    \end{deluxetable*}
}{
    \end{deluxetable}
}

%
\setcounter{planetcounter}{1}
%
\begin{figure}[ht]
\ifthenelse{\boolean{emulateapj}}{
	\plotone{img/\hatcurhtr{34}-rv.eps}
}{
	\includegraphics*[scale=0.5]{img/\hatcurhtr{34}-rv.eps}
}
\ifthenelse{\value{planetcounter}=1}{
\caption{
    {\em Top panel:} 
        High-precision RV measurements for \hbox{\hatcur{34}{}} shown
        as a function of orbital phase, along with our best-fit model
        (see \reftabl{planetparam}).  Open triangles show measurements
        from Subaru/HDS, filled circles show measurements from
        Keck/HIRES, and filled triangles show measurements from
        NOT/FIES. Zero phase corresponds to the time of mid-transit.
        The center-of-mass velocity and a linear trend have been subtracted.
    {\em Second panel:} 
        velocity $O\!-\!C$ residuals from the best-fit single
        Keplerian orbit model as a function of time. The residuals
        show a slight linear trend, possibly indicating a third body
        in the system. Note that the zero-points of the three separate
        instruments are independently free parameters.
    {\em Third panel:} 
        velocity $O\!-\!C$ residuals from the best fit including both
        the Keplerian orbit and linear trend, shown as a function of
        orbital phase. The error bars include a jitter term
        (\hatcurRVjitterA{34}\,\ms\ for the Keck/HIRES observations,
        and \hatcurRVjitterB{34}\,\ms\ for the Subaru/HDS
        observations; no jitter has been added to the NOT/FIES RV
        uncertainties) added in quadrature to the formal errors (see
        \refsecl{globmod}).
    {\em Fourth panel:} 
        bisector spans (BS) from Keck/HIRES, with the mean value
        subtracted. The measurement from the template spectrum is
        included.
    {\em Bottom panel:} Chromospheric activity index $S$
        measured from the Keck spectra.
    Note the different vertical scales of the panels. Observations
    shown twice are represented with open symbols.
}}{
\caption{
    Keck/HIRES observations of \hatcur{34}. The panels are as in
    \reffigl{rvbis34}.  The parameters used in the
    best-fit model are given in \reftabl{planetparam}.
}}
\label{fig:rvbis34}
\end{figure}
\setcounter{planetcounter}{2}
%
\begin{figure}[ht]
\ifthenelse{\boolean{emulateapj}}{
	\plotone{img/\hatcurhtr{35}-rv.eps}
}{
	\includegraphics*[scale=0.6]{img/\hatcurhtr{35}-rv.eps}
}
\ifthenelse{\value{planetcounter}=1}{
\caption{
    {\em Top panel:} Keck/HIRES RV measurements for
        \hbox{\hatcur{35}{}} shown as a function of orbital
        phase, along with our best-fit model (see
        \reftabl{planetparam}).  Zero phase corresponds to the
        time of mid-transit.  The center-of-mass velocity has been
        subtracted.
    {\em Second panel:} Velocity $O\!-\!C$ residuals from the best
        fit. The error bars include a jitter term
        (\hatcurRVjitter{35}\,\ms) added in quadrature to
        the formal errors (see \refsecl{globmod}).
    {\em Third panel:} Bisector spans (BS), with the mean value
        subtracted. The measurement from the template spectrum is
        included.
    {\em Bottom panel:} Chromospheric activity index $S$
        measured from the Keck spectra.
    Note the different vertical scales of the panels. Observations
    shown twice are represented with open symbols.
}}{
\caption{
    Keck/HIRES observations of \hatcur{35}. The panels are as in
    \reffigl{rvbis34}.  The parameters used in the
    best-fit model are given in \reftabl{planetparam}.
}}
\label{fig:rvbis35}
\end{figure}
\setcounter{planetcounter}{3}
%
\begin{figure}[ht]
\ifthenelse{\boolean{emulateapj}}{
	\plotone{img/\hatcurhtr{36}-rv.eps}
}{
	\includegraphics*[scale=0.6]{img/\hatcurhtr{36}-rv.eps}
}
\ifthenelse{\value{planetcounter}=1}{
\caption{
    {\em Top panel:} Keck/HIRES RV measurements for
        \hbox{\hatcur{36}{}} shown as a function of orbital
        phase, along with our best-fit model (see
        \reftabl{planetparam}).  Zero phase corresponds to the
        time of mid-transit.  The center-of-mass velocity has been
        subtracted.
    {\em Second panel:} Velocity $O\!-\!C$ residuals from the best
        fit. The error bars include a jitter term
        (\hatcurRVjitter{36}\,\ms) added in quadrature to
        the formal errors (see \refsecl{globmod}).
    {\em Third panel:} Bisector spans (BS), with the mean value
        subtracted. The measurement from the template spectrum is
        included.
    {\em Bottom panel:} Relative chromospheric activity index $S$
        measured from the Keck spectra.
    Note the different vertical scales of the panels. Observations
    shown twice are represented with open symbols.
}}{
\caption{
    FLWO~1.5\,m/TRES observations of \hatcur{36}. The panels are as in
    \reffigl{rvbis34}.  The $S$ index is not available for these
    observations. The parameters used in the best-fit model are given
    in \reftabl{planetparam}.
}}
\label{fig:rvbis36}
\end{figure}
\setcounter{planetcounter}{4}
%
\begin{figure}[ht]
\ifthenelse{\boolean{emulateapj}}{
	\plotone{img/\hatcurhtr{37}-rv.eps}
}{
	\includegraphics*[scale=0.6]{img/\hatcurhtr{37}-rv.eps}
}
\ifthenelse{\value{planetcounter}=1}{
\caption{
    {\em Top panel:} Keck/HIRES RV measurements for
        \hbox{\hatcur{37}{}} shown as a function of orbital
        phase, along with our best-fit model (see
        \reftabl{planetparam}).  Zero phase corresponds to the
        time of mid-transit.  The center-of-mass velocity has been
        subtracted.
    {\em Second panel:} Velocity $O\!-\!C$ residuals from the best
        fit. The error bars include a jitter term
        (\hatcurRVjitter{37}\,\ms) added in quadrature to
        the formal errors (see \refsecl{globmod}).
    {\em Third panel:} Bisector spans (BS), with the mean value
        subtracted. The measurement from the template spectrum is
        included.
    {\em Bottom panel:} Relative chromospheric activity index $S$
        measured from the Keck spectra.
    Note the different vertical scales of the panels. Observations
    shown twice are represented with open symbols.
}}{
\caption{
    FLWO~1.5\,m/TRES observations of \hatcur{37}. The panels are as in
    \reffigl{rvbis34}.  The $S$ index is not available for these
    observations. The parameters used in the best-fit model are given
    in \reftabl{planetparam}.
}}
\label{fig:rvbis37}
\end{figure}

For \hatcur{34} and \hatcur{35} we also show the $S$ index, which is a
measure of the chromospheric activity of the star derived from the
flux in the cores of the \ion{Ca}{2} H and K lines.  This index was
computed following \cite{isaacson:2010} and has been calibrated to the
scale of \citet{vaughan:1978}. A procedure for
obtaining calibrated $S$ index values from the TRES spectra has not
yet been developed, so we do not provide these measurements for
\hatcur{36} or \hatcur{37}.
We convert the $S$ index values to $\log R^{\prime}_{\rm HK}$
following \cite{noyes:1984} and find median values of $\log
R^{\prime}_{\rm HK} = -4.859$ and $\log R^{\prime}_{\rm HK} = -5.242$
for \hatcur{34} and \hatcur{35}, respectively. These values imply that
neither star has a particularly high level of chromospheric activity.

Following \cite{queloz:2001} and \cite{torres:2007}, we checked whether the
measured radial velocities are not real, but are instead caused by
distortions in the spectral line profiles due to contamination from a nearby
unresolved eclipsing binary.  A bisector (BS) analysis for each system based on
the Keck and TRES spectra was done as described in \S 5 of
\cite{bakos:2007a}.  For \hatcur{35}, which is relatively faint, we found
that the measured BSs were significantly affected by scattered moonlight and
applied an empirical correction for this effect following
\citet{hartman:2009} (see also \citet{kovacs:2010}). 
For \hatcur{34} the BS scatter is fairly high ($\sim
25$\,\ms), but this is in line with the high RV jitter ($\sim 60$\,\ms),
which is typical of an F star with $\vsini = \hatcurSMEiivsin{34}$\,\kms\
\citep{saar:2003,hartman:2011b}.

None of the systems show significant bisector span variations
(relative to the semi-amplitude of the RV variations) that phase with
the photometric ephemeris.  Such variations are generally expected if
the transit and RV signals were due to blends rather than planets.
While the lack of bisector span variations does not exclude {\em all}
blend scenarios, it does significantly limit the possible blend
scenarios that can reproduce our current data within the measurement
errors, i.e.~configurations that are compatible with the photometric
and spectroscopic observations, proper motions, color indices, and
moderately high resolution imaging.  We have found in the past that
invoking detailed blend modeling to exclude all possible blend
configurations and confirm the planet hypothesis \citep[e.g.][sections
  3.2.2, 3.2.3]{hartman:2011a} is rarely of any incremental value when
the ingress and egress durations are short relative to the total
transit duration, the RV variations exhibit a Keplerian orbit in phase
with the photometric ephemeris, and bisector spans show no correlation
with the orbit.  We conclude that the velocity variations detected for
all four stars are real, and that each star is orbited by a close-in
giant planet.

\subsection{Photometric follow-up observations}
\label{sec:phot}

%
\setcounter{planetcounter}{1}
%
\begin{figure}[!ht]
\ifthenelse{\boolean{emulateapj}}{
	\plotone{img/\hatcurhtr{34}-lc.eps}
}{
	\includegraphics*[scale=0.6]{img/\hatcurhtr{34}-lc.eps}
}
\ifthenelse{\value{planetcounter}=1}{
\caption{
    Unbinned transit \lcs{} for \hatcur{34}, acquired with
    KeplerCam at the \flwof{} telescope.  The light curves have been
    EPD- and TFA-processed, as described in \refsec{globmod}.
    The dates of the events are indicated.  Curves after the first are
    displaced vertically for clarity.  Our best fit from the global
    modeling described in \refsecl{globmod} is shown by the solid
    lines.  Residuals from the fits are displayed at the bottom, in the
    same order as the top curves.  The error bars represent the photon
    and background shot noise, plus the readout noise.
}}{
\caption{
    Similar to \reffigl{lc34}; here we show the follow-up
    \lcs{} for \hatcur{34}.
}}
\label{fig:lc34}
\end{figure}
\setcounter{planetcounter}{2}
%
\begin{figure}[!ht]
\ifthenelse{\boolean{emulateapj}}{
	\plotone{img/\hatcurhtr{35}-lc.eps}
}{
	\includegraphics*[scale=0.6]{img/\hatcurhtr{35}-lc.eps}
}
\ifthenelse{\value{planetcounter}=1}{
\caption{
    Unbinned transit \lcs{} for \hatcur{35}, acquired with
    KeplerCam at the \flwof{} telescope.  The light curves have been
    EPD- and TFA-processed, as described in \refsec{globmod}.
    The dates of the events are indicated.  Curves after the first are
    displaced vertically for clarity.  Our best fit from the global
    modeling described in \refsecl{globmod} is shown by the solid
    lines.  Residuals from the fits are displayed at the bottom, in the
    same order as the top curves.  The error bars represent the photon
    and background shot noise, plus the readout noise.
}}{
\caption{
    Similar to \reffigl{lc34}; here we show the follow-up
    \lcs{} for \hatcur{35}.
}}
\label{fig:lc35}
\end{figure}
\setcounter{planetcounter}{3}
%
\begin{figure}[!ht]
\ifthenelse{\boolean{emulateapj}}{
	\plotone{img/\hatcurhtr{36}-lc.eps}
}{
	\includegraphics*[scale=0.6]{img/\hatcurhtr{36}-lc.eps}
}
\ifthenelse{\value{planetcounter}=1}{
\caption{
    Unbinned transit \lcs{} for \hatcur{36}, acquired with
    KeplerCam at the \flwof{} telescope.  The light curves have been
    EPD- and TFA-processed, as described in \refsec{globmod}.
    The dates of the events are indicated.  Curves after the first are
    displaced vertically for clarity.  Our best fit from the global
    modeling described in \refsecl{globmod} is shown by the solid
    lines.  Residuals from the fits are displayed at the bottom, in the
    same order as the top curves.  The error bars represent the photon
    and background shot noise, plus the readout noise.
}}{
\caption{
    Similar to \reffigl{lc34}; here we show the follow-up
    \lcs{} for \hatcur{36}.
}}
\label{fig:lc36}
\end{figure}
\setcounter{planetcounter}{4}
%
\begin{figure}[!ht]
\ifthenelse{\boolean{emulateapj}}{
	\plotone{img/\hatcurhtr{37}-lc.eps}
}{
	\includegraphics*[scale=0.6]{img/\hatcurhtr{37}-lc.eps}
}
\ifthenelse{\value{planetcounter}=1}{
\caption{
    Unbinned transit \lcs{} for \hatcur{37}, acquired with
    KeplerCam at the \flwof{} telescope.  The light curves have been
    EPD- and TFA-processed, as described in \refsec{globmod}.
    The dates of the events are indicated.  Curves after the first are
    displaced vertically for clarity.  Our best fit from the global
    modeling described in \refsecl{globmod} is shown by the solid
    lines.  Residuals from the fits are displayed at the bottom, in the
    same order as the top curves.  The error bars represent the photon
    and background shot noise, plus the readout noise.
}}{
\caption{
    Similar to \reffigl{lc34}; here we show the follow-up
    \lcs{} for \hatcur{37}.
}}
\label{fig:lc37}
\end{figure}

In order to permit a more accurate modeling of the light curves, we
conducted additional photometric observations using the KeplerCam CCD
camera on the \flwof{} telescope, and the Spectral Instrument CCD on 
the 2.0\,m
Faulkes Telescope North (FTN) at Haleakala Observatory in Hawaii,
which is operated by the Las Cumbres Observatory Global Telescope
(LCOGT).  The observations for each target are summarized in
\reftabl{photobs}.

The reduction of these images was performed as described by
\citet{bakos:2010}.  We applied External Parameter Decorrelation
\citep[EPD;][]{bakos:2010} and the Trend Filtering Algorithm
\citep[TFA;][]{kovacs:2005} to remove trends simultaneously with the
light curve modeling.
The final time series, together with our
best-fit transit \lc{} models, are shown in the top portion of
\reffigls{lc34}{lc37} for \hatcur{34} through \hatcur{37}, respectively.
The individual measurements, permitting independent analysis
by other researchers, are reported in the Appendix, in 
\reftabls{phfu34}{phfu37} (the full data are available in electronic format).


\section{Analysis}
\label{sec:analysis}

\begin{figure*}[!ht]
\plottwo{img/\hatcurhtr{34}-iso-ar.eps}{img/\hatcurhtr{35}-iso-ar.eps}
\plottwo{img/\hatcurhtr{36}-iso-ar.eps}{img/\hatcurhtr{37}-iso-ar.eps}
\caption{
    Comparison of the measured values of $\teffstar$ and $\arstar$ for
    \hatcur{34} (upper left), \hatcur{35} (upper right), \hatcur{36}
    (lower left) and \hatcur{37} (lower right) to model isochrones
    from \cite{\hatcurisocite{34}}. The isochrones are generated for
    the measured metallicity of each star, and for ages of 0.5\,Gyr
    and 1 to 3\,Gyr in steps of 0.25\,Gyr for HAT-P-34, and of
    0.5\,Gyr and 1 to 14 Gyr in steps of 1\,Gyr for HAT-P-35, HAT-P-36
    and HAT-P-37 (ages increase from left to right in each plot). The
    lines show the 1$\sigma$ and 2$\sigma$ confidence ellipses for
    the measured parameters. The initial values of \teffstar\ and
    \arstar\ from the initial spectroscopic and \lc\ analyses are
    represented with a triangle in each panel.  }
\label{fig:iso}
\end{figure*}

\subsection{Properties of the parent star}
\label{sec:stelparam}

\ifthenelse{\boolean{emulateapj}}{
    \begin{deluxetable*}{lccccl}
}{
    \begin{deluxetable}{lccccl}
}
\tablewidth{0pc}
\tabletypesize{\scriptsize}
\tablecaption{
    Stellar parameters for \hatcur{34} through \hatcur{37}
    \label{tab:stellar}
}
\tablehead{
    \multicolumn{1}{c}{} &
    \multicolumn{1}{c}{{\bf HAT-P-34}} &
    \multicolumn{1}{c}{{\bf HAT-P-35}} &
    \multicolumn{1}{c}{{\bf HAT-P-36}} &
    \multicolumn{1}{c}{{\bf HAT-P-37}} &
    \multicolumn{1}{c}{} \\
    \multicolumn{1}{c}{~~~~~~~~Parameter~~~~~~~~} &
    \multicolumn{1}{c}{Value} &
    \multicolumn{1}{c}{Value} &
    \multicolumn{1}{c}{Value} &
    \multicolumn{1}{c}{Value} &
    \multicolumn{1}{c}{Source}
}
\startdata
\noalign{\vskip -3pt}
\sidehead{Spectroscopic properties}
~~~~$\teffstar$ (K)\dotfill         &  \hatcurSMEteff{34} &  \hatcurSMEteff{35} &  \hatcurSMEteff{36}  &  \hatcurSMEteff{37}  & Spec.~Analysis.\tablenotemark{a}\\
~~~~$\feh$\dotfill                  &  \hatcurSMEzfeh{34} &  \hatcurSMEzfeh{35} &  \hatcurSMEzfeh{36}  &  \hatcurSMEzfeh{37}  & Spec.~Analysis.                 \\
~~~~$\vsini$ (\kms)\dotfill         &  \hatcurSMEvsin{34} &  \hatcurSMEvsin{35} &  \hatcurSMEvsin{36}  &  \hatcurSMEvsin{37}  & Spec.~Analysis.                 \\
~~~~$\vmac$ (\kms)\dotfill          &  \hatcurSMEvmac{34} &  \hatcurSMEvmac{35} &  \hatcurSMEvmac{36}  &  \hatcurSMEvmac{37}  & Spec.~Analysis.                 \\
~~~~$\vmic$ (\kms)\dotfill          &  \hatcurSMEvmic{34} &  \hatcurSMEvmic{35} &  \hatcurSMEvmic{36}  &  \hatcurSMEvmic{37}  & Spec.~Analysis.                 \\
~~~~$\gamma_{\rm RV}$ (\kms)\dotfill &  \hatcurTRESgamma{34} &  \hatcurTRESgamma{35} &  \hatcurTRESgamma{36}  &  \hatcurTRESgamma{37}  & TRES                  \\
\sidehead{Photometric properties}
~~~~$V$ (mag)\dotfill               &  \hatcurCCtassmv{34} &  \hatcurCCtassmv{35} &  \hatcurCCtassmv{36} &  \hatcurCCtassmv{37} & TASS,GSC\tablenotemark{b}                \\
~~~~$\vic$ (mag)\dotfill            &  \hatcurCCtassvi{34} &  \hatcurCCtassvi{35} &  \hatcurCCtassvi{36} &  \hatcurCCtassvi{37} & TASS                \\
~~~~$J$ (mag)\dotfill               &  \hatcurCCtwomassJmag{34} &  \hatcurCCtwomassJmag{35} &  \hatcurCCtwomassJmag{36}&  \hatcurCCtwomassJmag{37}& 2MASS           \\
~~~~$H$ (mag)\dotfill               &  \hatcurCCtwomassHmag{34} &  \hatcurCCtwomassHmag{35} &  \hatcurCCtwomassHmag{36}&  \hatcurCCtwomassHmag{37}& 2MASS           \\
~~~~$K_s$ (mag)\dotfill             &  \hatcurCCtwomassKmag{34} &  \hatcurCCtwomassKmag{35} &  \hatcurCCtwomassKmag{36}&  \hatcurCCtwomassKmag{37}& 2MASS           \\
\sidehead{Derived properties}
~~~~$\mstar$ ($\msun$)\dotfill      &  \hatcurISOmlong{34} &  \hatcurISOmlong{35} &  \hatcurISOmlong{36}  &  \hatcurISOmlong{37}  & \hatcurisoshort{34}+\hatcurlumind{34}+Spec.~Analysis. \tablenotemark{c}\\
~~~~$\rstar$ ($\rsun$)\dotfill      &  \hatcurISOrlong{34} &  \hatcurISOrlong{35} &  \hatcurISOrlong{36}  &  \hatcurISOrlong{37}  & \hatcurisoshort{34}+\hatcurlumind{34}+Spec.~Analysis         \\
~~~~$\loggstar$ (cgs)\dotfill       &  \hatcurISOlogg{34} &  \hatcurISOlogg{35} &  \hatcurISOlogg{36}   &  \hatcurISOlogg{37}   & \hatcurisoshort{34}+\hatcurlumind{34}+Spec.~Analysis         \\
~~~~$\lstar$ ($\lsun$)\dotfill      &  \hatcurISOlum{34} &  \hatcurISOlum{35} &  \hatcurISOlum{36}    &  \hatcurISOlum{37}    & \hatcurisoshort{34}+\hatcurlumind{34}+Spec.~Analysis         \\
~~~~$M_V$ (mag)\dotfill             &  \hatcurISOmv{34} &  \hatcurISOmv{35} & \hatcurISOmv{36}   & \hatcurISOmv{37}   & \hatcurisoshort{34}+\hatcurlumind{34}+Spec.~Analysis         \\
~~~~$M_K$ (mag,\hatcurjhkfilset{34})\dotfill &  \hatcurISOMK{34} &  \hatcurISOMK{35} &  \hatcurISOMK{36}&  \hatcurISOMK{37}& \hatcurisoshort{34}+\hatcurlumind{34}+Spec.~Analysis         \\
~~~~Age (Gyr)\dotfill               &  \hatcurISOage{34} &  \hatcurISOage{35} &  \hatcurISOage{36}    &  \hatcurISOage{37}    & \hatcurisoshort{34}+\hatcurlumind{34}+Spec.~Analysis         \\
~~~~Distance (pc)\dotfill           &  \hatcurXdist{34}\phn &  \hatcurXdist{35}\phn &  \hatcurXdist{36}\phn  &  \hatcurXdist{37}\phn  & \hatcurisoshort{34}+\hatcurlumind{34}+Spec.~Analysis\\ [-1.5ex]
\enddata
\tablenotetext{a}{
    Based on the analysis of high resolution spectra. For \hatcur{34}
    and \hatcur{35} this corresponds to SME applied to iodine-free
    Keck/HIRES spectra, while for \hatcur{36} and \hatcur{37} this
    corresponds to SPC applied to the TRES spectra
    (\refsecl{stelparam}). These parameters also have a small
    dependence on the iterative analysis incorporating the isochrone
    search and global modeling of the data, as described in the text.
}
\tablenotetext{b}{For \hatcur{34} through \hatcur{36} the value is taken from the TASS catalog, while for \hatcur{37} the value is taken from the GSC version 2.3.2.}
\tablenotetext{c}{
    \hatcurisoshort{34}+\hatcurlumind{34}+Spec.~Analysis = Based on the
    \hatcurisoshort{34} isochrones \citep{\hatcurisocite{34}},
    \hatcurlumind{34} as a luminosity indicator, and the spectroscopic
    analysis results.
}
\ifthenelse{\boolean{emulateapj}}{
    \end{deluxetable*}
}{
    \end{deluxetable}
}

Stellar atmospheric parameters for \hatcur{34} and \hatcur{35} were
measured using our template spectra obtained with the Keck/HIRES
instrument, and the analysis package known as Spectroscopy Made Easy
\citep[SME;][]{valenti:1996}, along with the atomic line database of
\cite{valenti:2005}. For \hatcur{36} and \hatcur{37} the stellar
atmospheric parameters were determined by cross-correlating the TRES
observations against a finely sampled grid of synthetic spectra based
on \cite{kurucz:2005} model atmospheres. This procedure, known as
Stellar Parameter Classification (SPC), will be described in detail in
a forthcoming paper (Buchhave et al., in preparation). We note that SPC has
been performed in the past on numerous HATNet transiting planet candidates
(Buchhave, personal communication), and the results were consistent with 
those of SME.

For each star, we obtained the following {\em initial} spectroscopic
parameters and uncertainties:
\begin{itemize}
\item {\em \hatcur{34}} --
effective temperature $\teffstar=\hatcurSMEiteff{34}$\,K, 
metallicity $\feh=\hatcurSMEizfeh{34}$\,dex,
stellar surface gravity $\loggstar=\hatcurSMEilogg{34}$\,(cgs), and
projected rotational velocity $\vsini=\hatcurSMEivsin{34}$\,\kms.
\item {\em \hatcur{35}} --
effective temperature $\teffstar=\hatcurSMEiteff{35}$\,K, 
metallicity $\feh=\hatcurSMEizfeh{35}$\,dex,
stellar surface gravity $\loggstar=\hatcurSMEilogg{35}$\,(cgs), and
projected rotational velocity $\vsini=\hatcurSMEivsin{35}$\,\kms.
\item {\em \hatcur{36}} --
effective temperature $\teffstar=\hatcurSMEiteff{36}$\,K, 
metallicity $\feh=\hatcurSMEizfeh{36}$\,dex,
stellar surface gravity $\loggstar=\hatcurSMEilogg{36}$\,(cgs), and
projected rotational velocity $\vsini=\hatcurSMEivsin{36}$\,\kms.
\item {\em \hatcur{37}} --
effective temperature $\teffstar=\hatcurSMEiteff{37}$\,K, 
metallicity $\feh=\hatcurSMEizfeh{37}$\,dex,
stellar surface gravity $\loggstar=\hatcurSMEilogg{37}$\,(cgs), and
projected rotational velocity $\vsini=\hatcurSMEivsin{37}$\,\kms.
\end{itemize}

Following \cite{bakos:2010}, these initial values of \teffstar,
\loggstar, and \feh\ were used to determine the quadratic limb-darkening
coefficients needed in the global modeling of the follow-up photometry
(summarized in \refsecl{globmod}). This analysis yields \rhostar, the
mean stellar density, which is closely related to \arstar, the
normalized semimajor axis, and provides a tighter constraint on the
stellar parameters than does the spectroscopically determined
\loggstar\ \citep[e.g.][]{sozzetti:2007}.  We combined \rhostar,
\teffstar, and \feh\ with stellar evolution models from the
\hatcurisofull{34} series by \cite{\hatcurisocite{34}} to determine
probability distributions of other stellar properties, including
\loggstar.  For each system we carried out a second SME or SPC
iteration in which we adopted the new value of \loggstar\ so
determined and held it fixed in a new SME or SPC analysis, adjusting
only \teffstar, \feh, and \vsini, followed by a second global modeling
of the RV and light curves, together with improved limb darkening parameters. 
The {\em final} atmospheric parameters that we
adopt, together with stellar parameters inferred from the
\hatcurisoshort{34} models (such as the mass, radius and age) are
listed in \reftabl{stellar} for all four stars.

The inferred location of each star in a diagram of \arstar\ versus
\teffstar, analogous to the classical H-R diagram, is shown in \reffigl{iso}.
In each case the stellar properties and their 1$\sigma$ and 2$\sigma$
confidence ellipses are displayed against the backdrop of model
isochrones for a range of ages, and the appropriate stellar
metallicity.  For comparison, the locations implied by the initial SME
and SPC results are also shown (in each case with a triangle).

The stellar evolution modeling provides color indices that we
compared against the measured values, as a sanity check.  For each star
we used the near-infrared magnitudes from the 2MASS Catalogue
\citep{skrutskie:2006}, which are given in \reftabl{stellar}.  These
were converted to the photometric system of the models (ESO) using the
transformations by \citet{carpenter:2001}.  The resulting 2MASS-based color
indices were all consistent (within $1\sigma$) 
with the stellar model based color indices.

The distance for each star given in
\reftabl{stellar} was computed from the absolute $K$ magnitude from the
models and the 2MASS $K_s$ magnitudes, ignoring extinction.

\subsection{Global modeling of the data}
\label{sec:globmod}

\ifthenelse{\boolean{emulateapj}}{
    \begin{deluxetable*}{lcccc}
}{
    \begin{deluxetable}{lcccc}
}
\tabletypesize{\tiny}
\tablecaption{Orbital and planetary parameters for \hatcurb{34} through \hatcurb{37}\label{tab:planetparam}}
\tablehead{
    \multicolumn{1}{c}{} &
    \multicolumn{1}{c}{{\bf \hatcurb{34}}} &
    \multicolumn{1}{c}{{\bf \hatcurb{35}}} &
    \multicolumn{1}{c}{{\bf \hatcurb{36}}} &
    \multicolumn{1}{c}{{\bf \hatcurb{37}}} \\
    \multicolumn{1}{c}{~~~~~~~~~~~~~~~Parameter~~~~~~~~~~~~~~~} &
    \multicolumn{1}{c}{Value} &
    \multicolumn{1}{c}{Value} &
    \multicolumn{1}{c}{Value} &
    \multicolumn{1}{c}{Value}
}
\startdata
\noalign{\vskip -3pt}
\sidehead{\Lc{} parameters}
~~~$P$ (days)             \dotfill     & $\hatcurLCP{34}$              & $\hatcurLCP{35}$              & $\hatcurLCP{36}$              & $\hatcurLCP{37}$              \\
~~~$T_c$ (${\rm BJD}$)    
      \tablenotemark{a}   \dotfill     & $\hatcurLCT{34}$              & $\hatcurLCT{35}$              & $\hatcurLCT{36}$              & $\hatcurLCT{37}$              \\
~~~$T_{14}$ (days)
      \tablenotemark{a}   \dotfill     & $\hatcurLCdur{34}$            & $\hatcurLCdur{35}$            & $\hatcurLCdur{36}$            & $\hatcurLCdur{37}$            \\
~~~$T_{12} = T_{34}$ (days)
      \tablenotemark{a}   \dotfill     & $\hatcurLCingdur{34}$         & $\hatcurLCingdur{35}$         & $\hatcurLCingdur{36}$         & $\hatcurLCingdur{37}$         \\
~~~$\arstar$              \dotfill     & $\hatcurPPar{34}$             & $\hatcurPPar{35}$             & $\hatcurPPar{36}$             & $\hatcurPPar{37}$             \\
~~~$\zrstar$              \dotfill     & $\hatcurLCzeta{34}$\phn       & $\hatcurLCzeta{35}$\phn       & $\hatcurLCzeta{36}$\phn       & $\hatcurLCzeta{37}$\phn       \\
~~~$\rpl/\rstar$          \dotfill     & $\hatcurLCrprstar{34}$        & $\hatcurLCrprstar{35}$        & $\hatcurLCrprstar{36}$        & $\hatcurLCrprstar{37}$        \\
~~~$b^2$                  \dotfill     & $\hatcurLCbsq{34}$            & $\hatcurLCbsq{35}$            & $\hatcurLCbsq{36}$            & $\hatcurLCbsq{37}$            \\
~~~$b \equiv a \cos i/\rstar$
                          \dotfill     & $\hatcurLCimp{34}$            & $\hatcurLCimp{35}$            & $\hatcurLCimp{36}$            & $\hatcurLCimp{37}$            \\
~~~$i$ (deg)              \dotfill     & $\hatcurPPi{34}$\phn          & $\hatcurPPi{35}$\phn          & $\hatcurPPi{36}$\phn          & $\hatcurPPi{37}$\phn          \\

\sidehead{Quadratic limb-darkening coefficients \tablenotemark{b}}
~~~$c_1,i$ (linear term)  \dotfill     & $\hatcurLBii{34}$             & $\hatcurLBii{35}$             & $\hatcurLBii{36}$             & $\hatcurLBii{37}$             \\
~~~$c_2,i$ (quadratic term) \dotfill   & $\hatcurLBiii{34}$            & $\hatcurLBiii{35}$            & $\hatcurLBiii{36}$            & $\hatcurLBiii{37}$            \\
~~~$c_1,z$                \dotfill     & $\hatcurLBiz{34}$             & $\cdots$             & $\cdots$   & $\hatcurLBiz{37}$          \\
~~~$c_2,z$                \dotfill     & $\hatcurLBiiz{34}$            & $\cdots$            & $\cdots$    & $\hatcurLBiiz{37}$       \\

\sidehead{RV parameters}
~~~$K$ (\ms)              \dotfill     & $\hatcurRVK{34}$\phn\phn      & $\hatcurRVK{35}$\phn\phn      & $\hatcurRVK{36}$\phn\phn      & $\hatcurRVK{37}$\phn\phn      \\
~~~$e \cos \omega$\tablenotemark{c} 
                          \dotfill     & $\hatcurRVk{34}$\phs          & $\hatcurRVk{35}$\phs          & $\hatcurRVk{36}$\phs          & $\hatcurRVk{37}$\phs          \\
~~~$e \sin \omega$\tablenotemark{c}
                          \dotfill     & $\hatcurRVh{34}$              & $\hatcurRVh{35}$              & $\hatcurRVh{36}$              & $\hatcurRVh{37}$              \\
~~~$e$                     \dotfill    & $\hatcurRVeccen{34}$          & $\hatcurRVeccen{35}$          & $\hatcurRVeccen{36}$          & $\hatcurRVeccen{37}$          \\
~~~$\omega$ (deg)          \dotfill    & $\hatcurRVomega{34}$\phn      & $\hatcurRVomega{35}$\phn      & $\hatcurRVomega{36}$\phn      & $\hatcurRVomega{37}$\phn      \\
~~~$\dot{\gamma}$ (\msd)   \dotfill    & $\hatcurRVtrone{34}$\phn      & $\cdots$
                      & $\cdots$       & $\cdots$                \\
\sidehead{RV jitter}
~~~Keck/HIRES (\ms)        \dotfill    & \hatcurRVjitterA{34}           & \hatcurRVjitter{35}           & $\cdots$     & $\cdots$      \\
~~~Subaru/HDS (\ms)        \dotfill    & \hatcurRVjitterB{34}           & $\cdots$           & $\cdots$   & $\cdots$        \\
~~~NOT/FIES (\ms)          \dotfill    & \hatcurRVjitterC{34}           & $\cdots$           & $\cdots$     & $\cdots$      \\
~~~FLWO~1.5/TRES (\ms)     \dotfill    & $\cdots$           & $\cdots$           & \hatcurRVjitter{36}           & \hatcurRVjitter{37}           \\

\sidehead{Secondary eclipse parameters}
~~~$T_s$ (BJD)             \dotfill    & $\hatcurXsecondary{34}$       & $\hatcurXsecondary{35}$       & $\hatcurXsecondary{36}$       & $\hatcurXsecondary{37}$       \\
~~~$T_{s,14}$              \dotfill    & $\hatcurXsecdur{34}$          & $\hatcurXsecdur{35}$          & $\hatcurXsecdur{36}$          & $\hatcurXsecdur{37}$          \\
~~~$T_{s,12}$              \dotfill    & $\hatcurXsecingdur{34}$       & $\hatcurXsecingdur{35}$       & $\hatcurXsecingdur{36}$       & $\hatcurXsecingdur{37}$       \\

\sidehead{Planetary parameters}
~~~$\mpl$ ($\mjup$)        \dotfill    & $\hatcurPPmlong{34}$          & $\hatcurPPmlong{35}$          & $\hatcurPPmlong{36}$          & $\hatcurPPmlong{37}$          \\
~~~$\rpl$ ($\rjup$)        \dotfill    & $\hatcurPPrlong{34}$          & $\hatcurPPrlong{35}$          & $\hatcurPPrlong{36}$          & $\hatcurPPrlong{37}$          \\
~~~$C(\mpl,\rpl)$
    \tablenotemark{d}      \dotfill    & $\hatcurPPmrcorr{34}$         & $\hatcurPPmrcorr{35}$         & $\hatcurPPmrcorr{36}$         & $\hatcurPPmrcorr{37}$         \\
~~~$\rhopl$ (\gcmc)        \dotfill    & $\hatcurPPrho{34}$            & $\hatcurPPrho{35}$            & $\hatcurPPrho{36}$            & $\hatcurPPrho{37}$            \\
~~~$\log g_p$ (cgs)        \dotfill    & $\hatcurPPlogg{34}$           & $\hatcurPPlogg{35}$           & $\hatcurPPlogg{36}$           & $\hatcurPPlogg{37}$           \\
~~~$a$ (AU)                \dotfill    & $\hatcurPParel{34}$           & $\hatcurPParel{35}$           & $\hatcurPParel{36}$           & $\hatcurPParel{37}$           \\
~~~$T_{\rm eq}$ (K)        \dotfill    & $\hatcurPPteff{34}$           & $\hatcurPPteff{35}$           & $\hatcurPPteff{36}$           & $\hatcurPPteff{37}$           \\
~~~$\Theta$\tablenotemark{e}\dotfill   & $\hatcurPPtheta{34}$          & $\hatcurPPtheta{35}$          & $\hatcurPPtheta{36}$          & $\hatcurPPtheta{37}$          \\
~~~$\langle F \rangle$ ($10^{9}$\ergscmsq) \tablenotemark{f}
                          \dotfill     & $\hatcurPPfluxavg{34}$        & $\hatcurPPfluxavg{35}$        & $\hatcurPPfluxavg{36}$        & $\hatcurPPfluxavg{37}$        \\ [-1.5ex]
\enddata
\tablenotetext{a}{
	\footnotesize
    \ensuremath{T_c}: Reference epoch of mid transit that
    minimizes the correlation with the orbital period.
    \ensuremath{T_{14}}: total transit duration, time
    between first to last contact;
    \ensuremath{T_{12}=T_{34}}: ingress/egress time, time between first
    and second, or third and fourth contact.
    Barycentric Julian dates (BJD) throughout the paper are calculated
    from Coordinated Universal Time (UTC).
}
\tablenotetext{b}{
	\footnotesize
    Values for a quadratic law, adopted from the tabulations by
    \cite{claret:2004} according to the spectroscopic (SME) parameters
    listed in \reftabl{stellar}.
}
\tablenotetext{c}{
	\footnotesize
    ~Lagrangian orbital parameters derived from the global modeling, 
    and primarily determined by the RV data. 
}
\tablenotetext{d}{
	\footnotesize
    Correlation coefficient between the planetary mass \mpl\ and radius
    \rpl.
}
\tablenotetext{e}{
	\footnotesize
    The Safronov number is given by $\Theta = \frac{1}{2}(V_{\rm
    esc}/V_{\rm orb})^2 = (a/\rpl)(\mpl / \mstar )$
    \citep[see][]{hansen:2007}.
}
\tablenotetext{f}{
	\footnotesize
    Incoming flux per unit surface area, averaged over the orbit.
}
\ifthenelse{\boolean{emulateapj}}{
    \end{deluxetable*}
}{
    \end{deluxetable}
}

We modeled simultaneously the HATNet photometry, the follow-up
photometry, and the high-precision RV measurements using the procedures
described by \citet{bakos:2010}. Namely, the best fit was determined by a
downhill simplex minimization, and was followed by a Monte-Carlo Markov Chain
run to scan the parameter space around the minimum, and establish the
errors \citet{pal:2009b}.
For each system we used a
\cite{mandel:2002} transit model, together with the EPD and TFA
trend-filters, to describe the follow-up light curves, a
\cite{mandel:2002} transit model for the HATNet light curve(s), and a
Keplerian orbit using the formalism of \cite{pal:2009a} for the RV
curve(s).  For \hatcur{34} we included a linear trend in the RV model,
but find that it is only significant at the $\sim 2\sigma$ level; the
planet and stellar parameters are changed by less than $1\sigma$ when
the trend is not included in the fit. The parameters that we adopt for
each system are listed in \reftabl{planetparam}. In all cases we allow
the eccentricity to vary so that the uncertainty on this parameter is
propagated into the uncertainties on the other physical parameters,
such as the stellar and planetary masses and radii; the observations
of \hatcurb{35}, \hatcurb{36}, and \hatcurb{37} are consistent with
these planets being on circular orbits.



\section{Discussion}
\label{sec:discussion}

We have presented the discovery of four new transiting planets. Below
we briefly discuss their properties.

\subsection{\hatcurb{34}}
\label{sec:disc34}
\hatcurb{34} is a relatively massive $\mpl =
\hatcurPPmlong{34}\,\mjup$ planet on a relatively long period ($P =
\hatcurLCP{34}$\,d), eccentric ($e = \hatcurRVeccen{34}$) orbit. There
are only five known transiting planets with higher eccentricities
(HAT-P-2b, $e=0.5171 \pm 0.0033$, \citealp{pal:2010,bakos:2007a};
CoRoT-10b, $e=0.53 \pm 0.04$, \citealp{bonomo:2010}; 
CoRoT-20b, $e=0.562 \pm 0.013$, \citealp{deleuil:2012}; 
HD~17156b, $e =0.669 \pm 0.008$, \citealp{madhusudhan:2009}; 
and HD~80606b, $e = 0.9330 \pm 0.0005$, \citealp{hebrard:2010}), 
all of which have longer
orbital periods than \hatcurb{34}. Of these planets, HAT-P-2b is most
similar in orbital period to \hatcurb{34}, but it has a mass that is
more than two times larger than that of \hatcurb{34}. Two planets with
masses, radii and equilibrium temperatures within 10\% of the values
of \hatcurb{34} (assuming zero albedo and full heat redistribution) are 
CoRoT-18b \citep{hebrard:2011} and WASP-32b
\citep{maxted:2010}; however neither of these planets has a
significant eccentricity. 

\hatcurb{34} is a promising target for measuring the
Rossiter-McLaughlin effect \citep{rossiter:1924,mclaughlin:1924}, since
the host star is bright ($V = \hatcurCCtassmvshort{34}$), has a significant
spin (\vsini = \hatcurSMEvsin{34}\,\kms), and the transit is moderately
long ($T_{14} = \hatcurLCdur{34}$\,days).  Also, the transit is far
from equatorial ($b = \hatcurLCimp{34}$), a configuration that is
important for resolving the degeneracy between \vsini\ and $\lambda$,
which is the sky-plane projected angle between the planetary orbital
normal and the stellar spin axis.  \citet{winn:2010} pointed out that
hot Jupiters around stars with $\teffstar \gtrsim 6250\,K$ have a
higher chance of being misaligned. Based on the effective temperature of the
host star \hatcurSMEteff{34}\,K, we thus expect that \hatcurb{34} has
a higher chance of misalignment (note that this may not necessarily yield 
a non-zero
$\lambda$, if \hatcurb{34}'s orbit is tilted along the line of sight). 
Alternatively, \citet{schlaufman:2010} used a stellar rotation model
and observed $\vsini$ values to statistically identify TEP systems that
may be misaligned along the line of sight, and concluded these
preferentially occur at $\mstar > 1.2\msun$.  Based on the stellar mass
alone ($\hatcurISOmshort{34}$\,\msun) we the chances
for misalignment are increased.


\subsection{\hatcurb{35}}
\label{sec:disc35}
\hatcurb{35} is a very typical $\mpl = \hatcurPPmlong{35}\,\mjup$,
$\rpl = \hatcurPPrlong{35}\,\rjup$ planet on a $P = \hatcurLCP{35}$\,d
orbit and with an equilibrium temperature of $T_{\rm eq} =
\hatcurPPteff{35}$\,K (again, 
assuming zero albedo and full heat redistribution).  
There are four other planets with masses, radii
and equilibrium temperatures that are all within 10\% of the values for
\hatcurb{35}.  These are HAT-P-5b \citep{bakos:2007b}, HAT-P-6b
\citep{noyes:2008}, OGLE-TR-211b \citep{udalski:2008}, and WASP-26b
\citep{smalley:2010}.  The stellar effective temperature
(\hatcurSMEteff{35}\,K) is close to the assumed border-line between
well-aligned and misaligned systems, making it an interesting system
for testing the RM effect (with the caveat that $\vsini$, and thus the expected
amplitude of the anomaly, is low).

\subsection{\hatcurb{36}}
\label{sec:disc36}
\hatcurb{36} is a very short period ($P = \hatcurLCP{36}$\,d) planet
with a mass of $\mpl = \hatcurPPmlong{36}\,\mjup$, a radius of $\rpl =
\hatcurPPrlong{36}\,\rjup$, and an equilibrium temperature of $T_{\rm
  eq} = \hatcurPPteff{36}$\,K. There are two other planets with
masses, radii, and equilibrium temperatures within 10\% of the values
for \hatcurb{36}: TrES-3b \citep{odonovan:2007} and WASP-3b
\citep{pollacco:2008}. 

\subsection{\hatcurb{37}}
\label{sec:disc37}
Like the preceding planets, \hatcurb{37} also has very typical
physical properties, with $\mpl = \hatcurPPmlong{37}\,\mjup$, $\rpl =
\hatcurPPrlong{37}\,\rjup$, $P = \hatcurLCP{37}$\,d, and $T_{\rm eq} =
\hatcurPPteff{37}$\,K. Three planets with masses, radii and
equilibrium temperatures within 10\% of the values for \hatcurb{37}
are HD~189733b \citep{bouchy:2005}, OGLE-TR-113b \citep{bouchy:2004},
and XO-5b \citep{burke:2008}. \hatcur{37} lies just outside of the
field of view of the {\em Kepler} Space mission and is listed in the
Kepler Input Catalog
(KIC\footnote{http://www.cfa.harvard.edu/kepler/kic/kicindex.html}) as
KIC~12396036.

\subsection{On the eccentricity of \hatcurb{34}}
According to \citet{adams:2006}, the eccentricity of a hot
Jupiter's orbit decays both due to the tides on the star and due to the
tides on the planet, with the tides on the planet dominating the
circularization as long as the tidal quality factor of the planet
($Q_P$) is not much larger than the star's ($Q_{\star}$).  Both of
these factors are highly uncertain with various theoretical and
observational constraints ranging over several orders of magnitude.  In
particular tidal circularization of main sequence stars
\citep{claret:1997,meibom:2005,zahn:1989a,zahn:1989b} seem to indicate
$10^5\lesssim Q_{\star} \lesssim 10^6$.  On the other hand, the
discovery of extremely short period massive planets, the two most
dramatic being WASP-18b \citep{hellier:2009} and WASP-19b
\citep{hellier:2011}, seems to be inconsistent with such efficient
dissipation (Penev et al.~in preparation), requiring much larger values
$Q_{\star}\gtrsim 10^8$, which coincide well with the theoretical values
derived by \citet{penev:2011}, who argue that binary stars and
star-planet systems are subject to different modes of dissipation in
the star.  The tidal dissipation parameter in the planet has also been
the subject of many studies attempting to constrain it either from
theory \citep{bodenheimer:2003,ogilvie:2004} or from
the observed configuration of Jupiter's satellites
\citep{goldreich:1966} giving $10^5\lesssim Q_P \lesssim 10^7$.

\renewcommand{\textfloatsep}{10mm}
\begin{figure*}[!ht]
\begin{center}
\includegraphics[scale=0.4]{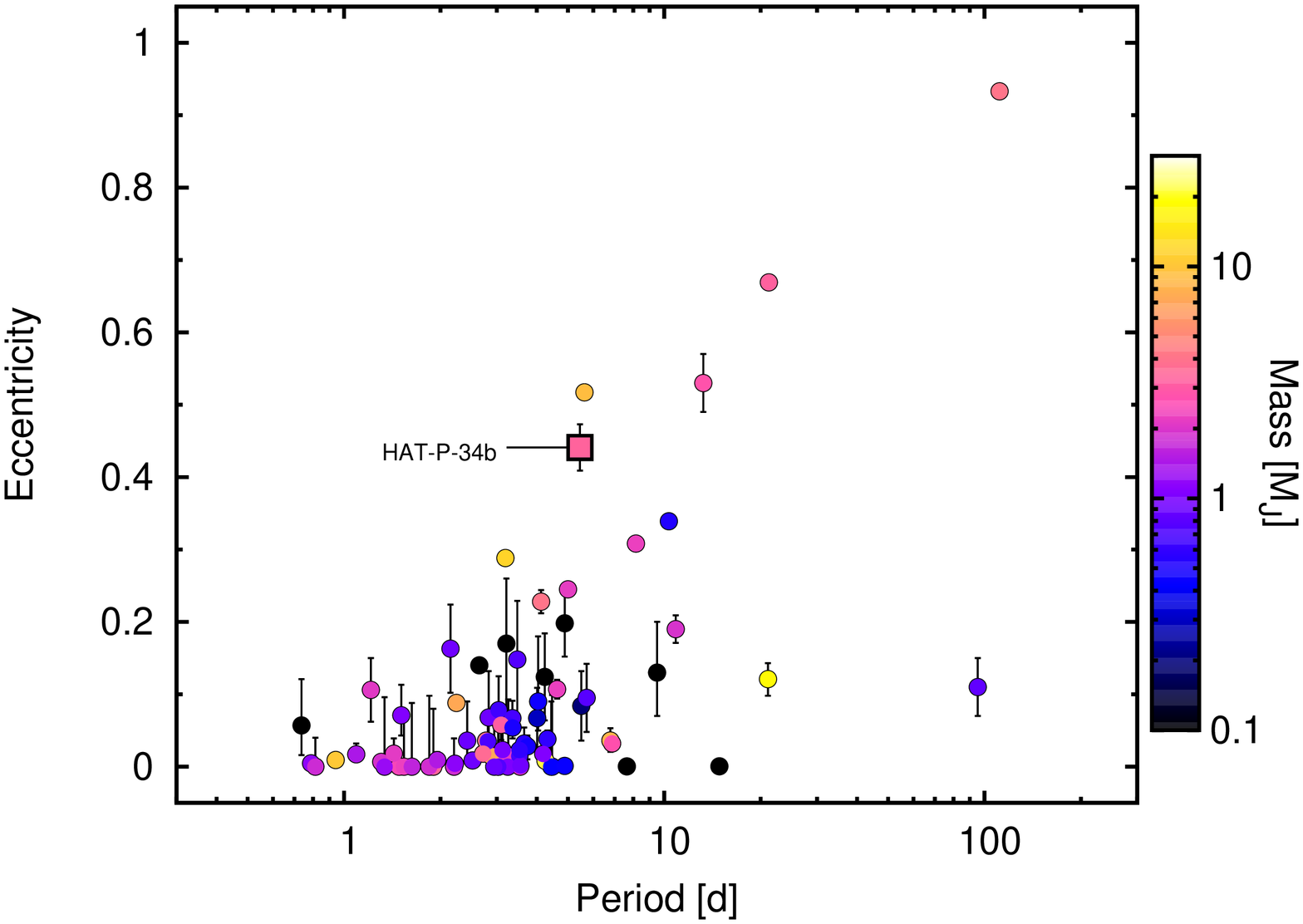}
\caption{
   Orbital-period--eccentricity diagram of TEPs with eccentricity
   uncertainty less than $0.1$. The color (greyscale shade) 
   of the symbols indicates the
   mass of each planet. \hatcurb{34} is labelled. As expected from
   tidal evolution theory, high eccentricity planets tend to have
   longer orbital periods and greater masses.
}
\label{fig:plpereccen}
\end{center}
\end{figure*}
\renewcommand{\textfloatsep}{4mm}

\begin{figure*}[!ht]
\begin{center}
\includegraphics[scale=0.4]{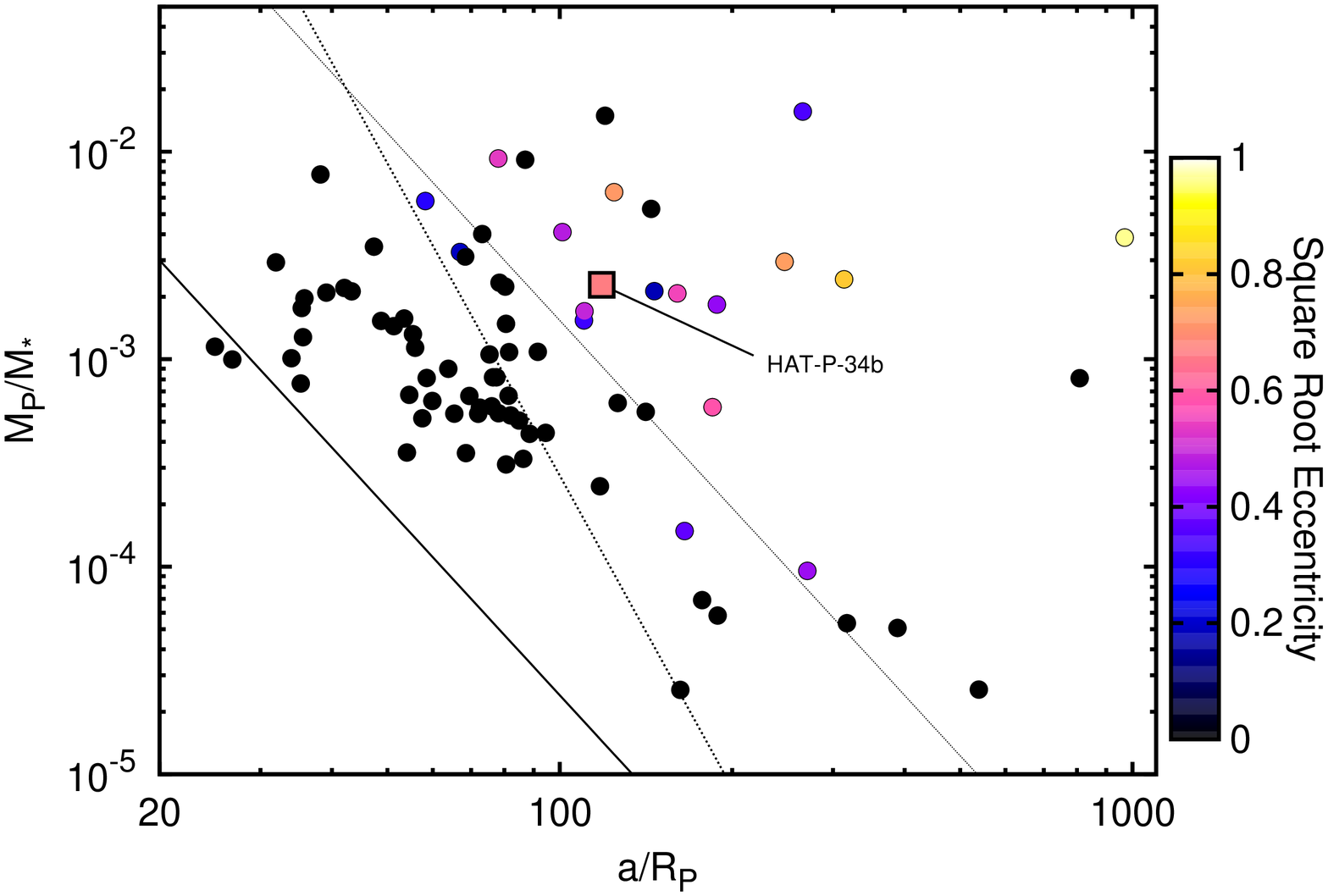}
\caption{
   A ``tidal'' diagram following Fig.~3 of \cite{pont:2011}. The color
   (greyscale shade)
   of the symbols indicates the eccentricity, we assume zero
   eccentricity for planets that have a measured eccentricity within
   4$\sigma$ of zero. The dotted line shows the locus of points with a
   circularization time-scale of 1\,Gyr assuming small eccentricity,
   $Q_{P} = 10^6$, and $P = 3$\,d. The thick solid line shows the relation
   $a = 2 a_{H}$ where $a_{H}$ is the semi-major axis at which the
   radius of the planet equals its Hill radius. The thin solid line
   shows $a = 4 a_{H}$. 
}
\label{fig:eccen}
\end{center}
\end{figure*}

With this in mind we conclude that the circularization of HAT-P-34b's
orbit is likely dominated by the tidal dissipation in the planet and
using $Q_P=10^6$ and the expression for the tidal circularization
timescale from \citet{adams:2006}, we estimate the eccentricity of
HAT-P-34b should decay on the scale of 2Gyr, i.e.~it is not in
conflict with theoretical expectations. The possible outer companion
indicated by the RV trend may also be responsible for pumping the
eccentricity of the inner planet \hatcurb{34} (see
\citealp{correia:2011} for a discussion).

\reffigl{plpereccen} shows \hatcurb{34} on the orbital
period--eccentricity plane of TEPs with well determined parameters
(using our own compilation that attempts to keep up with various
refinements to these parameters).  It is apparent that eccentricity is
correlated with orbital period and with planet mass, as expected from
tidal theory.  \hatcurb{34} lies in a sparse position in these
diagrams; for example, it has a high eccentricity for its period, the
only similar planet being HAT-P-2b.

\reffigl{eccen} is a ``tidal'' plot \citep[see Fig.~3
  of][]{pont:2011}, showing TEPs with well measured properties in the
$a/\rpl$--$\mpl/\mstar$ plane, using more data points (including the
present discoveries) than \citet{pont:2011}.  Since $\tau_c =
(4/63)Q_P\sqrt{(a^3/GM_{\star}}(a/R_P)^5\mpl/\mstar$, we expect
planets with small relative semi-major axis ($a/R_P$) or planets with
small relative mass ($\mpl/\mstar$) to be circularized.  This is
indeed the case, as shown by the intensity (color) scale representing
eccentricity. For hot Jupiters that migrate in by circularization of
an initially very eccentric orbit, the expected ``parking distance''
is $\sim 2 a_H$ \citep{ford:2006}, where $a_H$ is the semi-major axis
at which the radius of the planet equals its Hill radius. The thick solid
line in \reffigl{eccen} shows this relation. A fairly good match for the
dividing line between the circularized (denoted by black points) and
eccentric (grey or color) points is at $a \approx 4 a_H$ (marked with
a thin solid line). This relation now includes very small mass Kepler
discoveries.  \hatcurb{34} belongs to the sparse group of high
relative semi-major axis ($a/\rpl$) and massive extrasolar planets.


\acknowledgements 
HATNet operations have been funded by NASA grants NNG04GN74G,
NNX08AF23G. We acknowledge partial funding of the HATNet follow-up effort
from NSF AST-1108686. We acknowledge partial support also from the
Kepler Mission under NASA Cooperative Agreement NCC2-1390 (D.W.L., PI). 
G.K.~thanks the Hungarian Scientific Research Foundation (OTKA) for
support through grant K-81373.  This research has made use of Keck
telescope time granted through NOAO (program A289Hr) and NASA (N167Hr,
N029Hr).  This paper uses observations obtained with facilities of the
Las Cumbres Observatory Global Telescope.  Data presented in this paper
are based on observations obtained at the HAT station at the
Submillimeter Array of SAO, and HAT station at the Fred Lawrence
Whipple Observatory of SAO. The authors wish to recognize and
acknowledge the very significant cultural role and reverence that the
summit of Mauna Kea has always had within the indigenous Hawaiian
community.  We are most fortunate to have the opportunity to conduct
observations {\mbox from this mountain.}
\clearpage



\appendix
\section{Spectroscopic and Photometric Data}

The following tables present the spectroscopic data (radial velocities,
bisector spans, and activity index measurements) and high precision
photometric data for the four planets presented in this paper.

\ifthenelse{\boolean{emulateapj}}{
    \begin{deluxetable*}{lrrrrrrr}
}{
    \begin{deluxetable}{lrrrrrrr}
}
\tablewidth{0pc}
\tablecaption{
    Relative radial velocities, bisector spans, and activity index
    measurements of \hatcur{34}.
    \label{tab:rvs34}
}
\tablehead{
    \colhead{BJD} &
    \colhead{RV\tablenotemark{a}} &
    \colhead{\ensuremath{\sigma_{\rm RV}}\tablenotemark{b}} &
    \colhead{BS} &
    \colhead{\ensuremath{\sigma_{\rm BS}}} &
    \colhead{S\tablenotemark{c}} &
    \colhead{Phase} &
    \colhead{Instrument}\\
    \colhead{\hbox{(2,454,000$+$)}} &
    \colhead{(\ms)} &
    \colhead{(\ms)} &
    \colhead{(\ms)} &
    \colhead{(\ms)} &
    \colhead{} &
    \colhead{} &
    \colhead{}
}
\startdata
\ifthenelse{\boolean{rvtablelong}}{
    \input{data/\hatcurhtr{34}_rvtable.tex}
}{
    \input{data/\hatcurhtr{34}_rvtable_short.tex}
}
\enddata
\tablenotetext{a}{
    The zero-point of these velocities is arbitrary. An overall offset
    $\gamma_{\rm rel}$ fitted to these velocities in \refsecl{globmod}
    has {\em not} been subtracted.
}
\tablenotetext{b}{
    Internal errors excluding the component of astrophysical jitter
    considered in \refsecl{globmod}.
}
\tablenotetext{c}{
    Chromospheric activity index, calibrated to the scale
    of \citet{vaughan:1978}.
}
\ifthenelse{\boolean{rvtablelong}}{
    \tablecomments{
        Note that for the iodine-free template exposures we do not
        measure the RV but do measure the BS and S index.  Such
        template exposures can be distinguished by the missing RV
        value.
    }
}{
    \tablecomments{
        Note that for the iodine-free template exposures we do not
        measure the RV but do measure the BS and S index.  Such
        template exposures can be distinguished by the missing RV
        value.  This table is presented in its entirety in the
        electronic edition of the Astrophysical Journal.  A portion is
        shown here for guidance regarding its form and content.
    }
} 
\ifthenelse{\boolean{emulateapj}}{
    \end{deluxetable*}
}{
    \end{deluxetable}
}

\ifthenelse{\boolean{emulateapj}}{
    \begin{deluxetable*}{lrrrrrrr}
}{
    \begin{deluxetable}{lrrrrrrr}
}
\tablewidth{0pc}
\tablecaption{
    Relative radial velocities, bisector spans, and activity index
    measurements of \hatcur{35}.
    \label{tab:rvs35}
}
\tablehead{
    \colhead{BJD} &
    \colhead{RV\tablenotemark{a}} &
    \colhead{\ensuremath{\sigma_{\rm RV}}\tablenotemark{b}} &
    \colhead{BS} &
    \colhead{\ensuremath{\sigma_{\rm BS}}} &
    \colhead{S\tablenotemark{c}} &
    \colhead{Phase} &
    \colhead{Instrument}\\
    \colhead{\hbox{(2,454,000$+$)}} &
    \colhead{(\ms)} &
    \colhead{(\ms)} &
    \colhead{(\ms)} &
    \colhead{(\ms)} &
    \colhead{} &
    \colhead{} &
    \colhead{}
}
\startdata
\ifthenelse{\boolean{rvtablelong}}{
    \input{data/\hatcurhtr{35}_rvtable.tex}
}{
    \input{data/\hatcurhtr{35}_rvtable_short.tex}
}
\enddata
\tablenotetext{a}{
    The zero-point of these velocities is arbitrary. An overall offset
    $\gamma_{\rm rel}$ fitted to these velocities in \refsecl{globmod}
    has {\em not} been subtracted.
}
\tablenotetext{b}{
    Internal errors excluding the component of astrophysical jitter
    considered in \refsecl{globmod}.
}
\tablenotetext{c}{
    Chromospheric activity index, calibrated to the scale
    of \citet{vaughan:1978}.
}
\tablenotetext{d}{
    The FIES/NOT observations of \hatcur{35} were not used in the
    analysis, see the footnote to \reftabl{highsnspecobs}. Transit
    ingress began during the hour-long exposure obtained at phase
    $0.973$, and the exposure obtained at phase $0.344$ has a low S/N
    ratio and was obtained during morning twilight.
}
\ifthenelse{\boolean{rvtablelong}}{
    \tablecomments{
        Note that for the iodine-free template exposures we do not
        measure the RV but do measure the BS and S index.  Such
        template exposures can be distinguished by the missing RV
        value.
    }
}{
    \tablecomments{
        Note that for the iodine-free template exposures we do not
        measure the RV but do measure the BS and S index.  Such
        template exposures can be distinguished by the missing RV
        value.  This table is presented in its entirety in the
        electronic edition of the Astrophysical Journal.  A portion is
        shown here for guidance regarding its form and content.
    }
} 
\ifthenelse{\boolean{emulateapj}}{
    \end{deluxetable*}
}{
    \end{deluxetable}
}

\ifthenelse{\boolean{emulateapj}}{
    \begin{deluxetable*}{lrrrrrr}
}{
    \begin{deluxetable}{lrrrrrr}
}
\tablewidth{0pc}
\tablecaption{
    Relative radial velocities, bisector spans, and activity index
    measurements of \hatcur{36}.
    \label{tab:rvs36}
}
\tablehead{
    \colhead{BJD} &
    \colhead{RV\tablenotemark{a}} &
    \colhead{\ensuremath{\sigma_{\rm RV}}\tablenotemark{b}} &
    \colhead{BS} &
    \colhead{\ensuremath{\sigma_{\rm BS}}} &
    \colhead{Phase}\\
    \colhead{\hbox{(2,454,000$+$)}} &
    \colhead{(\ms)} &
    \colhead{(\ms)} &
    \colhead{(\ms)} &
    \colhead{(\ms)} &
    \colhead{}
}
\startdata
\ifthenelse{\boolean{rvtablelong}}{
    \input{data/\hatcurhtr{36}_rvtable.tex}
}{
    \input{data/\hatcurhtr{36}_rvtable_short.tex}
}
\enddata
\tablenotetext{a}{
    The zero-point of these velocities is arbitrary. An overall offset
    $\gamma_{\rm rel}$ fitted to these velocities in \refsecl{globmod}
    has {\em not} been subtracted.
}
\tablenotetext{b}{
    Internal errors excluding the component of astrophysical jitter
    considered in \refsecl{globmod}.
}
\ifthenelse{\boolean{emulateapj}}{
    \end{deluxetable*}
}{
    \end{deluxetable}
}

\ifthenelse{\boolean{emulateapj}}{
    \begin{deluxetable*}{lrrrrrr}
}{
    \begin{deluxetable}{lrrrrrr}
}
\tablewidth{0pc}
\tablecaption{
    Relative radial velocities, bisector spans, and activity index
    measurements of \hatcur{37}.
    \label{tab:rvs37}
}
\tablehead{
    \colhead{BJD} &
    \colhead{RV\tablenotemark{a}} &
    \colhead{\ensuremath{\sigma_{\rm RV}}\tablenotemark{b}} &
    \colhead{BS} &
    \colhead{\ensuremath{\sigma_{\rm BS}}} &
    \colhead{Phase}\\
    \colhead{\hbox{(2,454,000$+$)}} &
    \colhead{(\ms)} &
    \colhead{(\ms)} &
    \colhead{(\ms)} &
    \colhead{(\ms)} &
    \colhead{}
}
\startdata
\ifthenelse{\boolean{rvtablelong}}{
    \input{data/\hatcurhtr{37}_rvtable.tex}
}{
    \input{data/\hatcurhtr{37}_rvtable_short.tex}
}
\enddata
\tablenotetext{a}{
    The zero-point of these velocities is arbitrary. An overall offset
    $\gamma_{\rm rel}$ fitted to these velocities in \refsecl{globmod}
    has {\em not} been subtracted.
}
\tablenotetext{b}{
    Internal errors excluding the component of astrophysical jitter
    considered in \refsecl{globmod}.
}
\ifthenelse{\boolean{emulateapj}}{
    \end{deluxetable*}
}{
    \end{deluxetable}
}

\begin{deluxetable}{lrrrr}
\tablewidth{0pc}
\tablecaption{
    High-precision differential photometry of
    \hatcur{34}\label{tab:phfu34}.
}
\tablehead{
    \colhead{BJD} & 
    \colhead{Mag\tablenotemark{a}} & 
    \colhead{\ensuremath{\sigma_{\rm Mag}}} &
    \colhead{Mag(orig)\tablenotemark{b}} & 
    \colhead{Filter} \\
    \colhead{\hbox{~~~~(2,400,000$+$)~~~~}} & 
    \colhead{} & 
    \colhead{} &
    \colhead{} & 
    \colhead{}
}
\startdata
\input{data/\hatcurhtr{34}_phfu_tab_short.tex}
\enddata
\tablenotetext{a}{
    The out-of-transit level has been subtracted. These magnitudes have
    been subjected to the EPD and TFA procedures, carried out
    simultaneously with the transit fit.
}
\tablenotetext{b}{
    Raw magnitude values without application of the EPD and TFA
    procedures.
}
\tablecomments{
    This table is available in a machine-readable form in the online
    journal.  A portion is shown here for guidance regarding its form
    and content.
}
\end{deluxetable}

\begin{deluxetable}{lrrrr}
\tablewidth{0pc}
\tablecaption{
    High-precision differential photometry of
    \hatcur{35}\label{tab:phfu35}.
}
\tablehead{
    \colhead{BJD} & 
    \colhead{Mag\tablenotemark{a}} & 
    \colhead{\ensuremath{\sigma_{\rm Mag}}} &
    \colhead{Mag(orig)\tablenotemark{b}} & 
    \colhead{Filter} \\
    \colhead{\hbox{~~~~(2,400,000$+$)~~~~}} & 
    \colhead{} & 
    \colhead{} &
    \colhead{} & 
    \colhead{}
}
\startdata
\input{data/\hatcurhtr{35}_phfu_tab_short.tex}
\enddata
\tablenotetext{a}{
    The out-of-transit level has been subtracted. These magnitudes have
    been subjected to the EPD and TFA procedures, carried out
    simultaneously with the transit fit.
}
\tablenotetext{b}{
    Raw magnitude values without application of the EPD and TFA
    procedures.
}
\tablecomments{
    This table is available in a machine-readable form in the online
    journal.  A portion is shown here for guidance regarding its form
    and content.
}
\end{deluxetable}

\begin{deluxetable}{lrrrr}
\tablewidth{0pc}
\tablecaption{
    High-precision differential photometry of
    \hatcur{36}\label{tab:phfu36}.
}
\tablehead{
    \colhead{BJD} & 
    \colhead{Mag\tablenotemark{a}} & 
    \colhead{\ensuremath{\sigma_{\rm Mag}}} &
    \colhead{Mag(orig)\tablenotemark{b}} & 
    \colhead{Filter} \\
    \colhead{\hbox{~~~~(2,400,000$+$)~~~~}} & 
    \colhead{} & 
    \colhead{} &
    \colhead{} & 
    \colhead{}
}
\startdata
\input{data/\hatcurhtr{36}_phfu_tab_short.tex}
\enddata
\tablenotetext{a}{
    The out-of-transit level has been subtracted. These magnitudes have
    been subjected to the EPD and TFA procedures, carried out
    simultaneously with the transit fit.
}
\tablenotetext{b}{
    Raw magnitude values without application of the EPD and TFA
    procedures.
}
\tablecomments{
    This table is available in a machine-readable form in the online
    journal.  A portion is shown here for guidance regarding its form
    and content.
}
\end{deluxetable}

\begin{deluxetable}{lrrrr}
\tablewidth{0pc}
\tablecaption{
    High-precision differential photometry of
    \hatcur{37}\label{tab:phfu37}.
}
\tablehead{
    \colhead{BJD} & 
    \colhead{Mag\tablenotemark{a}} & 
    \colhead{\ensuremath{\sigma_{\rm Mag}}} &
    \colhead{Mag(orig)\tablenotemark{b}} & 
    \colhead{Filter} \\
    \colhead{\hbox{~~~~(2,400,000$+$)~~~~}} & 
    \colhead{} & 
    \colhead{} &
    \colhead{} & 
    \colhead{}
}
\startdata
\input{data/\hatcurhtr{37}_phfu_tab_short.tex}
\enddata
\tablenotetext{a}{
    The out-of-transit level has been subtracted. These magnitudes have
    been subjected to the EPD and TFA procedures, carried out
    simultaneously with the transit fit.
}
\tablenotetext{b}{
    Raw magnitude values without application of the EPD and TFA
    procedures.
}
\tablecomments{
    This table is available in a machine-readable form in the online
    journal.  A portion is shown here for guidance regarding its form
    and content.
}
\end{deluxetable}


\begin{thebibliography}{}

\bibitem[Adams and Laughlin(2006)]{adams:2006}
    Adams, F.~C., \& Laughlin, G.\ 2006, \apj, 649, 1004

\bibitem[Bakos et al.(2004)]{bakos:2004}
 Bakos, G.~\'A., Noyes, R.~W., Kov\'acs, G., Stanek, K.~Z.,
 Sasselov, D.~D., \& Domsa, I.~2004, \pasp, 116, 266

\bibitem[Bakos et al.(2007a)]{bakos:2007a}
 Bakos, G.~\'A., et al.~2007a, \apj, 670, 826

\bibitem[Bakos et al.(2007b)]{bakos:2007b}
 Bakos, G.~\'A., et al.~2007b, \apj, 671, L173

\bibitem[Bakos et al.(2010)]{bakos:2010} Bakos, G.~{\'A}., et al.~2010,
\apj, 710, 1724

\bibitem[Bodenheimer, Laughlin, \& Lin(2003)]{bodenheimer:2003}
Bodenheimer, P., Laughlin, G., \& Lin, D.~N.~C.~2003, \apj, 592, 555


\bibitem[Bonomo et al.(2010)]{bonomo:2010} Bonomo, A.~S., et al.\ 2010, \aap, submitted


\bibitem[Borucki et al.(2011)]{borucki:2011} Borucki, W.~J., et al.\ 2011, \apj, 736, 19

\bibitem[Bouchy et  al.(2004)]{bouchy:2004} Bouchy, F., Pont, F., Santos, N.~C., Melo, C., Mayor, M., Queloz, D., \& Udry, S.\ 2004, \aap, 421, L13

\bibitem[Bouchy et  al.(2005)]{bouchy:2005} Bouchy, F., et al.\ 2005, \aap, 444, L15

\bibitem[Buchhave et al.(2010)]{buchhave:2010}
Buchhave, L.~A., et al.\ 2010, \apj, 720, 1118

\bibitem[Burke et al.(2008)]{burke:2008} Burke, C.~J., et al.\  2008, \apj, 686, 1331

\bibitem[Butler et al.(1996)]{butler:1996} 
Butler, R.~P.~et al.~1996, \pasp, 108, 500

\bibitem[Carpenter(2001)]{carpenter:2001} Carpenter, J.~M.~2001, \aj, 121, 2851 
\bibitem[Claret(2004)]{claret:2004}
 Claret, A.~2004, \aap, 428, 1001

\bibitem[Cegla et al.(2012)]{cegla:2012} Cegla, H.~M., et al.~2012, 
\mnras, 421, L54 


\bibitem[Claret \& Cunha(1997)]{claret:1997} 
Claret, A.,\& Cunha, N.~C.~S.~1997, \aap, 318, 187 

\bibitem[Correia et al.(2011)]{correia:2011}
Correia, A.~C.~M., Bou\'e, G., \& Laskar, J.\ 2011, ApJL in press, arXiv:1111.5486

\bibitem[Deleuil et al.(2012)]{deleuil:2012} Deleuil, M., et al.~2012, 
\aap, 538, A145 

\bibitem[Djupvik \& Andersen(2010)]{djupvik:2010} Djupvik, A.~A., \&
  Andersen, J.\ 2010, in ``Highlights of Spanish Astrophysics V''
  eds. J.~M.~Diego, L.~J.~Goicoechea, J.~I.~Gonz\'alez-Serrano, \&
  J.~Gorgas (Springer: Berlin), p.\ 211

\bibitem[Droege et al.(2006)]{droege:2006}
Droege, T.~F., Richmond, M.~W., \& Sallman, M.~2006, \pasp, 118, 1666

\bibitem[Ford and Rasio(2006)]{ford:2006}
Ford, E.~B., \& Rasio, F.~A.\ 2006, \apj, 638, L45

\bibitem[F\H{u}r\'{e}sz(2008)]{furesz:2008} F\H{u}r\'esz, G.\ 2008, Ph.D. thesis, University of Szeged, Hungary

\bibitem[Goldreich \& Soter(1966)]{goldreich:1966}
Goldreich, P., \& Soter, S.~1966, Icarus, 5, 375 

\bibitem[Hansen \& Barman(2007)]{hansen:2007} Hansen, B.~M.~S., \& Barman, T.~2007, \apj, 671, 861 

\bibitem[Hartman et al.(2009)]{hartman:2009} Hartman, J.~D., et al.~2009, 
\apj, 706, 785 

\bibitem[Hartman et al.(2011a)]{hartman:2011a}
Hartman, J.~D., Bakos, G.~\'A., Kipping, D.~M., et al.\ 2011a, \apj, 728, 138

\bibitem[Hartman et al.(2011b)]{hartman:2011b}
Hartman, J.~D., Bakos, G.~\'A., Torres, G., et al.\ 2011b, \apj, 742, 59

\bibitem[H\'ebrard et al.(2010)]{hebrard:2010} H\'ebrard, G., D\'esert, J.-M., D{\'{\i}}az, R.~F., et al.\ 2010, \aap, 516, A95

\bibitem[H\'ebrard et al.(2011)]{hebrard:2011} H\'ebrard, G., Evans, T.~M., Alonso, R.\ 2011, \aap, 533, A130

\bibitem[Hellier et al.(2009)]{hellier:2009} Hellier, C., et al.~2009, 
\nat, 460, 1098 

\bibitem[Hellier et al.(2011)]{hellier:2011} Hellier, C., Anderson, D.~R., 
Collier-Cameron, A., Miller, G.~R.~M., Queloz, D., Smalley, B., Southworth, 
J., \& Triaud, A.~H.~M.~J.~2011, \apjl, 730, L31 

\bibitem[Howard et al.(2010)]{howard:2010}
Howard, A.~W., et al.\ 2010, \apj, 721, 1467

\bibitem[Howard et al.(2011)]{howard:2011}
Howard, A.~W., Marcy, G.~W., Bryson, S.~T., et al.\ 2011, arXiv:1103.2541

\bibitem[Isaacson \& Fischer(2010)]{isaacson:2010}
Isaacson, H., \& Fischer, D.\ 2010, \apj, 725, 875

\bibitem[Johnson et al.(2009)]{johnson:2009} Johnson, J.~A., Winn,
  J.~N., Albrecht, S., Howard, A.~W., Marcy, G.~W., \& Gazak,
  J.~Z.\ 2009, \pasp, 121, 1104

\bibitem[Johnson et al.(2011)]{johnson:2011} Johnson, J.~A., Winn, J.~N., Bakos, G.~\'A., et al.\ 2011, \apj, 735, 24

\bibitem[Kov\'acs et al.(2002)]{kovacs:2002}
Kov\'acs, G., Zucker, S., \& Mazeh, T.~2002, \aap, 391, 369

\bibitem[Kov\'acs et al.(2005)]{kovacs:2005}
Kov\'acs, G., Bakos, G.~\'A., \& Noyes, R.~W.~2005, \mnras, 356, 557

\bibitem[Kov\'acs et al.(2010)]{kovacs:2010}
Kov\'acs, G., Bakos, G.~\'a., Hartman, J.~D., et al.\ 2010, \apj, 724, 866

\bibitem[Kurucz(2005)]{kurucz:2005} Kurucz, R.~L.\ 2005, Memorie 
della Societa Astronomica Italiana Supplementi, 8, 14

\bibitem[Lasker et al.(2008)]{lasker:2008}
 Lasker, B.~M., et al.\ 2008, \aj, 136, 735

\bibitem[Latham et al.(2009)]{latham:2009} Latham, D.~W., et al.~2009, 
\apj, 704, 1107

\bibitem[Latham et al.(2011)]{latham:2011} Latham, D.~W., Rowe, J.~F., Quinn, S.~N., et al.\ 2011, \apj, 732, L24

\bibitem[Lissauer et al.(2011)]{lissauer:2011}
Lissauer, J.~J., Ragozzine, D., Fabrycky, D.~C., et al.\ 2011, \apjs, 197, 8

\bibitem[Madhusudhan \& Winn(2009)]{madhusudhan:2009} Madhusudhan, N., \& Winn, J.~N.\ 2009, \apj, 693, 784

\bibitem[Makarov et al.(2009)]{makarov:2009} Makarov, V.~V., Beichman, 
C.~A., Catanzarite, J.~H., Fischer, D.~A., Lebreton, J., Malbet, F., 
\& Shao, M.~2009, \apjl, 707, L73 


\bibitem[Mandel \& Agol(2002)]{mandel:2002}
 Mandel, K., \& Agol, E.~2002, \apjl, 580, L171

\bibitem[Marcy \& Butler(1992)]{marcy:1992}
 Marcy, G.~W., \& Butler, R.~P.~1992, \pasp, 104, 270

\bibitem[Mart{\'{\i}}nez-Arn{\'a}iz et 
al.(2010)]{martinez:2010} Mart{\'{\i}}nez-Arn{\'a}iz, 
R., Maldonado, J., Montes, D., Eiroa, C., 
\& Montesinos, B.~2010, \aap, 520, A79 


\bibitem[Maxted et al.(2010)]{maxted:2010} Maxted, P.~F.~L, Anderson, D.~R., Collier Cameron, A., et al.\ 2010, \pasp, 122, 1465

\bibitem[McLaughlin(1924)]{mclaughlin:1924} McLaughlin, D.~B.~1924,
\apj, 60, 22 

\bibitem[Meibom \& Mathieu(2005)]{meibom:2005} Meibom, S., \& Mathieu, R.~D.~2005, 
\apj, 620, 970 

\bibitem[Noguchi et al.(2002)]{noguchi:2002}
Noguchi, K., et al.\ 2002, \pasj, 54, 855

\bibitem[Noyes et al.(1984)]{noyes:1984} Noyes, R.~W., Hartmann, L.~W.,
Baliunas, S.~L., Duncan, D.~K., \& Vaughan, A.~H.~1984, \apj, 279, 763

\bibitem[Noyes et al.(2008)]{noyes:2008} Noyes, R.~W., et al.\  2008, \apjl, 673, L79

\bibitem[O'Donovan et al.(2007)]{odonovan:2007} O'Donovan, F.~T., et  al.\ 2007, \apjl, 663, L37

\bibitem[Ogilvie \& Lin(2004)]{ogilvie:2004} 
Ogilvie, G.~I., \& Lin, D.~N.~C.~2004, \apj, 610, 477 

\bibitem[P\'al(2009a)]{pal:2009a}
P\'al, A.\ 2009a, \mnras, 396, 1737

\bibitem[P\'al(2009b)]{pal:2009b}
P\'al, A.\ 2009b, PhD thesis, Department of Astronomy, E\H{o}tv\H{o}s Lor\'and University, arXiv:0906.3486

\bibitem[P\'al et al.(2010)]{pal:2010}
P\'al, A., et al.\ 2010, \mnras, 401, 2665

\bibitem[Penev \& Sasselov(2011)]{penev:2011} 
Penev, K., \& Sasselov, D.~2011, \apj, 731, 67 

\bibitem[Pollacco et al.(2008)]{pollacco:2008} Pollacco, D., et al.\  2008, \mnras, 385, 1576

\bibitem[Pont et al.(2011)]{pont:2011} Pont, F., Husnoo, N., Mazeh, T., \& Fabrycky, D.\ 2011, \mnras, 414, 1278

\bibitem[Queloz et al.(2001)]{queloz:2001}
Queloz, D.~et al.~2001, \aap, 379, 279

\bibitem[Quinn et al.(2010)]{quinn:2010} Quinn, S.~N., et al.\ 2010, \apj, submitted, arXiv:1008.3565

\bibitem[Rauer(2011)]{rauer:2011} Rauer, H.\ 2011, in ``Detection and 
Dynamics of Transiting Exoplanets, St.~Michel l'Observatoire, France'', 
Edited by F.~Bouchy; R.~D{\'{\i}}az; C.~Moutou; EPJ Web of Conferences, 
Volume 11, id.07001, 11, 7001

\bibitem[Rossiter(1924)]{rossiter:1924} Rossiter, R.~A.~1924, \apj,
60, 15 

\bibitem[Saar et al.(2003)]{saar:2003}
Saar, S.~H., Hatzes, A., Cochran, W., \& Paulson, D.\ 2003, The Future of Cool-Star Astrophysics: 12th Cambridge Workshop on Cool Stars, Stellar Systems, and the Sun , 12, 694

\bibitem[Sato et al.(2002)]{sato:2002}
Sato, B., Kambe, E., Takeda, Y., Izumiura, H., \& Ando, H.\ 2002, \pasj, 54, 873

\bibitem[Sato et al.(2005)]{sato:2005}
Sato, B., et al.\ 2005, \apj, 633, 465

\bibitem[Schlaufman(2010)]{schlaufman:2010}
Schlaufman, K.~C.\ 2010, \apj, 719, 602

\bibitem[Schneider et al.(2011)]{schneider:2011} Schneider, J., Dedieu, 
C., Le Sidaner, P., Savalle, R., \& Zolotukhin, I.~2011, \aap, 532, A79

\bibitem[Skrutskie et al.(2006)]{skrutskie:2006} Skrutskie, M.~F., et 
al.~2006, \aj, 131, 1163

\bibitem[Smalley et al.(2010)]{smalley:2010} Smalley, B., Anderson, D.~R., Collier Cameron, A., et al.\ 2010, \aap, 520, A56

\bibitem[Sozzetti et al.(2007)]{sozzetti:2007}
 Sozzetti, A.~et al.~2007, \apj, 664, 1190

\bibitem[Torres et al.(2007)]{torres:2007}
 Torres, G.~et al.~2007, \apjl, 666, 121

\bibitem[Udalski et  al.(2008)]{udalski:2008} Udalski, A., et al.\ 2008, \aap, 482, 299

\bibitem[Valenti \& Fischer(2005)]{valenti:2005}
 Valenti, J.~A., \& Fischer, D.~A. 2005, \apjs, 159, 141

\bibitem[Valenti \& Piskunov(1996)]{valenti:1996}
 Valenti, J.~A., \& Piskunov, N.~1996, \aaps, 118, 595

\bibitem[Vaughan, Preston \& Wilson(1978)]{vaughan:1978}
Vaughan, A.~H., Preston, G.~W., \& Wilson, O.~C.~1978, \pasp, 90, 267

\bibitem[Vogt et al.(1994)]{vogt:1994}
 Vogt, S.~S.~et al.~1994, Proc.~SPIE, 2198, 362

\bibitem[Winn et al.(2010)]{winn:2010} Winn, J.~N., Fabrycky, D., 
Albrecht, S., \& Johnson, J.~A.~2010, \apjl, 718, L145 

\bibitem[Wright et al.(2011)]{wright:2011} Wright, J.~T., et al.~2011, 
\pasp, 123, 412 

\bibitem[Yi et al.(2001)]{yi:2001}
 Yi, S.~K.~et al.~2001, \apjs, 136, 417

\bibitem[Zahn \& Bouchet(1989)]{zahn:1989a} Zahn, J.-P., \& Bouchet, L.~1989, \aap, 223, 112 

\bibitem[Zahn(1989)]{zahn:1989b} Zahn, J.-P.~1989, \aap, 220, 112 


\bibitem[Wright(2005)]{wright:2005} Wright, J.~T.~2005, \pasp, 117, 657 

\end{thebibliography}
\end{document}